\documentclass[a4paper,11pt]{article}
\usepackage{jheppub} 

\usepackage{tikz}
\usepackage{makecell}
\usepackage{graphicx}

\usepackage{float}

\def\beq#1\eeq{\begin{align}#1\end{align}}

\title{On classification of rank two theories with eight supercharges Part III: Seiberg-Witten geometry}

\author[~]{Dan Xie}

\affiliation[a]{Department of Mathematics, Tsinghua University, Beijing, 100084, China}

\abstract{
We study Seiberg-Witten (SW) geometries for rank-two theories, encompassing 4D field theories as well as  5D and 6D Kaluza-Klein (KK) theories. The singular model for each SW geometry is derived from a one-parameter family of algebraic curves
$y^2 = f(x,t)$,
where $t$ parametrizes one dimension of Coulomb branch moduli space. The functional form of $f(x,t)$ is systematically determined through analysis of singular fibers at $t=\infty$. Two powerful computational methods enable this determination:
a): Liu's algorithm for determining singular fibers from local equation; b): The canonical resolution method for fiber degeneration.
Our construction provides not only a complete description of known solutions but also establishes a robust framework for generating new theories. This methodology proves particularly valuable for the systematic exploration of 5D and 6D theories.}

\begin{document} 
\maketitle
\flushbottom

\section{Introduction}
\begin{figure}
\begin{center}

\tikzset{every picture/.style={line width=0.75pt}} 

\begin{tikzpicture}[x=0.45pt,y=0.45pt,yscale=-1,xscale=1]

\draw   (208,643.36) .. controls (208,588.48) and (311.42,544) .. (439,544) .. controls (566.58,544) and (670,588.48) .. (670,643.36) .. controls (670,698.23) and (566.58,742.72) .. (439,742.72) .. controls (311.42,742.72) and (208,698.23) .. (208,643.36) -- cycle ;
\draw    (408,607) -- (421,620.72) ;
\draw    (406,620.72) -- (420,605.72) ;

\draw    (204,635) -- (217,648.72) ;
\draw    (202,648.72) -- (216,633.72) ;

\draw   (289,472.86) .. controls (289,435.93) and (298.4,406) .. (310,406) .. controls (321.6,406) and (331,435.93) .. (331,472.86) .. controls (331,509.78) and (321.6,539.72) .. (310,539.72) .. controls (298.4,539.72) and (289,509.78) .. (289,472.86) -- cycle ;
\draw    (306,468.72) .. controls (299,463.72) and (293,444.72) .. (314,435.72) ;
\draw    (302,468.72) .. controls (317,464.72) and (322,442.72) .. (310,435.72) ;
\draw    (310,514.72) .. controls (303,509.72) and (297,490.72) .. (318,481.72) ;
\draw    (305,514.72) .. controls (320,510.72) and (325,488.72) .. (313,481.72) ;
\draw  [dash pattern={on 0.84pt off 2.51pt}]  (311,542) -- (312,615.72) ;
\draw  [dash pattern={on 0.84pt off 2.51pt}]  (412,537) -- (413,610.72) ;
\draw   (398,484) .. controls (398,476.27) and (404.27,470) .. (412,470) .. controls (419.73,470) and (426,476.27) .. (426,484) .. controls (426,491.73) and (419.73,498) .. (412,498) .. controls (404.27,498) and (398,491.73) .. (398,484) -- cycle ;
\draw   (427,478) .. controls (427,471.37) and (432.37,466) .. (439,466) .. controls (445.63,466) and (451,471.37) .. (451,478) .. controls (451,484.63) and (445.63,490) .. (439,490) .. controls (432.37,490) and (427,484.63) .. (427,478) -- cycle ;
\draw   (450,473) .. controls (450,466.37) and (455.37,461) .. (462,461) .. controls (468.63,461) and (474,466.37) .. (474,473) .. controls (474,479.63) and (468.63,485) .. (462,485) .. controls (455.37,485) and (450,479.63) .. (450,473) -- cycle ;
\draw   (397,459) .. controls (397,452.37) and (402.37,447) .. (409,447) .. controls (415.63,447) and (421,452.37) .. (421,459) .. controls (421,465.63) and (415.63,471) .. (409,471) .. controls (402.37,471) and (397,465.63) .. (397,459) -- cycle ;
\draw   (394,435) .. controls (394,428.37) and (399.37,423) .. (406,423) .. controls (412.63,423) and (418,428.37) .. (418,435) .. controls (418,441.63) and (412.63,447) .. (406,447) .. controls (399.37,447) and (394,441.63) .. (394,435) -- cycle ;
\draw   (372,486) .. controls (372,479.37) and (377.37,474) .. (384,474) .. controls (390.63,474) and (396,479.37) .. (396,486) .. controls (396,492.63) and (390.63,498) .. (384,498) .. controls (377.37,498) and (372,492.63) .. (372,486) -- cycle ;
\draw   (403,510) .. controls (403,503.37) and (408.37,498) .. (415,498) .. controls (421.63,498) and (427,503.37) .. (427,510) .. controls (427,516.63) and (421.63,522) .. (415,522) .. controls (408.37,522) and (403,516.63) .. (403,510) -- cycle ;

\draw   (190,510) .. controls (190,502.27) and (196.27,496) .. (204,496) .. controls (211.73,496) and (218,502.27) .. (218,510) .. controls (218,517.73) and (211.73,524) .. (204,524) .. controls (196.27,524) and (190,517.73) .. (190,510) -- cycle ;
\draw   (219,504) .. controls (219,497.37) and (224.37,492) .. (231,492) .. controls (237.63,492) and (243,497.37) .. (243,504) .. controls (243,510.63) and (237.63,516) .. (231,516) .. controls (224.37,516) and (219,510.63) .. (219,504) -- cycle ;
\draw   (242,499) .. controls (242,492.37) and (247.37,487) .. (254,487) .. controls (260.63,487) and (266,492.37) .. (266,499) .. controls (266,505.63) and (260.63,511) .. (254,511) .. controls (247.37,511) and (242,505.63) .. (242,499) -- cycle ;
\draw   (189,485) .. controls (189,478.37) and (194.37,473) .. (201,473) .. controls (207.63,473) and (213,478.37) .. (213,485) .. controls (213,491.63) and (207.63,497) .. (201,497) .. controls (194.37,497) and (189,491.63) .. (189,485) -- cycle ;
\draw   (186,461) .. controls (186,454.37) and (191.37,449) .. (198,449) .. controls (204.63,449) and (210,454.37) .. (210,461) .. controls (210,467.63) and (204.63,473) .. (198,473) .. controls (191.37,473) and (186,467.63) .. (186,461) -- cycle ;
\draw   (164,512) .. controls (164,505.37) and (169.37,500) .. (176,500) .. controls (182.63,500) and (188,505.37) .. (188,512) .. controls (188,518.63) and (182.63,524) .. (176,524) .. controls (169.37,524) and (164,518.63) .. (164,512) -- cycle ;
\draw   (195,536) .. controls (195,529.37) and (200.37,524) .. (207,524) .. controls (213.63,524) and (219,529.37) .. (219,536) .. controls (219,542.63) and (213.63,548) .. (207,548) .. controls (200.37,548) and (195,542.63) .. (195,536) -- cycle ;

\draw  [dash pattern={on 0.84pt off 2.51pt}]  (210,562) -- (211,635.72) ;
\draw   (243,261.36) .. controls (243,206.48) and (346.42,162) .. (474,162) .. controls (601.58,162) and (705,206.48) .. (705,261.36) .. controls (705,316.23) and (601.58,360.72) .. (474,360.72) .. controls (346.42,360.72) and (243,316.23) .. (243,261.36) -- cycle ;
\draw    (443,225) -- (456,238.72) ;
\draw    (441,238.72) -- (455,223.72) ;

\draw    (239,253) -- (252,266.72) ;
\draw    (237,266.72) -- (251,251.72) ;

\draw   (330,73.86) .. controls (330,36.93) and (339.4,7) .. (351,7) .. controls (362.6,7) and (372,36.93) .. (372,73.86) .. controls (372,110.78) and (362.6,140.72) .. (351,140.72) .. controls (339.4,140.72) and (330,110.78) .. (330,73.86) -- cycle ;
\draw    (346,69.72) .. controls (339,64.72) and (333,45.72) .. (354,36.72) ;
\draw    (342,69.72) .. controls (357,65.72) and (362,43.72) .. (350,36.72) ;
\draw    (350,115.72) .. controls (343,110.72) and (337,91.72) .. (358,82.72) ;
\draw    (345,115.72) .. controls (360,111.72) and (365,89.72) .. (353,82.72) ;
\draw  [dash pattern={on 0.84pt off 2.51pt}]  (346,160) -- (347,233.72) ;
\draw  [dash pattern={on 0.84pt off 2.51pt}]  (447,155) -- (448,228.72) ;
\draw  [dash pattern={on 0.84pt off 2.51pt}]  (245,180) -- (246,253.72) ;
\draw    (424,43.22) -- (449,131) ;
\draw    (479,46.22) -- (448,131) ;
\draw    (424,44) .. controls (444,16.22) and (438,67.22) .. (479,45.22) ;
\draw    (219,58.22) -- (244,146) ;
\draw    (274,61.22) -- (243,146) ;
\draw    (219,59) .. controls (239,31.22) and (233,82.22) .. (274,60.22) ;

\draw (405,477.4) node [anchor=north west][inner sep=0.75pt]  [font=\tiny]  {$3$};
\draw (436,473.4) node [anchor=north west][inner sep=0.75pt]  [font=\tiny]  {$2$};
\draw (458,467.4) node [anchor=north west][inner sep=0.75pt]  [font=\tiny]  {$1$};
\draw (410,506.4) node [anchor=north west][inner sep=0.75pt]  [font=\tiny]  {$1$};
\draw (380,481.4) node [anchor=north west][inner sep=0.75pt]  [font=\tiny]  {$1$};
\draw (402,453.4) node [anchor=north west][inner sep=0.75pt]  [font=\tiny]  {$2$};
\draw (400,428.4) node [anchor=north west][inner sep=0.75pt]  [font=\tiny]  {$1$};
\draw (165,655.4) node [anchor=north west][inner sep=0.75pt]    {$\infty $};
\draw (192,454.4) node [anchor=north west][inner sep=0.75pt]  [font=\tiny]  {$1$};
\draw (194,479.4) node [anchor=north west][inner sep=0.75pt]  [font=\tiny]  {$2$};
\draw (172,507.4) node [anchor=north west][inner sep=0.75pt]  [font=\tiny]  {$1$};
\draw (202,532.4) node [anchor=north west][inner sep=0.75pt]  [font=\tiny]  {$1$};
\draw (250,493.4) node [anchor=north west][inner sep=0.75pt]  [font=\tiny]  {$1$};
\draw (228,499.4) node [anchor=north west][inner sep=0.75pt]  [font=\tiny]  {$2$};
\draw (197,503.4) node [anchor=north west][inner sep=0.75pt]  [font=\tiny]  {$3$};
\draw (206,241.4) node [anchor=north west][inner sep=0.75pt]    {$\infty $};
\draw (37,217.4) node [anchor=north west][inner sep=0.75pt]    [font=\tiny] {$Singular\ model$};
\draw (31,582.4) node [anchor=north west][inner sep=0.75pt]   [font=\tiny]  {$Smooth\ model$};

\end{tikzpicture}

\end{center}
\caption{Upper: the singular model of Coulomb branch solution; Bottom: the smooth model for the Coulomb branch solution by resolving the singularity of the singular SW geometry.}
\label{singular-smooth}
\end{figure}
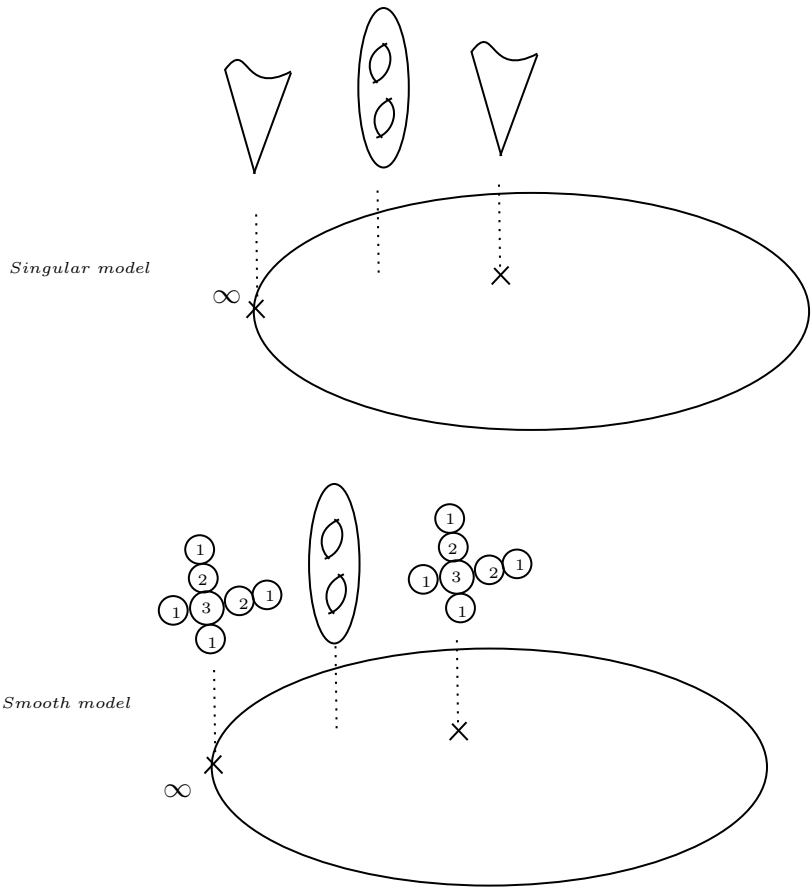

This is the third in a series of papers aimed at classifying rank-two theories with eight supercharges in dimension $d\geq 4$. Our primary strategy is to classify the rank-two Coulomb branch solution, which is encoded by a genus-two Riemann surface fibered over the parameter space of the physical theories. 
In previous studies \cite{Xie:2022aad,Xie:2023wqx}, we mainly studied the so-called \textbf{smooth compact} model for the one parameter Coulomb branch solution. Specifically, we first compactify the parameter space so that the base of the fibration is compact, and then we resolve the singular points on the singular fiber to obtain the so-called relatively minimal model, ensuring no $-1$ curves appear in the fibers. The result is a smooth and compact total space $S$ of the genus-two fibration $f: S \to B$. The physical properties of the theory are derived from this smooth model as follows:
\begin{enumerate}
    \item The generic fiber $F_{s}=f^{-1}(s)$, where $s \in B$, is a smooth genus-two Riemann surface, and the low-energy physics corresponds to a $U(1)^2$ abelian gauge theory. The coupling constants of low energy theory are determined by the complex structure moduli of $f^{-1}(s)$.
    \item The singular fiber $F_{s_0}=f^{-1}(s_0)$ is expressed as $F = \sum n_i C_i$, where $C_i$ are the irreducible components of $F$. From this configuration, one can associate a quiver gauge theory \cite{Xie:2022aad}, which serves as the 3D mirror of the low-energy theory. One could find the 4d low energy theory 
    by looking at 3d mirror.
    \item The singular fiber $F_\infty=f^{-1}(\infty)$ at infinity of $B$ provides information about the UV theory, see table. \ref{UVfiber} for the summary.
\end{enumerate}
See Figure~\ref{singular-smooth} for an illustration of the smooth model of SW solution.

The initial step in our classification program is to categorize the local singular fiber. In Part I~\cite{Xie:2022aad}, we leverage known results \cite{namikawa1973complete, matsumoto2011pseudo} on the one-parameter degeneration of genus-two fibrations to classify possible IR theories. This degeneration is described combinatorially by the weighted graph of a cut system on the Riemann surface, from which one can deduce either a superconformal field theory (SCFT) or an IR-free gauge theory. The possible UV theories are determined by looking at the property of the singular fiber at $\infty$,  see the results listed in Table~\ref{UVfiber}.

\begin{table}[htp]
\begin{center}
\begin{tabular}{|c|c|} \hline
Dimension & Property of dual graph of $F_\infty$ \\ \hline
4d theory & a tree of rational curves \\ \hline
5d theory & a chain of rational curves with one loop \\ \hline
6d theory & a chain of rational curves with two loops \\ \hline
\end{tabular}
\end{center}
\caption{The UV space-time dimensions and the properties of the dual graph of $F_\infty$.}
\label{UVfiber}
\end{table}

In Part II of this series \cite{Xie:2023wqx}, we investigate the global Coulomb branch configuration, ensuring that the singular fibers in the bulk are the simplest $I_1$ singularities. Such configurations are termed Lefschetz pencils. Physically, this correspond to the generic deformations of the physical theory so that one get simplest low energy theory at various singularities. 
By utilizing the fact that local degeneration is classified by the conjugacy classes of the mapping class group, the global geometry reduces to the factorization of the identity element within the mapping class group. Factorizations for many known theories are documented in~\cite{Xie:2023wqx}.

While the previous two papers primarily employed topological methods to study smooth models, providing a holomorphic description of these models proved challenging. For a holomorphic description, it is more effective to use the so-called singular model, obtained by blowing down the divisors of the smooth model. In fact, previous studies on 
the Seiberg-Witten solution is mostly written in the form of those singular models \cite{Seiberg:1994aj, Seiberg:1994rs, Nekrasov:1996cz, Ganor:1996pc}. While it is generally difficult to classify singular model for the SW solution of general genus,
A key result for genus-two fiberations, due to Horikawa \cite{horikawa1976deformations,horikawaalgebraic}, states that every smooth genus-two fibration $f: S \to B$ is birational to a double covering $f: S \to F_e$, branched over a curve $B_0\subset F_e$, where $F_e$ is a $\mathbb{P}^1$ bundle over $B$. When $B = \mathbb{P}^1$,  $F_e$ becomes the Hirzebruch surface. This double covering framework significantly simplifies the 
study of the full SW geometry for rank two theory. Our strategy is following:
\begin{enumerate}
    \item We still start with one parameter family of  SW solution, so that the base $B=\mathbb{P}^1$ for the SW fibration. Now according to Horikawa's result, the singular model for the SW solution is a doubling covering over the surface 
 $F_0 = \mathbb{P}^1 \times \mathbb{P}^1$ \footnote{We will show later that other Hizebruch surface seems not good for the physical reasons.}. Thus, the singular model can be described by the equation
    \begin{equation*}
        y^2 = f(x, t),
    \end{equation*}
    where $x$ and $t$ are local coordinates around $(0,0)$ on $F_0$. The branch locus $B_0$ for the double covering is defined by $f(x, t) = 0$, with divisor class $B_0 \in 6F + 2S$.\footnote{Here, $F$ is the divisor class of the fiber $\mathbb{P}^1$, and $S$ is the divisor class of the base $\mathbb{P}^1$, with intersection numbers $F^2 = 0$, $S^2 = 0$, and $F \cdot S = 1$. Together, they generate the Picard group of $F_0$. Another choice $B_0=6F+4S$ is also possible.} The  fiberation at $t = \infty$ is determined by the local equation
    \begin{equation*}
        y^2 = x'^6 s^2 f(1/x', 1/s),
    \end{equation*}
  where $x^{'}, s$ are the local coordinate near $x,t=\infty$ point, the above transformation automatically determines the singular fiber for the UV theory. 
    \item Numerous one-parameter models remain possible. The special Kähler condition imposes additional constraints. We utilize a result from~\cite{Xie:2023zxn} showing that the full Seiberg-Witten geometry can be derived from the one-parameter family as
    \begin{equation*}
        y^2 = f(x, P) + \text{def},
    \end{equation*}
    where $P$ is a polynomial encoding the Coulomb branch parameters, such as $P = \dots + xv + t$ with $v,t$ the two  Coulomb branch operators, and ``def'' denotes the miniversal deformation of $f(x, t)$, yielding a non-abelian flavor symmetry group (Only ADE type in this paper). 
    \item There are still a large number of possible one parameter models (i.e. the choice of the function $f(x,t)$), and we take a method so that the singularities at the bulk are as singular as possible. This is the opposite limit as taken in part II of this series \cite{Xie:2023wqx}. It is then possible to write down 
    a large number of possible SW geometries which can be identified with the known theories.  
    \item An algorithm to go from the singular model to the smooth model is desirable. Fortunately, Horikawa’s canonical resolution \cite{horikawa1976deformations,horikawaalgebraic} provides such a method. Given the local equation, one can explicitly determine the smooth model for the singular fiber at $\infty$ using the resolution method, with numerous examples available in \ref{double}.
    This method then link the singular models in this paper and the smooth models studied in \cite{Xie:2022aad,Xie:2023wqx}.
    \item In the genus-one case, Tate’s algorithm efficiently identifies singular fiber types from the local equation. For genus two, Liu’s algorithm \cite{liu1993courbes,liu1994conducteur}, along with a Sage package, proves invaluable in determining the singular fiber type, aiding in the construction of the global Seiberg-Witten geometry.
\end{enumerate}
Applying this algorithm to 4D, 5D, and 6D theories, we uncover many intriguing new Seiberg-Witten geometries, detailed in tables~[\ref{4da}, \ref{4db}, \ref{4dc}] for 4d theories (see also \cite{Argyres:2022lah, Argyres:2022puv,Argyres:2022fwy} for the SW geometry of 4d SCFT cases), tables~[\ref{5da},\ref{5db}] for 5d theories, and tables. [\ref{6da},\ref{6db}] for 6d theories.

This paper is organized as follows: Section 2 reviews the fundamentals of double coverings of algebraic surfaces and solutions to the special Kähler condition; Section 3 explores the Seiberg-Witten geometry for rank-one theories; Section 4 addresses the Seiberg-Witten geometry for rank-two theories; and Section 5 provides a conclusion.

\section{Double covering and Seiberg-Witten geometry}

The Seiberg-Witten geometry of a four-dimensional $\mathcal{N}=2$ supersymmetric theory is typically derived using string/M-theory constructions \cite{Witten:1997sc} or its connection to integrable systems \cite{Martinec:1995by,Donagi:1995cf}. Once the Seiberg-Witten (SW) geometry is identified, the corresponding SW differential must be carefully computed. In \cite{Xie:2023zxn}, we propose a method to determine the SW geometry from a 
\textbf{one-parameter} family of algebraic curve, where the parameter often corresponds to the Coulomb branch operator with the maximal scaling dimension. We assume the SW geometry is described by hyperelliptic families in this paper.

By compactifying the one-parameter Coulomb branch and incorporating the singular fiber at infinity, one obtains an algebraic surface $S$ fibered over $\mathbb{P}^1$. Such fibered algebraic surfaces \cite{horikawa1976deformations,horikawaalgebraic}, defined by hyperelliptic families, have been extensively studied, particularly in terms of their birational geometry, which is now well understood. Several useful models arise in this context:

\begin{enumerate}
    \item \textbf{Relative Minimal Model}: Here, the surface $S$ is non-singular, and the singular fiber $F = \sum n_i C_i$ contains no $-1$ curves.
    \item \textbf{Normally Minimal Model}: In this case, the surface $S$ is also non-singular, and moreover the components of the singular fiber intersect normally. Unlike relative minimal model, one often need to introduce $-1$ curve in the singular fiber.
    \item \textbf{Singular Model}: This model describes the full family via a single algebraic equation, $y^2 = f(x,t)$, where $t$ parameterizes the base $\mathbb{P}^1$.
\end{enumerate}

Each model has its strengths and weaknesses. The relative minimal model is valuable for analyzing the low-energy theory via three-dimensional mirror symmetry, though it resists a simple global description. The normally minimal model, closely related to the relative minimal model through simple blow-ups or contractions of $-1$ curves, facilitates certain computations, such as determining Euler numbers. The singular model, while straightforward to express globally, complicates efforts to extract the low-energy theory.

Linking the singular and smooth models is generally challenging. However, for hyperelliptic families, two key results enable systematic study \cite{horikawa1976deformations,horikawaalgebraic}: (a) Every smooth model can be related to a singular model through blow-ups and contractions, with the singular model expressible as a double covering over the Hirzebruch surface $F_e$. The double covering allows the singular model to be written explicitly. (b) Given a singular model defined by a double covering of a Hirzebruch surface, one can apply the canonical resolution to recover the relative or normally minimal model. In the following sections, we will give many explicit examples illustrating how to go from singular model to smooth model, and vice versa.

Leveraging these results, we classify the rank two SW geometry as follows:
\begin{enumerate}
    \item We classify algebraic surfaces defined by a double covering over the Hirzebruch surface $F_0$.\footnote{While double coverings over general Hirzebruch surfaces $F_e$ are possible, detailed analysis for SW geometry reveals that $F_0$ suffices.} Here, $F_0 \cong \mathbb{P}^1 \times \mathbb{P}^1$. Using local coordinates $(x,t)$ around the point $(0,0)$ of $F_0$, the double covering is given by
    \begin{equation*}
        y^2 = f(x,t).
    \end{equation*}
    The singular fiber at $t = \infty$ is determined by the divisor class of the branched locus $B$, defined by $f = 0$ in the $(x,t)$ patch.
    \item To analyze the singular fiber at infinity, one may employ Horikawa’s canonical resolution or, more conveniently, Liu’s arithmetic method \cite{liu1993courbes, liu1994conducteur}, which can be implemented in Sage \cite{sagemath}. We will try to find all the families whose fiber at $\infty$ will recover those analyzed in \cite{Xie:2022aad,Xie:2023wqx}.
    \item  For a given singular fiber type at $\infty$, there might be many possible one parameter family; we further require that the singular fiber at the bulk to be as simple as enough.
\end{enumerate}
Using above strategy, we can determine the SW geometry for almost all the rank two theories in various dimensions.

\subsection{Singular model from double covering}

Consider a genus two fibered algebraic surface $\pi: S \to \Delta$. After a birational transformation, it can be described as a double covering $f: S' \to W_e$, branched over a curve $B$ in $W_e$, where $W_e$ is a Hirzebruch surface if the base $\Delta$ of the fibered surface is $\mathbb{P}^1$. The surface $S'$ is realized as a hypersurface in the total space of the line bundle defined by the branch curve $B$, with the equation given by
\[
y^2 = f(x, t),
\]
where $f = 0$ defines the branch curve $B\subset W_e$. Here, $(x, t)$ are local coordinates around $(0, 0)$ in $W_e$, and $t$ denotes the coordinate on the base $B$, so that the fibration $\Gamma: W_e \to B$ is given by the trivial projection.

We will take $W_e=F_0$ as this will be the main case for us. $F_0$ is simply the product $\mathbb{P}^1\times \mathbb{P}^1$, and here we review some of its properties.
The divisor class of $F_0$ is spanned by $F$ and $S$, where $F$ is the fiber and $S$ is a horizontal section. Their intersection numbers are $F^2 = S^2 = 0$ and $S \cdot F = 1$. The divisor class of a general curve $B$ is expressed as $[B] = nF + mS$, and we use $(n,m)$ to denote its divisor class. 
The coordinate $y$ is a section of the line bundle associated with the divisor $\frac{[B]}{2}$, implying that $[B]$ must be an even divisor.

For applications to Seiberg-Witten geometry, the surface $S$ must be rational, i.e., $q(S) = p_g(S) = 0$. The Euler number and Chern numbers $(c_1^2, c_2)$ can be computed from the data of $F_0$ and $[B]$ (see Appendix \ref{canonical}). The rationality condition imposes strong constraints on the divisor class $[B]$. As we will see, the most common cases are $[B] = (6, 2)$ or $[B] = (6, 4)$ (see Appendix \ref{double}) . Below, we illustrate how to determine the singular fiber at $t = \infty$ and thereby deduce properties of the UV theory.

\textbf{Case $[B] = (6, 2)$}: Consider the case where $B$ is a $(6, 2)$ curve. The equation for $S'$ becomes
\[
y^2 = h_0(x)t^2 + h_1(x)t + h_2(x),
\]
where $(x, t)$ are local coordinates around $(0, 0)$ in $F_0$, the polynomials $h_i(x)$ satisfy $\deg(h_i) \leq 6$, and $t$ is regarded as the Coulomb branch coordinate. In the patch around $(\infty, \infty)$, with coordinates $(x', s)$ which is related to the original local coordinates by $x' = \frac{1}{x}$ and $s = \frac{1}{t}$, the curve transforms to
\[
y^2 = h_0'(x') + h_1'(x')s + h_2'(x')s^2,
\]
where $h_i'(x') = x'^6 h_i\left(\frac{1}{x'}\right)$.

\textbf{Example}: Take $f = x^6 + t$. At infinity, the curve becomes $f = s^2 + s x'^6$.

\textbf{Case $[B] = (2m, 2n)$}: For a general branch curve with divisor class $[B] = 2mF + 2nS$, the equation in the $(0, 0)$ coordinate patch is
\[
\boxed{f(x, t) = \sum_{i=0}^{2n} h_i(x) t^i},
\]
and in the $(\infty, \infty)$ patch, it becomes
\[
\boxed{f(x^{'}, s) = \sum_{i=0}^{2n} x^{'2m} h_i\left(\frac{1}{x^{'}}\right) s^{2n - i}},
\]
where we use the letters $(x^{'}, s)$ to denote coordinates around $(\infty, \infty)$ for simplicity.

To determine the singular fiber type at $t = \infty$, one can employ either the canonical resolution method (discussed in the next subsection) or an arithmetic approach. In the genus one case, Tate’s algorithm uses the discriminant and $j$-invariant to classify singular fibers. For genus two, the situation is more complex, but a similar algorithm, developed by Liu \cite{liu1993courbes, liu1994conducteur}, generalizes Tate's algorithm using Igusa invariants (analogs of the discriminant and $j$-invariant). The singular fiber type is determined from the exponents of these invariants. Although computing Igusa invariants is challenging, a Sage package facilitates this process.

\subsection{Smooth model from canonical resolution}
Once the singular model is established, the relative minimal model for the singular fiber can be determined using the canonical resolution method. Below, we outline the detailed procedure. Consider a hyperelliptic family defined locally by $y^2 = f(x,t)$, with our focus on the singular fiber at $t = 0$. The double covering is given by $f: S \to \Delta$, where $\Delta$ is the locus defined by $f(x,t) = 0$. The second projection to the Coulomb branch operator is $\Gamma: \Delta \to B$, defined by $(x,t) \mapsto t$. The smooth model for the singular fiber $F_{t=0}$ is obtained from the double covering as follows: first, identify the divisor $\Gamma_0 = \Gamma^{-1}(0)$, defined by $t = 0$ in $\Delta$, and then compute the singular fiber $F_{t=0}$ as the pullback $f^{-1}(\Gamma^{-1}(0))$. The canonical resolution proceeds as follows:

\begin{enumerate}
    \item \textbf{Identify the bad poinst}: The fiber is singular if and only if $f(x,t)$ has multiple roots in $x$ at $t = 0$: equation $f(x,0)=0$ has multiple roots. Define the bad points as $(x_i, 0)$, where $x_i$ are the multiple roots of $f(x,0)$. These points are singular points on curve $y^2=f(x,0) $, notice that bad points may arise even if the surface is smooth at those locations.
    \item \textbf{Perform the blow-up}: For each bad point $(x_i, 0)$, perform a blow-up $\sigma$ centered at that point (This is the process performed at the base space $\Delta$.). The new double covering is defined with the divisor class of the  branched locus given by:
    \[
    [B'] = \sigma^*[B] - 2\left\lfloor \frac{m_p}{2} \right\rfloor E_p,
    \]
    where $\sigma^*[B]$ is the total transform of the branch locus $B$, $E_p$ is the exceptional locus, and $m_p$ is the multiplicity of $B$ at the bad point $P$. If $m_p$ is even, the new branched locus is the proper transform of $B$; if $m_p$ is odd, it includes the proper transform of $B$ plus the exceptional locus $E_p$.
    \item \textbf{Iterate the process}: Repeat the above steps until no bad points remain.A bad point is determined by looking at the intersection number of the proper transform of $\Gamma_0$ and the new branch locus (which might include the exceptional locus).
    \end{enumerate}

\begin{figure}
\begin{center}

\tikzset{every picture/.style={line width=0.75pt}} 

\begin{tikzpicture}[x=0.55pt,y=0.55pt,yscale=-1,xscale=1]

\draw    (480,160) -- (558,160) ;
\draw [shift={(560,160)}, rotate = 180] [color={rgb, 255:red, 0; green, 0; blue, 0 }  ][line width=0.75]    (10.93,-3.29) .. controls (6.95,-1.4) and (3.31,-0.3) .. (0,0) .. controls (3.31,0.3) and (6.95,1.4) .. (10.93,3.29)   ;
\draw    (376,160) -- (454,160) ;
\draw [shift={(456,160)}, rotate = 180] [color={rgb, 255:red, 0; green, 0; blue, 0 }  ][line width=0.75]    (10.93,-3.29) .. controls (6.95,-1.4) and (3.31,-0.3) .. (0,0) .. controls (3.31,0.3) and (6.95,1.4) .. (10.93,3.29)   ;
\draw    (579,175) -- (579,238) ;
\draw [shift={(579,240)}, rotate = 270] [color={rgb, 255:red, 0; green, 0; blue, 0 }  ][line width=0.75]    (10.93,-3.29) .. controls (6.95,-1.4) and (3.31,-0.3) .. (0,0) .. controls (3.31,0.3) and (6.95,1.4) .. (10.93,3.29)   ;
\draw    (480,252) -- (558,252) ;
\draw [shift={(560,252)}, rotate = 180] [color={rgb, 255:red, 0; green, 0; blue, 0 }  ][line width=0.75]    (10.93,-3.29) .. controls (6.95,-1.4) and (3.31,-0.3) .. (0,0) .. controls (3.31,0.3) and (6.95,1.4) .. (10.93,3.29)   ;
\draw    (468,176) -- (468,239) ;
\draw [shift={(468,241)}, rotate = 270] [color={rgb, 255:red, 0; green, 0; blue, 0 }  ][line width=0.75]    (10.93,-3.29) .. controls (6.95,-1.4) and (3.31,-0.3) .. (0,0) .. controls (3.31,0.3) and (6.95,1.4) .. (10.93,3.29)   ;
\draw    (373,255) -- (451,255) ;
\draw [shift={(453,255)}, rotate = 180] [color={rgb, 255:red, 0; green, 0; blue, 0 }  ][line width=0.75]    (10.93,-3.29) .. controls (6.95,-1.4) and (3.31,-0.3) .. (0,0) .. controls (3.31,0.3) and (6.95,1.4) .. (10.93,3.29)   ;
\draw    (360,178) -- (360,241) ;
\draw [shift={(360,243)}, rotate = 270] [color={rgb, 255:red, 0; green, 0; blue, 0 }  ][line width=0.75]    (10.93,-3.29) .. controls (6.95,-1.4) and (3.31,-0.3) .. (0,0) .. controls (3.31,0.3) and (6.95,1.4) .. (10.93,3.29)   ;
\draw [line width=1.5]  [dash pattern={on 5.63pt off 4.5pt}]  (240,160) -- (331,160) ;
\draw [line width=1.5]  [dash pattern={on 5.63pt off 4.5pt}]  (247,256) -- (338,256) ;
\draw    (216,176) -- (216,239) ;
\draw [shift={(216,241)}, rotate = 270] [color={rgb, 255:red, 0; green, 0; blue, 0 }  ][line width=0.75]    (10.93,-3.29) .. controls (6.95,-1.4) and (3.31,-0.3) .. (0,0) .. controls (3.31,0.3) and (6.95,1.4) .. (10.93,3.29)   ;

\draw (573,150.4) node [anchor=north west][inner sep=0.75pt]    {$S$};
\draw (570,248.4) node [anchor=north west][inner sep=0.75pt]    {$\Delta $};
\draw (458,255.4) node [anchor=north west][inner sep=0.75pt]    {$\Delta _{1}$};
\draw (461,151.4) node [anchor=north west][inner sep=0.75pt]    {$S_{1}$};
\draw (346,260.4) node [anchor=north west][inner sep=0.75pt]    {$\Delta _{2}$};
\draw (349,152.4) node [anchor=north west][inner sep=0.75pt]    {$S_{2}$};
\draw (210,151.4) node [anchor=north west][inner sep=0.75pt]    {$S_{n}$};
\draw (207,249.4) node [anchor=north west][inner sep=0.75pt]    {$\Delta _{n}$};

\end{tikzpicture}

\end{center}
\caption{Canonical resolution of double covered surface.}
\label{cresolution}
\end{figure}
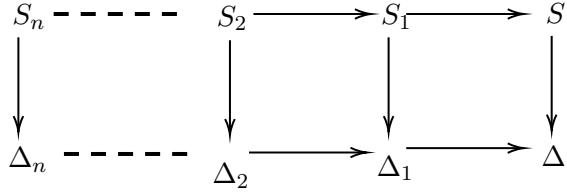

After completing the canonical resolution, the divisor $\Gamma_0^*$ and branched locus $B^*$ are expressed as:
\begin{align*}
    B^* &= \tilde{B} + \sum \alpha_i E_i, \\
    \Gamma_0^* &= \tilde{\Gamma_0} + \sum \beta_i E_i,
\end{align*}
where $\alpha_i \in \{0, 1\}$, and $\beta_i$ is a non-negative integer, both to be determined carefully. The normal minimal model for the singular fiber is the double covering over $\Gamma_0^*$, with its multiplicity and intersection numbers derived from the configuration of exceptional curves. The rules for  $\pi^{-1}(E)$ ($E$ is a component of $(\Gamma_0^*)$ with multiplicity  $n$) are:

\begin{itemize}
    \item If $E$ is not a component of $B^*$ and does not intersect $B^*$, then $\pi^{-1}(E) = C_1 + C_2$, with $C_1 \cdot C_2 = 0$ and $C_1^2 = C_2^2 = E^2$. So one get two components in the singular fiber.
    \item If $E$ is not a component of $B^*$ and intersects $B^*$ transversely at odd divisor \footnote{The intersection is not equal to $2Q$, with $Q$ a divisor on the curve.}, then $\pi^{-1}(E) = C$, with self-intersection $C^2 = 2E^2$. 
        \item If $E$ is not a component of $B^*$ and intersects $B^*$ transversely at even divisor \footnote{The intersection is equal to $2Q$, with $Q$ a divisor on the curve.}, then $\pi^{-1}(E) = D_1+D_2$, with self-intersection $D_i^2=E^2-deg(Q),~D_1\cdot D_2=deg(Q)$. 
    \item If $E$ is a component of $B^*$, then $\pi^{-1}(E) = 2D$, with intersection number $D^2 = \frac{E^2}{2}$.
\end{itemize}

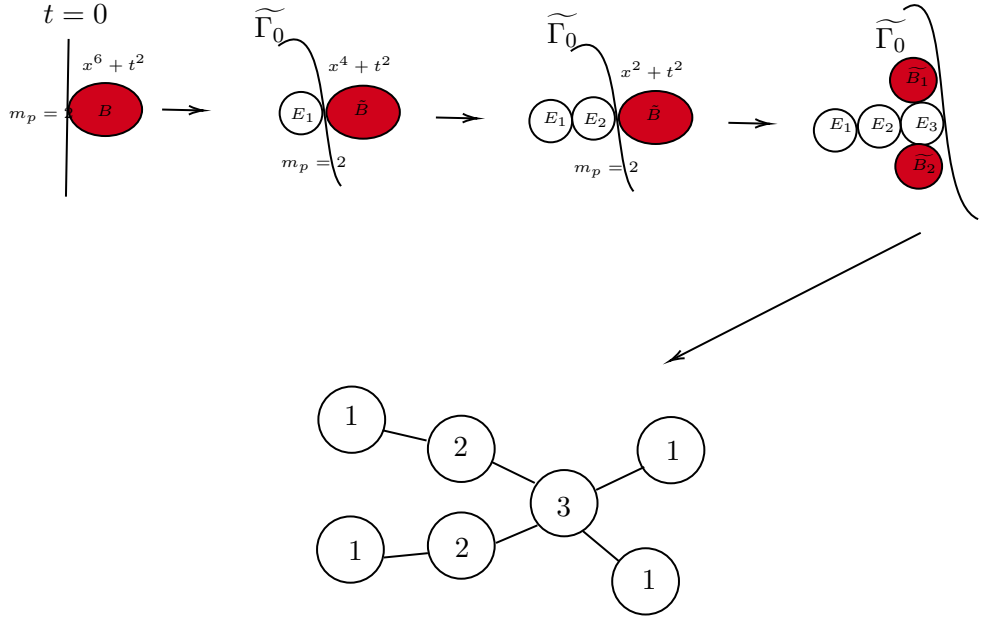
\begin{figure}[H]
    \centering
  \tikzset{every picture/.style={line width=0.75pt}} 

\begin{tikzpicture}[x=0.50pt,y=0.50pt,yscale=-1,xscale=1]

\draw    (142,60.72) -- (140,179.72) ;
\draw  [fill={rgb, 255:red, 208; green, 2; blue, 27 }  ,fill opacity=1 ] (142,114) .. controls (142,102.95) and (154.31,94) .. (169.5,94) .. controls (184.69,94) and (197,102.95) .. (197,114) .. controls (197,125.05) and (184.69,134) .. (169.5,134) .. controls (154.31,134) and (142,125.05) .. (142,114) -- cycle ;
\draw    (212,114) -- (241,114.67) ;
\draw [shift={(243,114.72)}, rotate = 181.33] [color={rgb, 255:red, 0; green, 0; blue, 0 }  ][line width=0.75]    (10.93,-3.29) .. controls (6.95,-1.4) and (3.31,-0.3) .. (0,0) .. controls (3.31,0.3) and (6.95,1.4) .. (10.93,3.29)   ;
\draw  [fill={rgb, 255:red, 208; green, 2; blue, 27 }  ,fill opacity=1 ] (335,116) .. controls (335,104.95) and (347.31,96) .. (362.5,96) .. controls (377.69,96) and (390,104.95) .. (390,116) .. controls (390,127.05) and (377.69,136) .. (362.5,136) .. controls (347.31,136) and (335,127.05) .. (335,116) -- cycle ;
\draw    (417,120) -- (446,120.67) ;
\draw [shift={(448,120.72)}, rotate = 181.33] [color={rgb, 255:red, 0; green, 0; blue, 0 }  ][line width=0.75]    (10.93,-3.29) .. controls (6.95,-1.4) and (3.31,-0.3) .. (0,0) .. controls (3.31,0.3) and (6.95,1.4) .. (10.93,3.29)   ;
\draw   (300,116.72) .. controls (300,107.88) and (307.16,100.72) .. (316,100.72) .. controls (324.84,100.72) and (332,107.88) .. (332,116.72) .. controls (332,125.56) and (324.84,132.72) .. (316,132.72) .. controls (307.16,132.72) and (300,125.56) .. (300,116.72) -- cycle ;
\draw    (299,66) .. controls (339,36) and (329,152.72) .. (346,171.72) ;
\draw  [fill={rgb, 255:red, 208; green, 2; blue, 27 }  ,fill opacity=1 ] (554,120) .. controls (554,108.95) and (566.31,100) .. (581.5,100) .. controls (596.69,100) and (609,108.95) .. (609,120) .. controls (609,131.05) and (596.69,140) .. (581.5,140) .. controls (566.31,140) and (554,131.05) .. (554,120) -- cycle ;
\draw    (636,124) -- (665,124.67) ;
\draw [shift={(667,124.72)}, rotate = 181.33] [color={rgb, 255:red, 0; green, 0; blue, 0 }  ][line width=0.75]    (10.93,-3.29) .. controls (6.95,-1.4) and (3.31,-0.3) .. (0,0) .. controls (3.31,0.3) and (6.95,1.4) .. (10.93,3.29)   ;
\draw   (519,120.72) .. controls (519,111.88) and (526.16,104.72) .. (535,104.72) .. controls (543.84,104.72) and (551,111.88) .. (551,120.72) .. controls (551,129.56) and (543.84,136.72) .. (535,136.72) .. controls (526.16,136.72) and (519,129.56) .. (519,120.72) -- cycle ;
\draw    (518,70) .. controls (558,40) and (548,156.72) .. (565,175.72) ;
\draw   (487,122.72) .. controls (487,113.88) and (494.16,106.72) .. (503,106.72) .. controls (511.84,106.72) and (519,113.88) .. (519,122.72) .. controls (519,131.56) and (511.84,138.72) .. (503,138.72) .. controls (494.16,138.72) and (487,131.56) .. (487,122.72) -- cycle ;
\draw  [fill={rgb, 255:red, 208; green, 2; blue, 27 }  ,fill opacity=1 ] (757,91.86) .. controls (757,82.55) and (764.84,75) .. (774.5,75) .. controls (784.16,75) and (792,82.55) .. (792,91.86) .. controls (792,101.17) and (784.16,108.72) .. (774.5,108.72) .. controls (764.84,108.72) and (757,101.17) .. (757,91.86) -- cycle ;
\draw   (765,124.72) .. controls (765,115.88) and (772.16,108.72) .. (781,108.72) .. controls (789.84,108.72) and (797,115.88) .. (797,124.72) .. controls (797,133.56) and (789.84,140.72) .. (781,140.72) .. controls (772.16,140.72) and (765,133.56) .. (765,124.72) -- cycle ;
\draw    (767,38.72) .. controls (807,8.72) and (785,180.72) .. (823,196.72) ;
\draw   (733,126.72) .. controls (733,117.88) and (740.16,110.72) .. (749,110.72) .. controls (757.84,110.72) and (765,117.88) .. (765,126.72) .. controls (765,135.56) and (757.84,142.72) .. (749,142.72) .. controls (740.16,142.72) and (733,135.56) .. (733,126.72) -- cycle ;
\draw   (700,129.72) .. controls (700,120.88) and (707.16,113.72) .. (716,113.72) .. controls (724.84,113.72) and (732,120.88) .. (732,129.72) .. controls (732,138.56) and (724.84,145.72) .. (716,145.72) .. controls (707.16,145.72) and (700,138.56) .. (700,129.72) -- cycle ;
\draw  [fill={rgb, 255:red, 208; green, 2; blue, 27 }  ,fill opacity=1 ] (761,156.86) .. controls (761,147.55) and (768.84,140) .. (778.5,140) .. controls (788.16,140) and (796,147.55) .. (796,156.86) .. controls (796,166.17) and (788.16,173.72) .. (778.5,173.72) .. controls (768.84,173.72) and (761,166.17) .. (761,156.86) -- cycle ;
\draw    (780,207) -- (593.78,302.8) ;
\draw [shift={(592,303.72)}, rotate = 332.78] [color={rgb, 255:red, 0; green, 0; blue, 0 }  ][line width=0.75]    (10.93,-3.29) .. controls (6.95,-1.4) and (3.31,-0.3) .. (0,0) .. controls (3.31,0.3) and (6.95,1.4) .. (10.93,3.29)   ;
\draw   (329,348) .. controls (329,334.19) and (340.19,323) .. (354,323) .. controls (367.81,323) and (379,334.19) .. (379,348) .. controls (379,361.81) and (367.81,373) .. (354,373) .. controls (340.19,373) and (329,361.81) .. (329,348) -- cycle ;
\draw   (328,446) .. controls (328,432.19) and (339.19,421) .. (353,421) .. controls (366.81,421) and (378,432.19) .. (378,446) .. controls (378,459.81) and (366.81,471) .. (353,471) .. controls (339.19,471) and (328,459.81) .. (328,446) -- cycle ;
\draw   (411,370) .. controls (411,356.19) and (422.19,345) .. (436,345) .. controls (449.81,345) and (461,356.19) .. (461,370) .. controls (461,383.81) and (449.81,395) .. (436,395) .. controls (422.19,395) and (411,383.81) .. (411,370) -- cycle ;
\draw   (488,411) .. controls (488,397.19) and (499.19,386) .. (513,386) .. controls (526.81,386) and (538,397.19) .. (538,411) .. controls (538,424.81) and (526.81,436) .. (513,436) .. controls (499.19,436) and (488,424.81) .. (488,411) -- cycle ;
\draw    (527,432) -- (554,454.72) ;
\draw    (537,401) -- (573,383.72) ;
\draw    (377,356) -- (410,363.72) ;
\draw   (568,371) .. controls (568,357.19) and (579.19,346) .. (593,346) .. controls (606.81,346) and (618,357.19) .. (618,371) .. controls (618,384.81) and (606.81,396) .. (593,396) .. controls (579.19,396) and (568,384.81) .. (568,371) -- cycle ;
\draw   (549,470) .. controls (549,456.19) and (560.19,445) .. (574,445) .. controls (587.81,445) and (599,456.19) .. (599,470) .. controls (599,483.81) and (587.81,495) .. (574,495) .. controls (560.19,495) and (549,483.81) .. (549,470) -- cycle ;
\draw   (411,443) .. controls (411,429.19) and (422.19,418) .. (436,418) .. controls (449.81,418) and (461,429.19) .. (461,443) .. controls (461,456.81) and (449.81,468) .. (436,468) .. controls (422.19,468) and (411,456.81) .. (411,443) -- cycle ;
\draw    (378,452) -- (411,448.72) ;
\draw    (462,441) -- (493,427.72) ;
\draw    (459,380) -- (491,395.72) ;

\draw (121,32.4) node [anchor=north west][inner sep=0.75pt]    {$t=0$};
\draw (150,70.4) node [anchor=north west][inner sep=0.75pt]  [font=\tiny]  {$x^{6} +t^{2}$};
\draw (161,108.4) node [anchor=north west][inner sep=0.75pt]    [font=\tiny]  {$B$};
\draw (95,110.4) node [anchor=north west][inner sep=0.75pt]  [font=\tiny]  {$m_{p} =2$};
\draw (299,147.4) node [anchor=north west][inner sep=0.75pt]  [font=\tiny]  {$m_{p} =2$};
\draw (306,110.4) node [anchor=north west][inner sep=0.75pt]    [font=\tiny]  {$E_{1}$};
\draw (279,35.4) node [anchor=north west][inner sep=0.75pt]    {$\widetilde{\Gamma _{0}}$};
\draw (353,104.4) node [anchor=north west][inner sep=0.75pt]     [font=\tiny] {$\tilde{B}$};
\draw (334,71.4) node [anchor=north west][inner sep=0.75pt]  [font=\tiny]  {$x^{4} +t^{2}$};
\draw (518,151.4) node [anchor=north west][inner sep=0.75pt]  [font=\tiny]  {$m_{p} =2$};
\draw (525,114.4) node [anchor=north west][inner sep=0.75pt]    [font=\tiny]  {$E_{2}$};
\draw (498,39.4) node [anchor=north west][inner sep=0.75pt]    {$\widetilde{\Gamma _{0}}$};
\draw (572,108.4) node [anchor=north west][inner sep=0.75pt]    [font=\tiny]  {$\tilde{B}$};
\draw (553,75.4) node [anchor=north west][inner sep=0.75pt]  [font=\tiny]  {$x^{2} +t^{2}$};
\draw (495,113.4) node [anchor=north west][inner sep=0.75pt]     [font=\tiny] {$E_{1}$};
\draw (740,118.4) node [anchor=north west][inner sep=0.75pt]    [font=\tiny]  {$E_{2}$};
\draw (744,43.4) node [anchor=north west][inner sep=0.75pt]    {$\widetilde{\Gamma _{0}}$};
\draw (766,79.4) node [anchor=north west][inner sep=0.75pt]    [font=\tiny]  {$\widetilde{B_{1}}$};
\draw (709,118.4) node [anchor=north west][inner sep=0.75pt]     [font=\tiny] {$E_{1}$};
\draw (773,117.4) node [anchor=north west][inner sep=0.75pt]    [font=\tiny]  {$E_{3}$};
\draw (771,145.4) node [anchor=north west][inner sep=0.75pt]    [font=\tiny]  {$\widetilde{B_{2}}$};
\draw (428,359.4) node [anchor=north west][inner sep=0.75pt]    {$2$};
\draw (346,335.4) node [anchor=north west][inner sep=0.75pt]    {$1$};
\draw (505,404.4) node [anchor=north west][inner sep=0.75pt]    {$3$};
\draw (429,435.4) node [anchor=north west][inner sep=0.75pt]    {$2$};
\draw (349,437.4) node [anchor=north west][inner sep=0.75pt]    {$1$};
\draw (587,362.4) node [anchor=north west][inner sep=0.75pt]    {$1$};
\draw (569,459.4) node [anchor=north west][inner sep=0.75pt]    {$1$};

\end{tikzpicture}

    \caption{Singular fiber configuration for $f(x,t) = x^6 + t^2$ after canonical resolution. a): Step 1: the multiplicity is $m_p=2$; b): the local equation for the singularity becomes $x^4+t^2=0$, and the multiplicity is $m_p=2$; c: the local equation for the singularity becomes $x^2+t^2=0$, and the multiplicity is $m_p=2$; d): There is no singularity left. 
      }
    \label{fig:ex1}
\end{figure}

\textbf{Example 1}:
Consider the branched locus $f(x,t) = x^6 + t^2$, with a bad point at $(0,0)$. The divisor $\Gamma_0$ is defined by $t = 0$ (see Figure~\ref{fig:ex1}). After the canonical resolution, the zero divisor and branched locus become:
\begin{align*}
    B^* &= \tilde{B}, \\
    \Gamma_0^* &= \tilde{\Gamma_0} + E_1 + 2E_2 + 3E_3,
\end{align*}
with the singular fiber configuration shown in Figure~\ref{fig:ex1}. $E_1, E_2,\tilde{\Gamma}_0$ do not belong to $B^*$ and do not intersect with $B^*$, and so each of them consists of two components in the pre-image, with the multiplicity $1,2,1$ respectively.  $E_3$ is not a component of $B^*$ and does intersect $B^*$ in two points, and 
so $k=1$, its preimage is a genus zero curve and the multiplicty is $3$. 

\begin{figure}[H]
    \centering
\tikzset{every picture/.style={line width=0.75pt}} 

\begin{tikzpicture}[x=0.50pt,y=0.50pt,yscale=-1,xscale=1]

\draw    (142,60.72) -- (140,179.72) ;
\draw  [fill={rgb, 255:red, 208; green, 2; blue, 27 }  ,fill opacity=1 ] (142,114) .. controls (142,102.95) and (154.31,94) .. (169.5,94) .. controls (184.69,94) and (197,102.95) .. (197,114) .. controls (197,125.05) and (184.69,134) .. (169.5,134) .. controls (154.31,134) and (142,125.05) .. (142,114) -- cycle ;
\draw    (212,114) -- (241,114.67) ;
\draw [shift={(243,114.72)}, rotate = 181.33] [color={rgb, 255:red, 0; green, 0; blue, 0 }  ][line width=0.75]    (10.93,-3.29) .. controls (6.95,-1.4) and (3.31,-0.3) .. (0,0) .. controls (3.31,0.3) and (6.95,1.4) .. (10.93,3.29)   ;
\draw  [fill={rgb, 255:red, 208; green, 2; blue, 27 }  ,fill opacity=1 ] (335,116) .. controls (335,104.95) and (347.31,96) .. (362.5,96) .. controls (377.69,96) and (390,104.95) .. (390,116) .. controls (390,127.05) and (377.69,136) .. (362.5,136) .. controls (347.31,136) and (335,127.05) .. (335,116) -- cycle ;
\draw    (417,120) -- (446,120.67) ;
\draw [shift={(448,120.72)}, rotate = 181.33] [color={rgb, 255:red, 0; green, 0; blue, 0 }  ][line width=0.75]    (10.93,-3.29) .. controls (6.95,-1.4) and (3.31,-0.3) .. (0,0) .. controls (3.31,0.3) and (6.95,1.4) .. (10.93,3.29)   ;
\draw  [fill={rgb, 255:red, 208; green, 2; blue, 27 }  ,fill opacity=1 ] (300,116.72) .. controls (300,107.88) and (307.16,100.72) .. (316,100.72) .. controls (324.84,100.72) and (332,107.88) .. (332,116.72) .. controls (332,125.56) and (324.84,132.72) .. (316,132.72) .. controls (307.16,132.72) and (300,125.56) .. (300,116.72) -- cycle ;
\draw    (299,66) .. controls (339,36) and (329,152.72) .. (346,171.72) ;
\draw  [fill={rgb, 255:red, 208; green, 2; blue, 27 }  ,fill opacity=1 ] (554,120) .. controls (554,108.95) and (566.31,100) .. (581.5,100) .. controls (596.69,100) and (609,108.95) .. (609,120) .. controls (609,131.05) and (596.69,140) .. (581.5,140) .. controls (566.31,140) and (554,131.05) .. (554,120) -- cycle ;
\draw    (636,124) -- (665,124.67) ;
\draw [shift={(667,124.72)}, rotate = 181.33] [color={rgb, 255:red, 0; green, 0; blue, 0 }  ][line width=0.75]    (10.93,-3.29) .. controls (6.95,-1.4) and (3.31,-0.3) .. (0,0) .. controls (3.31,0.3) and (6.95,1.4) .. (10.93,3.29)   ;
\draw   (519,120.72) .. controls (519,111.88) and (526.16,104.72) .. (535,104.72) .. controls (543.84,104.72) and (551,111.88) .. (551,120.72) .. controls (551,129.56) and (543.84,136.72) .. (535,136.72) .. controls (526.16,136.72) and (519,129.56) .. (519,120.72) -- cycle ;
\draw    (518,70) .. controls (558,40) and (548,156.72) .. (565,175.72) ;
\draw  [fill={rgb, 255:red, 208; green, 2; blue, 27 }  ,fill opacity=1 ] (487,122.72) .. controls (487,113.88) and (494.16,106.72) .. (503,106.72) .. controls (511.84,106.72) and (519,113.88) .. (519,122.72) .. controls (519,131.56) and (511.84,138.72) .. (503,138.72) .. controls (494.16,138.72) and (487,131.56) .. (487,122.72) -- cycle ;
\draw  [fill={rgb, 255:red, 208; green, 2; blue, 27 }  ,fill opacity=1 ] (799,124.86) .. controls (799,115.55) and (806.84,108) .. (816.5,108) .. controls (826.16,108) and (834,115.55) .. (834,124.86) .. controls (834,134.17) and (826.16,141.72) .. (816.5,141.72) .. controls (806.84,141.72) and (799,134.17) .. (799,124.86) -- cycle ;
\draw  [fill={rgb, 255:red, 208; green, 2; blue, 27 }  ,fill opacity=1 ] (765,124.72) .. controls (765,115.88) and (772.16,108.72) .. (781,108.72) .. controls (789.84,108.72) and (797,115.88) .. (797,124.72) .. controls (797,133.56) and (789.84,140.72) .. (781,140.72) .. controls (772.16,140.72) and (765,133.56) .. (765,124.72) -- cycle ;
\draw    (767,38.72) .. controls (807,8.72) and (785,180.72) .. (823,196.72) ;
\draw   (733,126.72) .. controls (733,117.88) and (740.16,110.72) .. (749,110.72) .. controls (757.84,110.72) and (765,117.88) .. (765,126.72) .. controls (765,135.56) and (757.84,142.72) .. (749,142.72) .. controls (740.16,142.72) and (733,135.56) .. (733,126.72) -- cycle ;
\draw  [fill={rgb, 255:red, 208; green, 2; blue, 27 }  ,fill opacity=1 ] (700,129.72) .. controls (700,120.88) and (707.16,113.72) .. (716,113.72) .. controls (724.84,113.72) and (732,120.88) .. (732,129.72) .. controls (732,138.56) and (724.84,145.72) .. (716,145.72) .. controls (707.16,145.72) and (700,138.56) .. (700,129.72) -- cycle ;
\draw    (738,167.72) -- (739.95,248.72) ;
\draw [shift={(740,250.72)}, rotate = 268.62] [color={rgb, 255:red, 0; green, 0; blue, 0 }  ][line width=0.75]    (10.93,-3.29) .. controls (6.95,-1.4) and (3.31,-0.3) .. (0,0) .. controls (3.31,0.3) and (6.95,1.4) .. (10.93,3.29)   ;
\draw  [fill={rgb, 255:red, 208; green, 2; blue, 27 }  ,fill opacity=1 ] (833,319.86) .. controls (833,310.55) and (840.84,303) .. (850.5,303) .. controls (860.16,303) and (868,310.55) .. (868,319.86) .. controls (868,329.17) and (860.16,336.72) .. (850.5,336.72) .. controls (840.84,336.72) and (833,329.17) .. (833,319.86) -- cycle ;
\draw  [fill={rgb, 255:red, 208; green, 2; blue, 27 }  ,fill opacity=1 ] (769,323.72) .. controls (769,314.88) and (776.16,307.72) .. (785,307.72) .. controls (793.84,307.72) and (801,314.88) .. (801,323.72) .. controls (801,332.56) and (793.84,339.72) .. (785,339.72) .. controls (776.16,339.72) and (769,332.56) .. (769,323.72) -- cycle ;
\draw    (801,233.72) .. controls (841,203.72) and (819,375.72) .. (857,391.72) ;
\draw   (737,325.72) .. controls (737,316.88) and (744.16,309.72) .. (753,309.72) .. controls (761.84,309.72) and (769,316.88) .. (769,325.72) .. controls (769,334.56) and (761.84,341.72) .. (753,341.72) .. controls (744.16,341.72) and (737,334.56) .. (737,325.72) -- cycle ;
\draw  [fill={rgb, 255:red, 208; green, 2; blue, 27 }  ,fill opacity=1 ] (704,328.72) .. controls (704,319.88) and (711.16,312.72) .. (720,312.72) .. controls (728.84,312.72) and (736,319.88) .. (736,328.72) .. controls (736,337.56) and (728.84,344.72) .. (720,344.72) .. controls (711.16,344.72) and (704,337.56) .. (704,328.72) -- cycle ;
\draw   (801,322.72) .. controls (801,313.88) and (808.16,306.72) .. (817,306.72) .. controls (825.84,306.72) and (833,313.88) .. (833,322.72) .. controls (833,331.56) and (825.84,338.72) .. (817,338.72) .. controls (808.16,338.72) and (801,331.56) .. (801,322.72) -- cycle ;
\draw  [fill={rgb, 255:red, 208; green, 2; blue, 27 }  ,fill opacity=1 ] (525,301.86) .. controls (525,292.55) and (532.84,285) .. (542.5,285) .. controls (552.16,285) and (560,292.55) .. (560,301.86) .. controls (560,311.17) and (552.16,318.72) .. (542.5,318.72) .. controls (532.84,318.72) and (525,311.17) .. (525,301.86) -- cycle ;
\draw  [fill={rgb, 255:red, 208; green, 2; blue, 27 }  ,fill opacity=1 ] (470,335.72) .. controls (470,326.88) and (477.16,319.72) .. (486,319.72) .. controls (494.84,319.72) and (502,326.88) .. (502,335.72) .. controls (502,344.56) and (494.84,351.72) .. (486,351.72) .. controls (477.16,351.72) and (470,344.56) .. (470,335.72) -- cycle ;
\draw    (541,237.72) .. controls (619,210.72) and (525,349.72) .. (595,391.72) ;
\draw   (438,337.72) .. controls (438,328.88) and (445.16,321.72) .. (454,321.72) .. controls (462.84,321.72) and (470,328.88) .. (470,337.72) .. controls (470,346.56) and (462.84,353.72) .. (454,353.72) .. controls (445.16,353.72) and (438,346.56) .. (438,337.72) -- cycle ;
\draw  [fill={rgb, 255:red, 208; green, 2; blue, 27 }  ,fill opacity=1 ] (405,340.72) .. controls (405,331.88) and (412.16,324.72) .. (421,324.72) .. controls (429.84,324.72) and (437,331.88) .. (437,340.72) .. controls (437,349.56) and (429.84,356.72) .. (421,356.72) .. controls (412.16,356.72) and (405,349.56) .. (405,340.72) -- cycle ;
\draw   (502,334.72) .. controls (502,325.88) and (509.16,318.72) .. (518,318.72) .. controls (526.84,318.72) and (534,325.88) .. (534,334.72) .. controls (534,343.56) and (526.84,350.72) .. (518,350.72) .. controls (509.16,350.72) and (502,343.56) .. (502,334.72) -- cycle ;
\draw  [fill={rgb, 255:red, 208; green, 2; blue, 27 }  ,fill opacity=1 ] (534,334.72) .. controls (534,325.88) and (541.16,318.72) .. (550,318.72) .. controls (558.84,318.72) and (566,325.88) .. (566,334.72) .. controls (566,343.56) and (558.84,350.72) .. (550,350.72) .. controls (541.16,350.72) and (534,343.56) .. (534,334.72) -- cycle ;
\draw    (680,321.72) -- (600,321.72) ;
\draw [shift={(598,321.72)}, rotate = 360] [color={rgb, 255:red, 0; green, 0; blue, 0 }  ][line width=0.75]    (10.93,-3.29) .. controls (6.95,-1.4) and (3.31,-0.3) .. (0,0) .. controls (3.31,0.3) and (6.95,1.4) .. (10.93,3.29)   ;
\draw    (378,326.72) -- (298,326.72) ;
\draw [shift={(296,326.72)}, rotate = 360] [color={rgb, 255:red, 0; green, 0; blue, 0 }  ][line width=0.75]    (10.93,-3.29) .. controls (6.95,-1.4) and (3.31,-0.3) .. (0,0) .. controls (3.31,0.3) and (6.95,1.4) .. (10.93,3.29)   ;
\draw  [fill={rgb, 255:red, 208; green, 2; blue, 27 }  ,fill opacity=1 ] (149,269.86) .. controls (149,260.55) and (156.84,253) .. (166.5,253) .. controls (176.16,253) and (184,260.55) .. (184,269.86) .. controls (184,279.17) and (176.16,286.72) .. (166.5,286.72) .. controls (156.84,286.72) and (149,279.17) .. (149,269.86) -- cycle ;
\draw  [fill={rgb, 255:red, 208; green, 2; blue, 27 }  ,fill opacity=1 ] (112,331.72) .. controls (112,322.88) and (119.16,315.72) .. (128,315.72) .. controls (136.84,315.72) and (144,322.88) .. (144,331.72) .. controls (144,340.56) and (136.84,347.72) .. (128,347.72) .. controls (119.16,347.72) and (112,340.56) .. (112,331.72) -- cycle ;
\draw    (182,235.72) .. controls (260,208.72) and (166,347.72) .. (236,389.72) ;
\draw   (80,333.72) .. controls (80,324.88) and (87.16,317.72) .. (96,317.72) .. controls (104.84,317.72) and (112,324.88) .. (112,333.72) .. controls (112,342.56) and (104.84,349.72) .. (96,349.72) .. controls (87.16,349.72) and (80,342.56) .. (80,333.72) -- cycle ;
\draw  [fill={rgb, 255:red, 208; green, 2; blue, 27 }  ,fill opacity=1 ] (47,336.72) .. controls (47,327.88) and (54.16,320.72) .. (63,320.72) .. controls (71.84,320.72) and (79,327.88) .. (79,336.72) .. controls (79,345.56) and (71.84,352.72) .. (63,352.72) .. controls (54.16,352.72) and (47,345.56) .. (47,336.72) -- cycle ;
\draw   (144,330.72) .. controls (144,321.88) and (151.16,314.72) .. (160,314.72) .. controls (168.84,314.72) and (176,321.88) .. (176,330.72) .. controls (176,339.56) and (168.84,346.72) .. (160,346.72) .. controls (151.16,346.72) and (144,339.56) .. (144,330.72) -- cycle ;
\draw  [fill={rgb, 255:red, 208; green, 2; blue, 27 }  ,fill opacity=1 ] (176,330.72) .. controls (176,321.88) and (183.16,314.72) .. (192,314.72) .. controls (200.84,314.72) and (208,321.88) .. (208,330.72) .. controls (208,339.56) and (200.84,346.72) .. (192,346.72) .. controls (183.16,346.72) and (176,339.56) .. (176,330.72) -- cycle ;
\draw   (165,299.72) .. controls (165,290.88) and (172.16,283.72) .. (181,283.72) .. controls (189.84,283.72) and (197,290.88) .. (197,299.72) .. controls (197,308.56) and (189.84,315.72) .. (181,315.72) .. controls (172.16,315.72) and (165,308.56) .. (165,299.72) -- cycle ;
\draw   (228,545) .. controls (228,531.19) and (239.19,520) .. (253,520) .. controls (266.81,520) and (278,531.19) .. (278,545) .. controls (278,558.81) and (266.81,570) .. (253,570) .. controls (239.19,570) and (228,558.81) .. (228,545) -- cycle ;
\draw   (540,463) .. controls (540,449.19) and (551.19,438) .. (565,438) .. controls (578.81,438) and (590,449.19) .. (590,463) .. controls (590,476.81) and (578.81,488) .. (565,488) .. controls (551.19,488) and (540,476.81) .. (540,463) -- cycle ;
\draw   (305,542) .. controls (305,528.19) and (316.19,517) .. (330,517) .. controls (343.81,517) and (355,528.19) .. (355,542) .. controls (355,555.81) and (343.81,567) .. (330,567) .. controls (316.19,567) and (305,555.81) .. (305,542) -- cycle ;
\draw   (547,543) .. controls (547,529.19) and (558.19,518) .. (572,518) .. controls (585.81,518) and (597,529.19) .. (597,543) .. controls (597,556.81) and (585.81,568) .. (572,568) .. controls (558.19,568) and (547,556.81) .. (547,543) -- cycle ;
\draw   (386,543) .. controls (386,529.19) and (397.19,518) .. (411,518) .. controls (424.81,518) and (436,529.19) .. (436,543) .. controls (436,556.81) and (424.81,568) .. (411,568) .. controls (397.19,568) and (386,556.81) .. (386,543) -- cycle ;
\draw   (464,543) .. controls (464,529.19) and (475.19,518) .. (489,518) .. controls (502.81,518) and (514,529.19) .. (514,543) .. controls (514,556.81) and (502.81,568) .. (489,568) .. controls (475.19,568) and (464,556.81) .. (464,543) -- cycle ;
\draw    (279,544) -- (306,543.72) ;
\draw    (356,544) -- (385,543.72) ;
\draw    (591,561) -- (630,589.22) ;
\draw    (436,543) -- (464,542.72) ;
\draw    (514,545) -- (546,544.72) ;
\draw   (628,602) .. controls (628,588.19) and (639.19,577) .. (653,577) .. controls (666.81,577) and (678,588.19) .. (678,602) .. controls (678,615.81) and (666.81,627) .. (653,627) .. controls (639.19,627) and (628,615.81) .. (628,602) -- cycle ;
\draw    (571,517) -- (567,488.22) ;
\draw    (489,604.22) -- (489,700.22) ;
\draw [shift={(489,702.22)}, rotate = 270] [color={rgb, 255:red, 0; green, 0; blue, 0 }  ][line width=0.75]    (10.93,-3.29) .. controls (6.95,-1.4) and (3.31,-0.3) .. (0,0) .. controls (3.31,0.3) and (6.95,1.4) .. (10.93,3.29)   ;
\draw   (540,699) .. controls (540,685.19) and (551.19,674) .. (565,674) .. controls (578.81,674) and (590,685.19) .. (590,699) .. controls (590,712.81) and (578.81,724) .. (565,724) .. controls (551.19,724) and (540,712.81) .. (540,699) -- cycle ;
\draw   (385,780) .. controls (385,766.19) and (396.19,755) .. (410,755) .. controls (423.81,755) and (435,766.19) .. (435,780) .. controls (435,793.81) and (423.81,805) .. (410,805) .. controls (396.19,805) and (385,793.81) .. (385,780) -- cycle ;
\draw   (547,779) .. controls (547,765.19) and (558.19,754) .. (572,754) .. controls (585.81,754) and (597,765.19) .. (597,779) .. controls (597,792.81) and (585.81,804) .. (572,804) .. controls (558.19,804) and (547,792.81) .. (547,779) -- cycle ;
\draw   (464,779) .. controls (464,765.19) and (475.19,754) .. (489,754) .. controls (502.81,754) and (514,765.19) .. (514,779) .. controls (514,792.81) and (502.81,804) .. (489,804) .. controls (475.19,804) and (464,792.81) .. (464,779) -- cycle ;
\draw    (591,797) -- (630,825.22) ;
\draw    (436,779) -- (464,778.72) ;
\draw    (514,781) -- (546,780.72) ;
\draw   (628,838) .. controls (628,824.19) and (639.19,813) .. (653,813) .. controls (666.81,813) and (678,824.19) .. (678,838) .. controls (678,851.81) and (666.81,863) .. (653,863) .. controls (639.19,863) and (628,851.81) .. (628,838) -- cycle ;
\draw    (571,753) -- (567,724.22) ;

\draw (121,32.4) node [anchor=north west][inner sep=0.75pt]    {$t=0$};
\draw (150,70.4) node [anchor=north west][inner sep=0.75pt]  [font=\tiny]  {$x^{5} +t$};
\draw (161,108.4) node [anchor=north west][inner sep=0.75pt]    [font=\tiny]{$B$};
\draw (90,110.4) node [anchor=north west][inner sep=0.75pt]  [font=\tiny]  {$m_{p} =1$};
\draw (298,143.4) node [anchor=north west][inner sep=0.75pt]  [font=\tiny]  {$m_{p} =2$};
\draw (306,110.4) node [anchor=north west][inner sep=0.75pt] [font=\tiny]   {$E_{1}$};
\draw (279,35.4) node [anchor=north west][inner sep=0.75pt]    {$\widetilde{\Gamma _{0}}$};
\draw (353,104.4) node [anchor=north west][inner sep=0.75pt]   [font=\tiny] {$\tilde{B}$};
\draw (334,72.4) node [anchor=north west][inner sep=0.75pt]  [font=\tiny]  {$x^{4} +t$};
\draw (518,151.4) node [anchor=north west][inner sep=0.75pt]  [font=\tiny]  {$m_{p} =1$};
\draw (525,114.4) node [anchor=north west][inner sep=0.75pt]   [font=\tiny] {$E_{2}$};
\draw (498,39.4) node [anchor=north west][inner sep=0.75pt]    {$\widetilde{\Gamma _{0}}$};
\draw (572,108.4) node [anchor=north west][inner sep=0.75pt]  [font=\tiny]  {$\tilde{B}$};
\draw (553,75.4) node [anchor=north west][inner sep=0.75pt]  [font=\tiny]  {$x^{3} +t$};
\draw (495,113.4) node [anchor=north west][inner sep=0.75pt]   [font=\tiny] {$E_{1}$};
\draw (740,118.4) node [anchor=north west][inner sep=0.75pt]  [font=\tiny]  {$E_{2}$};
\draw (744,43.4) node [anchor=north west][inner sep=0.75pt]    {$\widetilde{\Gamma _{0}}$};
\draw (805,113.4) node [anchor=north west][inner sep=0.75pt]    {$\widetilde{B}$};
\draw (802,86.4) node [anchor=north west][inner sep=0.75pt]  [font=\tiny]  {$x{^{2}} +t$};
\draw (709,118.4) node [anchor=north west][inner sep=0.75pt]   [font=\tiny] {$E_{1}$};
\draw (773,117.4) node [anchor=north west][inner sep=0.75pt]   [font=\tiny] {$E_{3}$};
\draw (810,153.4) node [anchor=north west][inner sep=0.75pt]  [font=\tiny]  {$m_{p} =2$};
\draw (744,317.4) node [anchor=north west][inner sep=0.75pt]   [font=\tiny] {$E_{2}$};
\draw (789,226.4) node [anchor=north west][inner sep=0.75pt]    {$\widetilde{\Gamma _{0}}$};
\draw (839,308.4) node [anchor=north west][inner sep=0.75pt]    {$\widetilde{B}$};
\draw (834,283.4) node [anchor=north west][inner sep=0.75pt]  [font=\tiny]  {$x{} +t$};
\draw (713,317.4) node [anchor=north west][inner sep=0.75pt]  [font=\tiny]  {$E_{1}$};
\draw (777,316.4) node [anchor=north west][inner sep=0.75pt]   [font=\tiny] {$E_{3}$};
\draw (845,354.4) node [anchor=north west][inner sep=0.75pt]  [font=\tiny]  {$m_{p} =1$};
\draw (809,313.4) node [anchor=north west][inner sep=0.75pt]    [font=\tiny]{$E_{4}$};
\draw (445,329.4) node [anchor=north west][inner sep=0.75pt]   [font=\tiny] {$E_{2}$};
\draw (509,250.4) node [anchor=north west][inner sep=0.75pt]    {$\widetilde{\Gamma _{0}}$};
\draw (534,289.4) node [anchor=north west][inner sep=0.75pt]   [font=\tiny] {$\widetilde{B}$};
\draw (414,329.4) node [anchor=north west][inner sep=0.75pt]  [font=\tiny]  {$E_{1}$};
\draw (478,328.4) node [anchor=north west][inner sep=0.75pt]   [font=\tiny] {$E_{3}$};
\draw (486,300.4) node [anchor=north west][inner sep=0.75pt]  [font=\tiny]  {$m_{p} =2$};
\draw (510,325.4) node [anchor=north west][inner sep=0.75pt]   [font=\tiny] {$E_{4}$};
\draw (541,326.4) node [anchor=north west][inner sep=0.75pt]  [font=\tiny]  {$E_{5}$};
\draw (87,325.4) node [anchor=north west][inner sep=0.75pt]   [font=\tiny] {$E_{2}$};
\draw (187,198.4) node [anchor=north west][inner sep=0.75pt]    {$\widetilde{\Gamma _{0}}$};
\draw (160,257.4) node [anchor=north west][inner sep=0.75pt]    {$\widetilde{B}$};
\draw (56,325.4) node [anchor=north west][inner sep=0.75pt]    [font=\tiny]{$E_{1}$};
\draw (120,324.4) node [anchor=north west][inner sep=0.75pt]   [font=\tiny] {$E_{3}$};
\draw (152,321.4) node [anchor=north west][inner sep=0.75pt] [font=\tiny]   {$E_{4}$};
\draw (183,322.4) node [anchor=north west][inner sep=0.75pt]    [font=\tiny]{$E_{5}$};
\draw (172,290.4) node [anchor=north west][inner sep=0.75pt]   [font=\tiny] {$E_{6}$};
\draw (247,532.4) node [anchor=north west][inner sep=0.75pt]    {$2$};
\draw (486,535.4) node [anchor=north west][inner sep=0.75pt]    {$4$};
\draw (325,535.4) node [anchor=north west][inner sep=0.75pt]    {$2$};
\draw (405,532.4) node [anchor=north west][inner sep=0.75pt]    {$6$};
\draw (563,535.4) node [anchor=north west][inner sep=0.75pt]    {$10$};
\draw (649,597.4) node [anchor=north west][inner sep=0.75pt]    {$1$};
\draw (558,451.4) node [anchor=north west][inner sep=0.75pt]    {$5$};
\draw (486,771.4) node [anchor=north west][inner sep=0.75pt]    {$4$};
\draw (408,773.4) node [anchor=north west][inner sep=0.75pt]    {$2$};
\draw (563,771.4) node [anchor=north west][inner sep=0.75pt]    {$10$};
\draw (649,833.4) node [anchor=north west][inner sep=0.75pt]    {$1$};
\draw (558,687.4) node [anchor=north west][inner sep=0.75pt]    {$5$};

\end{tikzpicture}

    \caption{Singular fiber configuration for $f(x,t) = x^5 + t$ after canonical resolution. The singular fiber has $-1$ curve, and one can contract it to get the relative minimal model. The red components are the branched locus of the double covering, and 
   to compute the multiplicities for the resolution, one need to take the sum of the multiplicities of all the components of  the branch locus $\tilde{B}$. }
    \label{fig:ex2}
\end{figure}
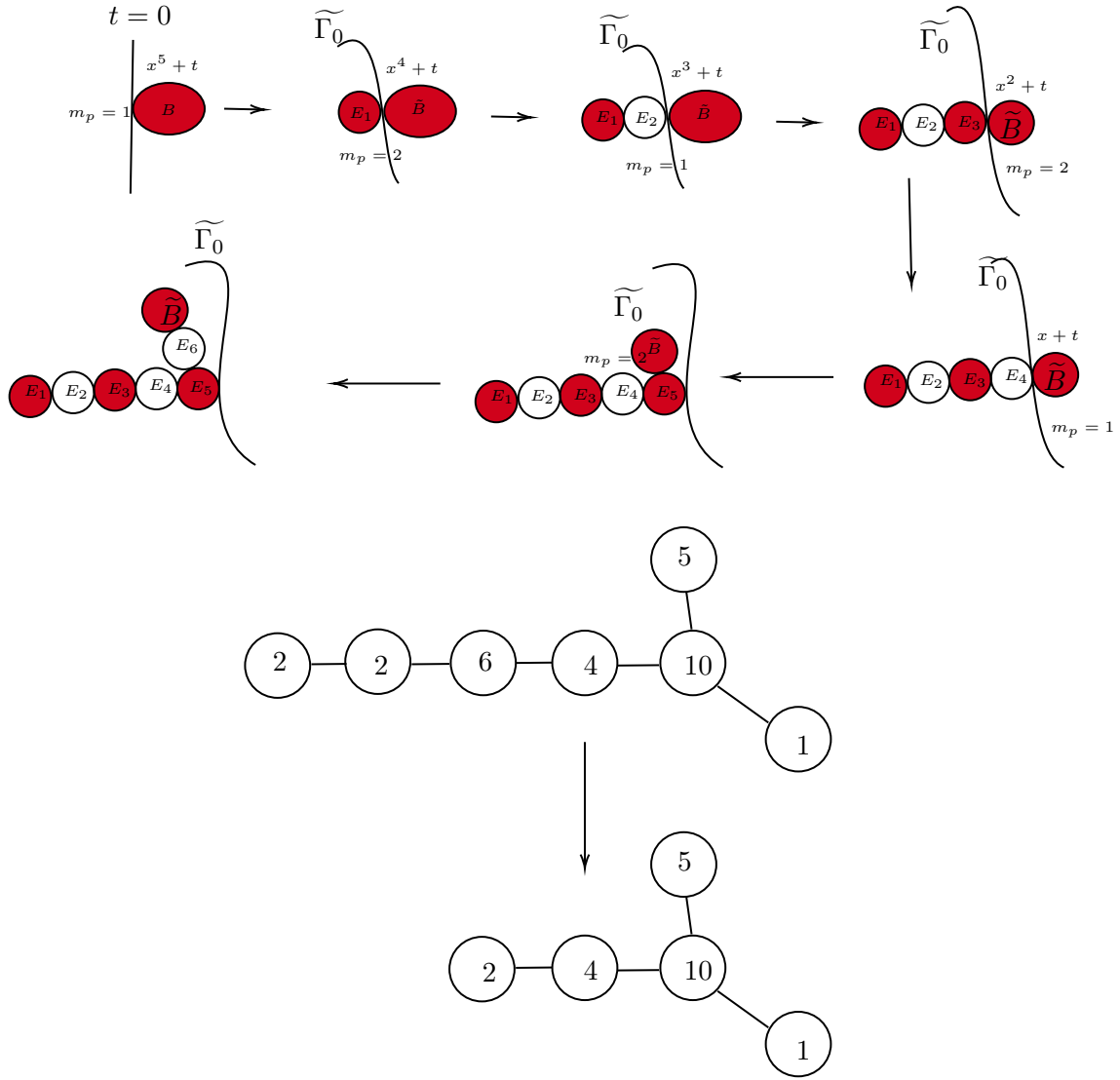

\textbf{Example 2}:
For the branched locus $f(x,t) = x^5 + t$, the bad point is at $(0,0)$, and $\Gamma_0$ is given by $t = 0$ (see Figure~\ref{fig:ex2}). 
After resolution, we obtain:
\begin{align*}
    \Gamma_0^* &= \tilde{\Gamma_0} + E_1 + 2E_2 + 3E_3 + 4E_4 + 5E_5 + 5E_6, \\
    B^* &= \tilde{B} + E_1 + E_3 + E_5,
\end{align*}
with the configuration depicted in Figure~\ref{fig:ex2}.  Notice that there is an intersection point at $x=\infty$ for the branched locus $B$ and the curve $\Gamma_0: t=0$.

\begin{figure}[H]
    \centering
\tikzset{every picture/.style={line width=0.75pt}} 

\begin{tikzpicture}[x=0.50pt,y=0.50pt,yscale=-1,xscale=1]

\draw    (142,60.72) -- (139,201.72) ;
\draw    (212,114) -- (241,114.67) ;
\draw [shift={(243,114.72)}, rotate = 181.33] [color={rgb, 255:red, 0; green, 0; blue, 0 }  ][line width=0.75]    (10.93,-3.29) .. controls (6.95,-1.4) and (3.31,-0.3) .. (0,0) .. controls (3.31,0.3) and (6.95,1.4) .. (10.93,3.29)   ;
\draw    (310,63.72) .. controls (388,36.72) and (294,166.72) .. (364,208.72) ;
\draw   (308,102.72) .. controls (308,93.88) and (315.16,86.72) .. (324,86.72) .. controls (332.84,86.72) and (340,93.88) .. (340,102.72) .. controls (340,111.56) and (332.84,118.72) .. (324,118.72) .. controls (315.16,118.72) and (308,111.56) .. (308,102.72) -- cycle ;
\draw   (532,414) .. controls (532,400.19) and (543.19,389) .. (557,389) .. controls (570.81,389) and (582,400.19) .. (582,414) .. controls (582,427.81) and (570.81,439) .. (557,439) .. controls (543.19,439) and (532,427.81) .. (532,414) -- cycle ;
\draw   (371,414) .. controls (371,400.19) and (382.19,389) .. (396,389) .. controls (409.81,389) and (421,400.19) .. (421,414) .. controls (421,427.81) and (409.81,439) .. (396,439) .. controls (382.19,439) and (371,427.81) .. (371,414) -- cycle ;
\draw   (449,414) .. controls (449,400.19) and (460.19,389) .. (474,389) .. controls (487.81,389) and (499,400.19) .. (499,414) .. controls (499,427.81) and (487.81,439) .. (474,439) .. controls (460.19,439) and (449,427.81) .. (449,414) -- cycle ;
\draw    (582,414) -- (639,414.72) ;
\draw    (421,414) -- (449,413.72) ;
\draw    (499,416) -- (531,415.72) ;
\draw   (639,415) .. controls (639,401.19) and (650.19,390) .. (664,390) .. controls (677.81,390) and (689,401.19) .. (689,415) .. controls (689,428.81) and (677.81,440) .. (664,440) .. controls (650.19,440) and (639,428.81) .. (639,415) -- cycle ;
\draw [color={rgb, 255:red, 208; green, 2; blue, 27 }  ,draw opacity=1 ][fill={rgb, 255:red, 255; green, 255; blue, 255 }  ,fill opacity=1 ]   (161,203.72) .. controls (158.01,201.58) and (135.04,180.28) .. (141,174.72) .. controls (146.96,169.16) and (177,163.72) .. (157,139.72) .. controls (137,115.72) and (127.68,75.07) .. (195,74.72) ;
\draw    (464,115) -- (493,115.67) ;
\draw [shift={(495,115.72)}, rotate = 181.33] [color={rgb, 255:red, 0; green, 0; blue, 0 }  ][line width=0.75]    (10.93,-3.29) .. controls (6.95,-1.4) and (3.31,-0.3) .. (0,0) .. controls (3.31,0.3) and (6.95,1.4) .. (10.93,3.29)   ;
\draw [color={rgb, 255:red, 208; green, 2; blue, 27 }  ,draw opacity=1 ]   (380,208.72) .. controls (376,205.72) and (333.04,178.28) .. (339,172.72) .. controls (344.96,167.16) and (375,161.72) .. (355,137.72) .. controls (335,113.72) and (326.68,68.07) .. (394,67.72) ;
\draw    (768,79.72) .. controls (851,49.72) and (748,176.72) .. (820,218.72) ;
\draw  [fill={rgb, 255:red, 208; green, 2; blue, 27 }  ,fill opacity=1 ] (737,104.72) .. controls (737,95.88) and (744.16,88.72) .. (753,88.72) .. controls (761.84,88.72) and (769,95.88) .. (769,104.72) .. controls (769,113.56) and (761.84,120.72) .. (753,120.72) .. controls (744.16,120.72) and (737,113.56) .. (737,104.72) -- cycle ;
\draw [color={rgb, 255:red, 208; green, 2; blue, 27 }  ,draw opacity=1 ]   (759,212.72) .. controls (738,175.72) and (738,173.72) .. (748,153.72) .. controls (758,133.72) and (770.02,136.63) .. (779,119.72) .. controls (787.98,102.81) and (799,202.72) .. (786,215.72) ;
\draw   (768,102.72) .. controls (768,93.88) and (775.16,86.72) .. (784,86.72) .. controls (792.84,86.72) and (800,93.88) .. (800,102.72) .. controls (800,111.56) and (792.84,118.72) .. (784,118.72) .. controls (775.16,118.72) and (768,111.56) .. (768,102.72) -- cycle ;
\draw    (549,63.72) .. controls (627,36.72) and (533,166.72) .. (603,208.72) ;
\draw  [fill={rgb, 255:red, 208; green, 2; blue, 27 }  ,fill opacity=1 ] (548,102.72) .. controls (548,93.88) and (555.16,86.72) .. (564,86.72) .. controls (572.84,86.72) and (580,93.88) .. (580,102.72) .. controls (580,111.56) and (572.84,118.72) .. (564,118.72) .. controls (555.16,118.72) and (548,111.56) .. (548,102.72) -- cycle ;
\draw [color={rgb, 255:red, 208; green, 2; blue, 27 }  ,draw opacity=1 ]   (619,208.72) .. controls (615,205.72) and (572.04,178.28) .. (578,172.72) .. controls (583.96,167.16) and (614,161.72) .. (594,137.72) .. controls (574,113.72) and (565.68,68.07) .. (633,67.72) ;
\draw   (516,104.72) .. controls (516,95.88) and (523.16,88.72) .. (532,88.72) .. controls (540.84,88.72) and (548,95.88) .. (548,104.72) .. controls (548,113.56) and (540.84,120.72) .. (532,120.72) .. controls (523.16,120.72) and (516,113.56) .. (516,104.72) -- cycle ;
\draw   (704,108.72) .. controls (704,99.88) and (711.16,92.72) .. (720,92.72) .. controls (728.84,92.72) and (736,99.88) .. (736,108.72) .. controls (736,117.56) and (728.84,124.72) .. (720,124.72) .. controls (711.16,124.72) and (704,117.56) .. (704,108.72) -- cycle ;
\draw    (771,240.72) -- (574.73,354.71) ;
\draw [shift={(573,355.72)}, rotate = 329.85] [color={rgb, 255:red, 0; green, 0; blue, 0 }  ][line width=0.75]    (10.93,-3.29) .. controls (6.95,-1.4) and (3.31,-0.3) .. (0,0) .. controls (3.31,0.3) and (6.95,1.4) .. (10.93,3.29)   ;
\draw    (559,454.72) -- (558.03,528.72) ;
\draw [shift={(558,530.72)}, rotate = 270.75] [color={rgb, 255:red, 0; green, 0; blue, 0 }  ][line width=0.75]    (10.93,-3.29) .. controls (6.95,-1.4) and (3.31,-0.3) .. (0,0) .. controls (3.31,0.3) and (6.95,1.4) .. (10.93,3.29)   ;
\draw   (541,573) .. controls (541,559.19) and (552.19,548) .. (566,548) .. controls (579.81,548) and (591,559.19) .. (591,573) .. controls (591,586.81) and (579.81,598) .. (566,598) .. controls (552.19,598) and (541,586.81) .. (541,573) -- cycle ;
\draw   (458,573) .. controls (458,559.19) and (469.19,548) .. (483,548) .. controls (496.81,548) and (508,559.19) .. (508,573) .. controls (508,586.81) and (496.81,598) .. (483,598) .. controls (469.19,598) and (458,586.81) .. (458,573) -- cycle ;
\draw    (591,573) -- (648,573.72) ;
\draw    (508,575) -- (540,574.72) ;
\draw   (648,574) .. controls (648,560.19) and (659.19,549) .. (673,549) .. controls (686.81,549) and (698,560.19) .. (698,574) .. controls (698,587.81) and (686.81,599) .. (673,599) .. controls (659.19,599) and (648,587.81) .. (648,574) -- cycle ;

\draw (121,32.4) node [anchor=north west][inner sep=0.75pt]    {$t=0$};
\draw (146,90.4) node [anchor=north west][inner sep=0.75pt]  [font=\tiny]  {$xt+x{^{4}}$};
\draw (175,57.4) node [anchor=north west][inner sep=0.75pt]    [font=\tiny] {$B$};
\draw (90,100.4) node [anchor=north west][inner sep=0.75pt]  [font=\tiny]  {$m_{p} =2$};
\draw (311,30.4) node [anchor=north west][inner sep=0.75pt]   [font=\tiny]  {$\widetilde{\Gamma _{0}}$};
\draw (316,94.4) node [anchor=north west][inner sep=0.75pt]  [font=\tiny]   {$E_{1}$};
\draw (471,406.4) node [anchor=north west][inner sep=0.75pt]    {$4$};
\draw (390,408.4) node [anchor=north west][inner sep=0.75pt]    {$1$};
\draw (547,405.4) node [anchor=north west][inner sep=0.75pt]    {$3$};
\draw (294,126.4) node [anchor=north west][inner sep=0.75pt]  [font=\tiny]  {$m_{p} =1$};
\draw (347,88.4) node [anchor=north west][inner sep=0.75pt]  [font=\tiny]  {$t+x{^{2}}$};
\draw (400,40.4) node [anchor=north west][inner sep=0.75pt]  [font=\tiny]   {$\tilde{B}$};
\draw (713,98.4) node [anchor=north west][inner sep=0.75pt]   [font=\tiny]  {$E_{1}$};
\draw (745,96.4) node [anchor=north west][inner sep=0.75pt]   [font=\tiny]  {$E_{2}$};
\draw (550,30.4) node [anchor=north west][inner sep=0.75pt]   [font=\tiny]  {$\widetilde{\Gamma _{0}}$};
\draw (533,126.4) node [anchor=north west][inner sep=0.75pt]  [font=\tiny]  {$m_{p} =2$};
\draw (587,98.4) node [anchor=north west][inner sep=0.75pt]  [font=\tiny]  {$t+x{}$};
\draw (638,41.4) node [anchor=north west][inner sep=0.75pt]   [font=\tiny]  {$\tilde{B}$};
\draw (521,95.4) node [anchor=north west][inner sep=0.75pt]   [font=\tiny]  {$E_{1}$};
\draw (554,95.4) node [anchor=north west][inner sep=0.75pt]  [font=\tiny]   {$E_{2}$};
\draw (787,40.4) node [anchor=north west][inner sep=0.75pt]   [font=\tiny]  {$\widetilde{\Gamma _{0}}$};
\draw (775,94.4) node [anchor=north west][inner sep=0.75pt]   [font=\tiny]  {$E_{3}$};
\draw (100,174.4) node [anchor=north west][inner sep=0.75pt]  [font=\tiny]  {$x=-1$};
\draw (719,177.4) node [anchor=north west][inner sep=0.75pt]   [font=\tiny]  {$\tilde{B}$};
\draw (656,405.4) node [anchor=north west][inner sep=0.75pt]    {$1$};
\draw (604,393.4) node [anchor=north west][inner sep=0.75pt]    {$2$};
\draw (480,565.4) node [anchor=north west][inner sep=0.75pt]    {$1$};
\draw (556,564.4) node [anchor=north west][inner sep=0.75pt]    {$3$};
\draw (665,564.4) node [anchor=north west][inner sep=0.75pt]    {$1$};
\draw (613,552.4) node [anchor=north west][inner sep=0.75pt]    {$2$};

\end{tikzpicture}

    \caption{Singular fiber configuration for $f(x,t) = xt + x^4(x + 1)$ after canonical resolution.}
    \label{fig:ex3}
\end{figure}
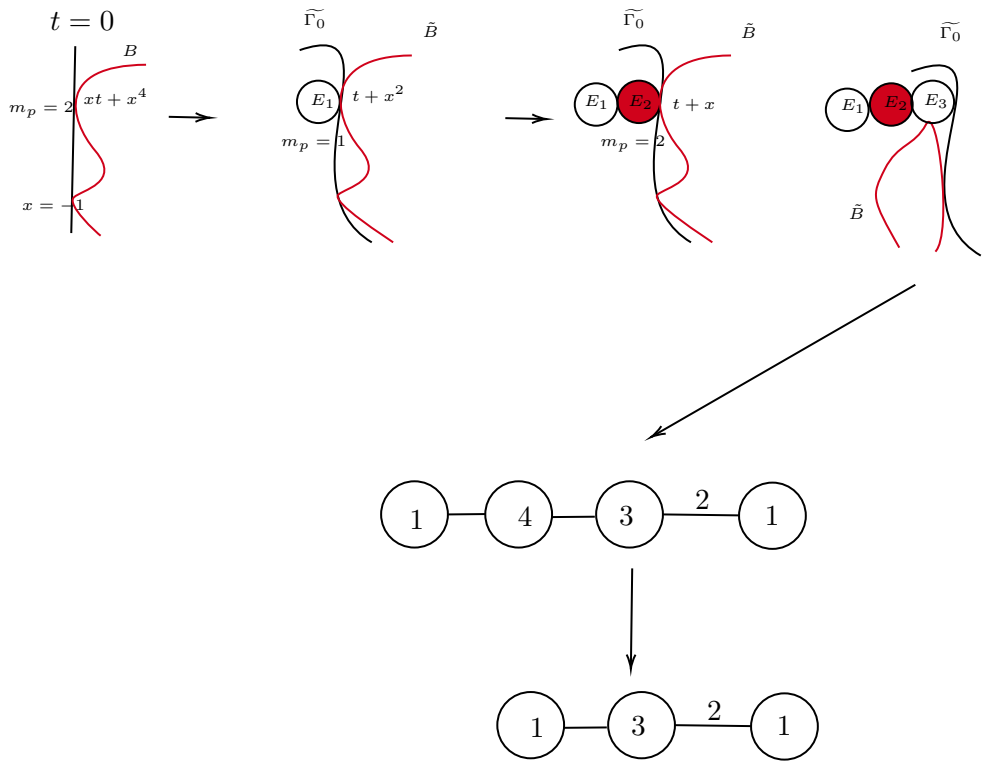

\textbf{Example 3}
For $f(x,t) = xt + x^4(x + 1)$, the bad point is at $(0,0)$, with $\Gamma_0$ defined by $t = 0$ (see Figure~\ref{fig:ex3}). The resolution process and resulting configuration are shown in Figure~\ref{fig:ex3}.

\begin{align*}
    \Gamma_0^* &= \tilde{\Gamma_0}+E_1+2E_2+3E_3 \\
    B^* &= \tilde{B} + E_2
\end{align*}

\newpage
\subsection{Full Seiberg-Witten Geometry from a One-Parameter Family}
\label{sec:full_sw}

Let us review how to construct the full Seiberg-Witten (SW) geometry from a one-parameter family curve \( y^2 = F(x,t) \) \cite{Xie:2023zxn}. The SW geometry containing all Coulomb branch operators is given by:
\begin{equation}
y^2 = F(x, P(x,\mathbf{u}))
\label{fullsw}
\end{equation}
where \( P \) is a polynomial encoding Coulomb branch operators (and potentially other couplings like mass deformations or  coupling couplings):
\[
P(x, \mathbf{u}) = \sum_{i} u_i x^i.
\]
The polynomial \( P \) is chosen such that the genus of the curve equals the number of Coulomb branch operators.

The Seiberg-Witten differential \( \lambda_{\text{SW}} \) must satisfy the special Kähler condition:
\[
\partial_{u_i} \lambda_{\text{SW}} \subset H^{1,0}(\Sigma_u, \mathbb{C}),
\]
where \( H^{1,0} \) denotes the space of holomorphic differentials on the Riemann surface \( \Sigma_u \) defined by Equation \eqref{fullsw}. A basis for \( H^{1,0} \) is given by (see appendix \ref{holodifferential}:)
\[
\left\{ \frac{x^i dx}{y} \ \bigg|\ i = 0, 1, \ldots, g-1 \right\}.
\]
We may further require:
\[
\partial_{u_i} \lambda_{\text{SW}} \subset \left[ \frac{x^i dx}{y} \right],
\]
meaning the derivative of \( \lambda_{\text{SW}} \) with respect to \( u_i \) lies in the cohomology class of \( \frac{x^i dx}{y} \).

Assume:
\[
\lambda_{\text{SW}} = R(x, \mathbf{u}) \frac{dx}{y},
\]
where \( R \) is a rational function of \( x \) and \( \mathbf{u} \). Taking the derivative with respect to \( u_i \) yields:
\begin{equation}
\partial_{u_i} \lambda_{\text{SW}} = a_i \frac{x^i dx}{y} + b_i d\left( \frac{x^{i+1}}{y} \right), \quad a_i \neq 0,
\label{swder}
\end{equation}
which must belong to the cohomology class of \( \frac{x^i dx}{y} \) if $a_i, b_i$ are constants. We have:
\[
d\left( \frac{x^{i+1}}{y} \right) = \frac{(i+1)x^i dx}{y} - \frac{x^{i+1}}{2y^2} \left( \frac{\partial F}{\partial P} P' + \frac{\partial F}{\partial x} \right) \frac{dx}{y},
\]
and we have used \( dy = \frac{1}{2y} \left( \frac{\partial F}{\partial P} P' + \frac{\partial F}{\partial x} \right) dx \). Using the form of SW differential, and the derivative $\partial_{u_i}( {R dx\over y})= [{\partial_{u_i} R}+{R\over 2y^2}{\partial F\over \partial P}x^i] {dx\over y}$, 
the equation \ref{swder} becomes:
\[
\frac{\partial R}{\partial u_i} + \frac{R}{2y^2} \frac{\partial F}{\partial P} x^i = a_i x^i + b_i \left[ (i+1)x^i - \frac{x^{i+1}}{2y^2} \left( \frac{\partial F}{\partial P} P' + \frac{\partial F}{\partial x} \right) \right].
\]

Assuming \( R \) depends on \( \mathbf{u} \) only through \( P \), \( x P' \), and \( \partial_{u_i} R = d_i x^i \), we derive:
\[
-\frac{1}{2} b_i x \left( \frac{\partial F}{\partial P} P' + \frac{\partial F}{\partial x} \right) = \frac{R}{2} \frac{\partial F}{\partial P} + \left( d_i - a_i - (i+1)b_i \right) F.
\]
A useful equation for \( R \) is:
\begin{equation}
\boxed{ \left( R +  b_i x P' \right) \frac{\partial F}{\partial P} = -b_i x \frac{\partial F}{\partial x} - 2c_i F }
\label{crucial}
\end{equation}
where \( b_i \), \( c_i \) are constants independent of $i$ (denoted as \( b \), \( c \)), and \( c = d_i - a_i - (i+1)b \). Let's find solution of $R$ for various form of $F$.

\textbf{\( F \) Linear in \( P \)}:
When \( F(x, P) = h_0(x) + h_1(x) P \), setting \( b_i = 0 \) simplifies Equation \eqref{crucial} to \( R h_1 = -2c F \). Choosing \( c = -1/2 \), we obtain:
\[
\boxed{ R = \frac{F}{h_1} = \frac{y^2}{h_1(x)} }
\label{linear}
\]
The SW differential $\lambda_{sw}={Rdx\over y}={ydx\over h_1}$ becomes:
- \( \lambda_{\text{SW}} = y dx \) if \( h_1 = 1 \),
- \( \lambda_{\text{SW}} = \frac{y dx}{x} \) if \( h_1 = x \).
These two cases are already considered in \cite{Xie:2015rpa}.

\textbf{\( F \) Quadratic in \( P \)}:
For \( F(x, P) = h_0(x) + h_1(x) P + h_2(x) P^2 \), $b$ is nonzero and could be scaled to one, and substituting into Equation \eqref{crucial} yields:
\[
\left( R +x P' \right) (h_1 + 2 h_2 P) = x \left( h_0' + h_1' P + h_2' P^2 \right) - 2c F.
\]
Assuming $R$ to have the same scaling as $P$, it may take the form \( R = b_0 P -  x P' \), the equation reduces to:
\[
b_0 h_1 P + 2 b_0 h_2 P^2 = -x \left( h_0' + h_1' P + h_2' P^2 \right) - 2c (h_0  + h_1 P + h_2 P^2).
\]
Equating coefficients for \( P^0 \), \( P^1 \), and \( P^2 \) gives:
\begin{align*}
0 &= - x h_0' -2 c h_0, \\
b_0 h_1 &= - x h_1' - 2c h_1, \\
2 b_0 h_2 &= - x h_2' - 2c h_2.
\end{align*}

For monomial \( h_i(x) = x^{k_i} \), solutions are:
- \( h_1 = 0 \): \( b_0 = \frac{k_0 - k_2}{2} \),
- \( h_0 = 0 \): \( b_0 = k_1 - k_2 \).

General solutions for \( h_0=0 \) gives:
\[
b_0 = \frac{(h_2 h_1'-h_1 h_2' ) x}{ h_1 h_2}, \quad c = \frac{ (h_1 h_2' -2 h_2 h_1')x}{2 h_1h_2}.
\]

General solutions for \( h_1= 0 \) gives:
\[
b_0 = \frac{(h_2 h_0' - h_0 h_2') x}{2 h_0 h_2}, \quad c = \frac{- h_0' x}{2 h_0}.
\]
In these cases, we notice that the coefficients $b_0$ is no longer constant, and so the SW differential would 
be a meromorphic differential. The position of the pole is determined by the coefficient $h_0$ and $h_2$. For example, if one of the coefficient $h_0(x)$ is monomial in $x$, 
the position of the pole is then determined by $h_2$.  This often means that one should  consider the Riemann surface with the pole removed. This is not surprising 
as the SW differential is meromorphic even for $F$ linear in $P$ (the pole could be at $x=\infty$ or $x=0$). The existence of the pole is related to the existence of 
the flavor symmetry.

\textbf{Homogeneous case}:

For general \( F(x, P) = \sum_i h_i(x) P^i \), assuming \( R = b_0 P - x P' \), we derive:
\[
\sum_i (i b_0 h_i) P^i = \sum_i \left( -\frac{1}{2} x h_i' - c h_i \right) P^i,
\]
leading to:
\[
i b_0 h_i = -\frac{1}{2} x h_i' - c h_i.
\]
Generally only two non-zero terms in \( F \) allow solutions. For nonzero \( h_i \) and \( h_j \), \( b_0, c \) are:
\[
b_0 = \frac{(h_j h_i' - h_i h_j')x}{ h_i h_j (j - i)}, \quad c = \frac{h_j' h_i i x - h_i' h_j j x}{2 h_i h_j (j - i)}.
\]

\begin{table}[htp]
\centering
\caption{Solutions for SW Differential \( R \)}
\label{tab:sw_solutions}
\begin{tabular}{|c|c|c|} \hline
\( F(x, P) \) & \( \lambda_{\text{SW}} = R \frac{dx}{y} \) & Coefficient \\ \hline
\( h_0 + h_1 P \) & \( R = \frac{y^2}{h_1(x)} \) & None \\ \hline
\( h_1 P + h_2 P^2 \) & \( R = b_0 P - x P' \) & \( b_0 = \frac{(h_2 h_1' - h_1 h_2')x}{ h_1 h_2} \) \\ \hline
\( h_0 + h_2 P^2 \) & \( R = b_0 P - x P' \) & \( b_0 = \frac{(h_2 h_0' - h_0 h_2')x}{2 h_0 h_2} \) \\ \hline
\( h_i P^i + h_j P^j \) & \( R = b_0 P - x P' \) & \( b_0 = \frac{(h_j h_i' - h_i h_j')x}{ h_i h_j (j-i)} \) \\ \hline
\end{tabular}
\end{table}

\textbf{Massive Case}
Let's now consider the massive case. The SW differential becomes (without the homogeneous constraint in $P$):
\[
\lambda_{\text{SW}} = \left( g + b_0 P -  x P' \right) \frac{dx}{y},
\]
If the function $F$ is quadratic in $P$: \( F(x, P) = h_0(x) + h_1(x) P + h_2(x) P^2 \), the general solutions are:
\begin{align*}
g &= \frac{ (h_0' h_1 h_2 - 2 h_0 h_1' h_2 + h_0 h_1 h_2') x }{ 2 h_2 (-h_1^2 + 4 h_0 h_2) }, \\
b_0 &= -\frac{ (h_1 h_1' h_2 - 2 h_0' h_2^2 - h_1^2 h_2' + 2 h_0 h_2 h_2')x }{ 2 h_2 (-h_1^2 + 4 h_0 h_2) }, \\
c &= \frac{ -2 h_1 h_1' h_2 + 4 h_0' h_2^2 + h_1^2 h_2' }{ 2 h_2 (-h_1^2 + 4 h_0 h_2) } x.
\end{align*}
Here, \( a_i \) in Equation \eqref{swder} may have poles reflecting flavor symmetry. It is straightforward to find solutions for $F$ of higher order polynomial in $P$.

In the following sections, we apply this formalism to construct SW geometries for specific theories.

\subsection{Strategy for Finding Hyperelliptic Seiberg-Witten Geometries}
Let us now explain our strategy for finding one-parameter hyperelliptic Seiberg-Witten geometries. The idea is to identify a family \( f(x,t) \) with the minimal number of singularities, which is the opposite of what is discussed in \cite{Xie:2023wqx}, where one assumes as many singularities as possible. We will elaborate on the theory in different dimensions below.

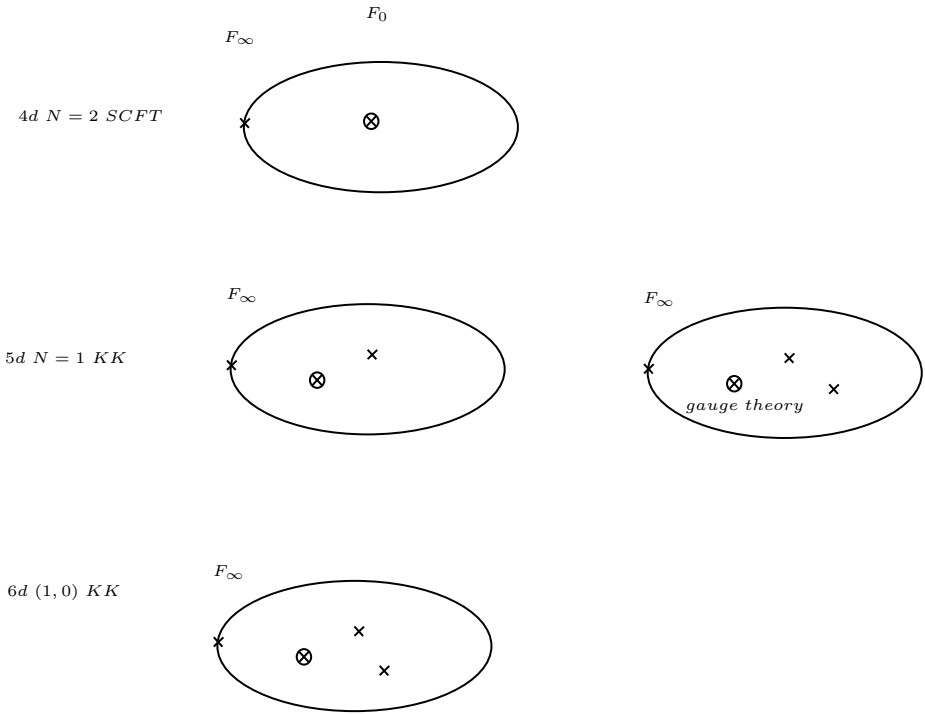
\begin{figure}
\begin{center}

\tikzset{every picture/.style={line width=0.75pt}} 

\begin{tikzpicture}[x=0.45pt,y=0.45pt,yscale=-1,xscale=1]

\draw   (297,142.61) .. controls (297,112.45) and (348.04,88) .. (411,88) .. controls (473.96,88) and (525,112.45) .. (525,142.61) .. controls (525,172.77) and (473.96,197.22) .. (411,197.22) .. controls (348.04,197.22) and (297,172.77) .. (297,142.61) -- cycle ;
\draw    (294,135.22) -- (301.2,143.22) ;
\draw    (294,143.08) -- (302,135.36) ;

\draw   (286,345.61) .. controls (286,315.45) and (337.04,291) .. (400,291) .. controls (462.96,291) and (514,315.45) .. (514,345.61) .. controls (514,375.77) and (462.96,400.22) .. (400,400.22) .. controls (337.04,400.22) and (286,375.77) .. (286,345.61) -- cycle ;
\draw    (400,329.22) -- (407.2,337.22) ;
\draw    (400,337.08) -- (408,329.36) ;

\draw    (283,338.22) -- (290.2,346.22) ;
\draw    (283,346.08) -- (291,338.36) ;

\draw   (352,354.72) .. controls (352,351.13) and (354.69,348.22) .. (358,348.22) .. controls (361.31,348.22) and (364,351.13) .. (364,354.72) .. controls (364,358.31) and (361.31,361.22) .. (358,361.22) .. controls (354.69,361.22) and (352,358.31) .. (352,354.72) -- cycle ; \draw   (353.76,350.12) -- (362.24,359.31) ; \draw   (362.24,350.12) -- (353.76,359.31) ;
\draw   (633,348.61) .. controls (633,318.45) and (684.04,294) .. (747,294) .. controls (809.96,294) and (861,318.45) .. (861,348.61) .. controls (861,378.77) and (809.96,403.22) .. (747,403.22) .. controls (684.04,403.22) and (633,378.77) .. (633,348.61) -- cycle ;
\draw    (747,332.22) -- (754.2,340.22) ;
\draw    (747,340.08) -- (755,332.36) ;

\draw    (630,341.22) -- (637.2,349.22) ;
\draw    (630,349.08) -- (638,341.36) ;

\draw   (699,357.72) .. controls (699,354.13) and (701.69,351.22) .. (705,351.22) .. controls (708.31,351.22) and (711,354.13) .. (711,357.72) .. controls (711,361.31) and (708.31,364.22) .. (705,364.22) .. controls (701.69,364.22) and (699,361.31) .. (699,357.72) -- cycle ; \draw   (700.76,353.12) -- (709.24,362.31) ; \draw   (709.24,353.12) -- (700.76,362.31) ;
\draw    (784,358.22) -- (791.2,366.22) ;
\draw    (784,366.08) -- (792,358.36) ;

\draw   (275,577.61) .. controls (275,547.45) and (326.04,523) .. (389,523) .. controls (451.96,523) and (503,547.45) .. (503,577.61) .. controls (503,607.77) and (451.96,632.22) .. (389,632.22) .. controls (326.04,632.22) and (275,607.77) .. (275,577.61) -- cycle ;
\draw    (389,561.22) -- (396.2,569.22) ;
\draw    (389,569.08) -- (397,561.36) ;

\draw    (272,570.22) -- (279.2,578.22) ;
\draw    (272,578.08) -- (280,570.36) ;

\draw   (341,586.72) .. controls (341,583.13) and (343.69,580.22) .. (347,580.22) .. controls (350.31,580.22) and (353,583.13) .. (353,586.72) .. controls (353,590.31) and (350.31,593.22) .. (347,593.22) .. controls (343.69,593.22) and (341,590.31) .. (341,586.72) -- cycle ; \draw   (342.76,582.12) -- (351.24,591.31) ; \draw   (351.24,582.12) -- (342.76,591.31) ;
\draw    (410,594.22) -- (417.2,602.22) ;
\draw    (410,602.08) -- (418,594.36) ;

\draw   (397,137.72) .. controls (397,134.13) and (399.69,131.22) .. (403,131.22) .. controls (406.31,131.22) and (409,134.13) .. (409,137.72) .. controls (409,141.31) and (406.31,144.22) .. (403,144.22) .. controls (399.69,144.22) and (397,141.31) .. (397,137.72) -- cycle ; \draw   (398.76,133.12) -- (407.24,142.31) ; \draw   (407.24,133.12) -- (398.76,142.31) ;

\draw (396,39.4) node [anchor=north west][inner sep=0.75pt]   [font=\tiny]  {$F_{0}$};
\draw (278,59.4) node [anchor=north west][inner sep=0.75pt]   [font=\tiny]  {$F_{\infty }$};
\draw (107,125.4) node [anchor=north west][inner sep=0.75pt]  [font=\tiny]  {$4d\ N=2\ SCFT$};
\draw (280,275.4) node [anchor=north west][inner sep=0.75pt]    [font=\tiny] {$F_{\infty }$};
\draw (96,328.4) node [anchor=north west][inner sep=0.75pt]    [font=\tiny] {$5d\ N=1\ KK$};
\draw (627,278.4) node [anchor=north west][inner sep=0.75pt]    [font=\tiny] {$F_{\infty }$};
\draw (662,368.4) node [anchor=north west][inner sep=0.75pt]   [font=\tiny]  {$gauge\ theory$};
\draw (98,523.4) node [anchor=north west][inner sep=0.75pt]   [font=\tiny]  {$6d\ ( 1,0) \ KK$};
\draw (269,507.4) node [anchor=north west][inner sep=0.75pt]   [font=\tiny]  {$F_{\infty }$};

\end{tikzpicture}

\end{center}
\caption{The combination of singularities of Coulomb branch solutions in various dimension.}
\label{combination}
\end{figure}

\textbf{4d \(\mathcal{N}=2\) SCFT:} This case has been discussed in \cite{Xie:2023zxn}. The classification relies on requiring a \(\mathbb{C}^*\) action that reflects the \(U(1)_R\) symmetry of \(\mathcal{N}=2\) SCFTs. By further imposing that the singularities defined by \(f(x,t)\) must be isolated, one achieves a complete classification. There are two singularities on \(\mathbb{P}^1\), which are dual to each other \cite{Xie:2023lko}. We notice several important aspects of the one-parameter family:
\begin{itemize}
\item First, only ADE singularities are found in all one-parameter families from \cite{Xie:2023zxn}, and we assume this restriction holds for subsequent studies. The singularity at \(\infty\), however, may allow non-ADE singularities, as they are not detected in the physical low-energy theory. For example, \(y^2 = x^5 + t^3\) has a non-ADE singularity at \(\infty\) given by \(y^2 = x t^4 + t x^6\).
\item A subtlety arises when the one-parameter family has genus less than \(g\). For instance, the rank-two SCFT with scaling dimensions \((4,3)\) uses the curve \(y^2 = x^4 t + t^2\), which has genus one for generic \(t\) \footnote{We assume the divisor class for the branch locus is $6F+4S$.}.
\item Conversely, substituting \(t\) with a polynomial \(P\) can increase the genus beyond \(g\). For example, the curve \(y^2 = t^2 + t x^6\) (rank two) becomes genus three when \(t\) is replaced by \(P = u x^2 + v x + t\), yielding scaling dimensions \((6,5,4)\).
\item In extreme cases, no quasi-homogeneous family of curves exists for the SCFT (e.g., \(y^2 = t^3\)). To obtain a genus-\(g\) family, one must include mass deformations and polynomial substitutions \(P\). This occurs in rank-two \(E_n\) theories.
\end{itemize}

\textbf{4d \(\mathcal{N}=2\) Asymptotically Free Theories:} These theories fall into two classes. The first class involves gauge groups coupled with free hypermultiplets. A SCFT limit is achievable by choosing the proper number of hypermultiplets (e.g., \(SU(n)\) with \(2n\) hypermultiplets in the fundamental representation). Starting from the SCFT curve, deformations yield asymptotically free theories. For example, the \(SU(3)\) SCFT with \(n_f=6\) hypermultiplets corresponds to the curve \(y^2 = t^2 + x^6\). The general \(n_f\) case uses \(y^2 = t^2 + x^{n_f}\). The second class involves gauge groups coupled with strongly coupled matter, where a SCFT limit is impossible. Here, the curve must be found by analyzing the singular fiber at \(\infty\). 
The curve is not quasi-homogeneous, but one may restore the homogeneous property by using the dynamical generated scale.

\textbf{5d \(\mathcal{N}=1\) KK Theories:} Without a \(\mathbb{C}^*\) action of SCFT or SCFT limit of asymptotical free theories, the strategy differs. The guideline is that the singular fiber (the smooth model) at \(\infty\) must be a chain of rational curves with a single loop. Singularities in the bulk are categorized:
\begin{enumerate}
\item One large singularity with large Euler number and one \(I_1\) singularity. The IR theory associated with the large singularity can not be described by a non-abelian gauge theory. The extra $I_1$ singularity comes from the KK charge.
\item One large singularity and two \(I_1\) singularities. The IR theory of the large singularity is a \textbf{non-abelian gauge theory},  and so the gauge coupling would give rise to an extra flavor symmetry in 5d, and an extra
$I_1$ singularity besides that associated with KK charge. 
\end{enumerate}

\textbf{6d \((1,0)\) KK Theories:} The singular fiber at \(\infty\) must be a chain of rational curves with two loops. We seek such fibers with small Euler number. The bulk contains a large singularity and two \(I_1\) singularities (see Figure~\ref{combination}).
The extra two $I_1$ singularities comes from two KK charges.

Once we find out the one-parameter family $y^2=f(x,t)$, the mass deformation can be found by finding the versal deformation of curve $f(x,t)$.

In the following sections, we will implement this strategy to construct Seiberg-Witten geometries for rank-one and rank-two theories in four, five and six dimensions.

\section{Seiberg-Witten Geometry for Rank One Theories}

In this section, we utilize the methods developed in the previous section to construct the Seiberg-Witten (SW) geometry for rank one theories. 
Some earlier results of SW geometry of rank one theories can be found in \cite{Seiberg:1994aj,Seiberg:1994rs,Argyres:1995xn,Minahan:1996cj}, which are expressed in terms of the Weierstrass form. By using the double covering trick, our method easily recovers all 
the old results. 

\subsection{SW Geometry from Double Covering}
As discussed earlier, the SW geometry arises from a one-parameter family of genus one fibrations, with the total space being an algebraic surface \( S \). The singular model of \( S \) is realized as a double cover \( f: S \to \mathbb{F}_0 \), branched over a divisor \( B \subset \mathbb{F}_0 (=\mathbb{P}^1\times \mathbb{P}^1)\). The divisor class of \( B \) is \( 4F + 2S \), where \( F \) and \( S \) are generators of the Picard group of \( \mathbb{F}_0 \).
This particular choice of $B$ would ensure that the total space $S$ is a rational surface, see appendix. \ref{double}.

Near the point \( (0, 0) \) on $\mathbb{F}_0 $ (with local coordinates \( (x, t) \)), the surface is given by:
\[
y^2 = a_0(x) + a_1(x)t + a_2(x)t^2
\]
Near \( (\infty, \infty) \), the surface takes the form:
\[
y^2 = x'^4 a_0\left(\frac{1}{x'}\right)s^2 + x'^4 a_1\left(\frac{1}{x'}\right)s + x'^4 a_2\left(\frac{1}{x'}\right)
\]
and now the local coordinate on $\mathbb{F}_0 $ is $(x^{'}, s)$.

The singular fiber type is determined via Tate's algorithm (see Table~\ref{tate}), which requires computing the discriminant \( \Delta \) and the \( j \)-invariant of the curve. For a quartic curve \( y^2 = ax^4 + bx^3 + cx^2 + dx + e \), the \( j \)-invariant is:
\begin{align*}
j &= \frac{I^3}{I^3 - 27J^2}, \\
I &= ae - \frac{bd}{4} + \frac{c^2}{12}, \\
J &= \frac{ace}{6} - \frac{ad^2}{16} - \frac{b^2e}{16} + \frac{bcd}{48} - \frac{c^3}{216}.
\end{align*}

\begin{table}[htp]
\caption{Classification of Singular Fibers via Tate's Algorithm}
\centering
\begin{tabular}{|c|c|c|c||c|c|c|c|} \hline
\textbf{Type} & \( \mathrm{ord}_t(\Delta) \) & \( \mathrm{ord}_t(j) \) & \textbf{Euler} & \textbf{Type} & \( \mathrm{ord}_t(\Delta) \) & \( \mathrm{ord}_t(j) \) & \textbf{Euler} \\ \hline
\( I_0 \) & 0 & \( \geq 0 \) & 0 & \( I_0^* \) & 6 & \( \geq 0 \) & 6 \\ \hline
\( I_m \) & \( m \) & \( -m \) & \( m \) & \( I_m^* \) & \( 6+m \) & \( -m \) & \( 6+m \) \\ \hline
\( II \) & 2 & \( \geq 0 \) & 2 & \( II^* \) & 10 & \( \geq 0 \) & 10 \\ \hline
\( III \) & 3 & \( \geq 0 \) & 3 & \( III^* \) & 9 & \( \geq 0 \) &  (9) \\ \hline
\( IV \) & 4 & \( \geq 0 \) & 4 & \( IV^* \) & 8 & \( \geq 0 \) & 8 \\ \hline
\end{tabular}
\label{tate}
\end{table}

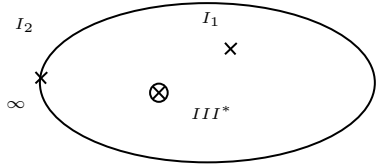
\begin{figure}[H]
\begin{center}

\tikzset{every picture/.style={line width=0.75pt}} 

\begin{tikzpicture}[x=0.55pt,y=0.55pt,yscale=-1,xscale=1]

\draw   (317,161.61) .. controls (317,131.45) and (368.04,107) .. (431,107) .. controls (493.96,107) and (545,131.45) .. (545,161.61) .. controls (545,191.77) and (493.96,216.22) .. (431,216.22) .. controls (368.04,216.22) and (317,191.77) .. (317,161.61) -- cycle ;
\draw    (314,154.22) -- (321.2,162.22) ;
\draw    (314,162.08) -- (322,154.36) ;

\draw   (392,168.47) .. controls (392,165.02) and (394.69,162.22) .. (398,162.22) .. controls (401.31,162.22) and (404,165.02) .. (404,168.47) .. controls (404,171.92) and (401.31,174.72) .. (398,174.72) .. controls (394.69,174.72) and (392,171.92) .. (392,168.47) -- cycle ; \draw   (393.76,164.05) -- (402.24,172.89) ; \draw   (402.24,164.05) -- (393.76,172.89) ;
\draw    (443,134.22) -- (450.2,142.22) ;
\draw    (443,142.08) -- (451,134.36) ;

\draw (298,115.4) node [anchor=north west][inner sep=0.75pt]    [font=\tiny]  {$I_{2}$};
\draw (292,172.4) node [anchor=north west][inner sep=0.75pt]    [font=\tiny]  {$\infty $};
\draw (418,175.4) node [anchor=north west][inner sep=0.75pt]  [font=\tiny]  {$III^{*}$};
\draw (425,111.4) node [anchor=north west][inner sep=0.75pt]   [font=\tiny]   {$I_{1}$};

\end{tikzpicture}

\end{center}
\caption{The singularity configurations for the curve $y^2= tx^4 + t^2(x + 1)^2$.}
\label{5de7}
\end{figure}

\textbf{Example}: Consider the curve \( y^2 = tx^4 + t^2(x + 1)^2 \). At \( t = 0 \):
\[
\Delta = 16 t^9 (16 + t), \quad j = 1 + \frac{3t}{16} + \frac{7t^2}{768} + \cdots
\]
We have $ord_t(\Delta)=9,~~ord_t  j=0$.
By Table~\ref{tate}, this corresponds to a \( III^* \) singularity. At \( t = -16 \):
\[
j = -\frac{16}{27(t + 16)} + \frac{4}{9} + \cdots,
\]
indicating an \( I_1 \) singularity (as $ord_{t^{'}}(\Delta)=1,~~ord_{t^{'}}  j=-1$). At \( t = \infty \):
\[
\Delta = 16t^{'2}(1 + 16t^{'}), \quad j = \frac{1}{108t^{'2}} + \frac{5}{27t^{'}} + \cdots,
\]
yielding an \( I_2 \) singularity  (as $ord_{t^{'}}(\Delta)=2,~~ord_{t^{'}}  j=-2$). The singular configuration is depicted in Figure~\ref{5de7} and corresponds to the SW geometry of a 5d theory.

The smooth model (relative minimal model) for singular fibers \( I_n, I_n^*, II^*, III^*, IV^* \)  are well known and corresponds to affine ADE Dynkin diagrams.

\subsubsection{4d Theory}
The SW geometry for 4d \( \mathcal{N}=2 \) SCFTs requires singular fibers at infinity that form a tree of rational curves. SCFTs arise from singularities \( II, III, IV, II^*, III^*, IV^*, I_0^* \). These geometries, admitting a \( \mathbb{C}^* \)-action, are classified (see \cite{Xie:2023zxn}), listed in Table~\ref{tab:4dSCFT}. Curves near \( (0,0) \) and \( (\infty, \infty) \) are shown:

\begin{table}[H]
\caption{4d \( \mathcal{N}=2 \) SCFT Curves}
\centering
\resizebox{5in}{!}{\begin{tabular}{|l|c|l|c|} \hline
\( 0 \) & Scaling Dim & \( \infty \) & Scaling Dim \\ \hline
\( y^2 = x^3 + t \) & \( 6/5 \) & \( y^2 = xt^2 + tx^4 \) & \( 6 \) \\ \hline
\( y^2 = x^4 + t \) & \( 4/3 \) & \( y^2 = t^2 + tx^4 \) & \( 4 \) \\ \hline
\( y^2 = x^3 + t^2 \) & \( 3/2 \) & \( y^2 = xt^2 + x^4 \) & \( 3 \) \\ \hline
\( y^2 = x^4 + t^2 \) & \( 2 \) & \( y^2 = x^4 + t^2 \) & \( 2 \) \\ \hline
\( y^2 = x^3 + xt \) & \( 4/3 \) & \( y^2 = xt^2 + x^3t \) & \( 4 \) \\ \hline
\( y^2 = x^4 + xt \) & \( 3/2 \) & \( y^2 = t^2 + x^3t \) & \( 3 \) \\ \hline
\( y^2 = x^3 + xt^2 \) & \( 2 \) & \( y^2 = xt^2 + x^3 \) & \( 2 \) \\ \hline
\( y^2 = t(x^3 + x^2 + x + 1) \) & \( 2 \) & \( y^2 = t(x^3 + x^2 + x + 1) \) & \( 2 \) \\ \hline
\end{tabular}}
\label{tab:4dSCFT}
\end{table}

For asymptotically free theories (e.g., \( I_n^* \) fibers), the one-parameter curve simplifies to \( y^2 = t^2 + x^n \). The mass deformed curve is:
\[
y^2 = \left( \Lambda x^2 + mx + t \right)^2 + \prod_{i=1}^n (x - m_i)
\]
where \( \sum m_i = 0 \). Here, \( t \) is the Coulomb branch operator, \( m \) is a mass parameter for \( U(1) \), and \( \Lambda \) is the dynamical scale. Table~\ref{tab:4dAF} lists these cases.

\begin{table}[H]
\caption{4d Asymptotically Free Theories}
\centering
\begin{tabular}{|l|c|l|c|} \hline
\( 0 \) & Singularities & \( \infty \) & Type \\ \hline
\( (x^2 + t)^2 + x^3 \) & \( IV + I_1 \) & \( (x^2 + t)^2 + x t^2 \) & \( I_1^* \) \\ \hline
\( (x^2 + t)^2 + x^2 \) & \( I_2 + 2I_1 \) & \( (x^2 + t)^2 + x^2 t^2 \) & \( I_2^* \) \\ \hline
\( (x^2 + t)^2 + x \) & \( 3I_1 \) & \( (x^2 + t)^2 + x^3 t^2 \) & \( I_3^* \) \\ \hline
\( (x^2 + t)^2 + 1 \) & \( 2I_1 \) & \( (x^2 + t)^2 + x^4 t^2 \) & \( I_4^* \) \\ \hline
\end{tabular}
\label{tab:4dAF}
\end{table}

\subsubsection{5d Theory}
For 5d theories, the singular fiber at infinity must form a chain of rational curves with one loop, requiring type \( I_n \) ( \( n > 0 \) ). Table~\ref{tab:5dTheories} lists examples with large flavor symmetries.

\begin{table}[H]
\caption{5d \( \mathcal{N}=1 \) Theories}
\centering
\resizebox{5in}{!}{\begin{tabular}{|c|c|c|c|c|} \hline
\( \infty \) & Type & \( 0 \) & Flavor Symmetry & Singularities \\ \hline
\( t + x^2(x+1) \) & \( I_1 \) & \( tx^4 + t^2x(x+1) \) & \( E_8 \) & \( II^* + I_1 \) \\ \hline
\( t + x^2(x+1)^2 \) & \( I_2 \) & \( tx^4 + t^2(x+1)^2 \) & \( E_7 \) & \( III^* + I_1 \) \\ \hline
\( (x+1)t + x^2(x+1)^2 \) & \( I_3 \) & \( tx^3(x+1) + t^2(x+1)^2 \) & \( E_6 \) & \( IV^* + I_1 \) \\ \hline
\( t^2 + x^2(x+1)^2 \) & \( I_4 \) & \( x^4 + t^2(x+1)^2 \) & \( E_5 \) & \( I_0^* + 2I_1 \) \\ \hline
\( xt^2 + x^2(x+1)^2 \) & \( I_6 \) & \( x^3 + t^2(x+1)^2 \) & \( SU(3) \times SU(2) \) & \( IV + 2I_1 \) \\ \hline
\( x(x+1)t^2 + x^2(x+1)^2 \) & \( I_8 \) & \( x^2(x+1) + t^2(x+1)^2 \) & \( SU(2) \) & \( I_2 + 2I_1 \) \\ \hline
\end{tabular}}
\label{tab:5dTheories}
\end{table}
Notice that the Seiberg-Witten (SW) differential is a meromorphic one-form with poles at ( $x = - 1$ ). 
The period integrals of the SW differential along cycles encircling $x=-1$ yields an additional charge vector corresponding to the winding modes of the circle compactification of 5d theory.

Additional examples with smaller flavor symmetries are provided in Table~\ref{tab:5dOther}:

\begin{table}[htp]
\caption{Other 5d \( \mathcal{N}=1 \) Theories}
\centering
\resizebox{5in}{!}{\begin{tabular}{|c|c|c|c|c|} \hline
\( 0 \) & Flavor Symmetry & Singularities & \( \infty \) Type \\ \hline
\( x^3 + (t^2 -6it +11)x^2 +40(1 - it)x + (48 -64it) \) & \( SU(5) \) & \( I_5 + 2I_1 \) & \( I_5 \) \\ \hline
\( x^3 + (-it^2 + 2t +3i)x^2 + (-3 + i2t)x -i \) & \( SU(3) \) & \( III + 2I_1 \) & \( I_7 \) \\ \hline
\( x^3 + \left(t^2 - \frac{3it}{\sqrt[3]{2}} - \frac{9}{2^{2/3}}\right)x^2 + \frac{1}{4}\left(3 \times 2^{2/3} +4it\right)x -\frac{1}{4} \) & \( U(1) \) & \( II + 2I_1 \) & \( I_8 \) \\ \hline
\( x^3 + (t^2 -3it - \frac{9}{4})x^2 + \left(\frac{3}{2} + it\right)x - \frac{1}{4} \) & None & \( 3I_1 \) & \( I_9 \) \\ \hline
\end{tabular}}
\label{tab:5dOther}
\end{table}

\subsubsection{6d Theory}
The SW geometry for 6d theories requires an \( I_0 \) fiber at infinity, implying an \( E_8 \) flavor symmetry. The curve is an irrelevant deformation of the 4d \( E_8 \) SCFT curve \( y^2 = tx^4 + t^2x \), taking the form:
\[
y^2 = tx^4 + t^2x f_2(x),
\]
where \( f_2(x) \) is a generic quadratic polynomial. Table~\ref{tab:6dTheories} summarizes this.

\begin{table}[H]
\caption{6d \( \mathcal{N}=(1,0) \) Theories}
\centering
\begin{tabular}{|c|c|c|c|c|} \hline
\( \infty \) & Type & \( 0 \) & Flavor Symmetry & Singularities \\ \hline
\( t + x(x+1)(x-1) \) & \( I_0 \) & \( tx^4 + t^2x(x-1)(x+1) \) & \( E_8 \) & \( II^* + 2I_1 \) \\ \hline
\end{tabular}
\label{tab:6dTheories}
\end{table}

\subsection{SW Geometry from \( \mathbb{P}^2 \) Bundles}
Most rank one SW geometries are expressed in Weierstrass form:
\[
y^2 = x^3 + a(t)x + b(t),
\]
embedded in \( \mathbb{P}^2 \)-bundles over \( \mathbb{P}^1 \) parameterized by $t$. Near \( t = \infty \), coordinates transform to \( x' = 1/x,~s=1/t \), yielding:
\[
y^2 = x'^3 + s^4 a(1/s)x' + s^6 b(1/s).
\]

For \( y^2 = x^3 + ax + b \), the discriminant and \( j \)-invariant simplify to:
\[
\Delta = 4a^3 + 27b^2, \quad j = \frac{4a^3}{\Delta}.
\]
From which one can easily find the singular fiber type by using Tate's algorithm.

4d SCFTs from \( \mathbb{P}^2 \)-bundles are listed in Table~\ref{tab:P2SCFTs}. For the SW geometries of asymptotically free and 5d theories in Weierstrass form, see \cite{Seiberg:1994aj,Seiberg:1994rs} and \cite{Yamada:1999xr}.

\begin{table}[H]
\caption{SW Geometry from \( \mathbb{P}^2 \)-bundles}
\centering
\begin{tabular}{|l|c|l|c|} \hline
\( 0 \) & Scaling Dim & \( \infty \) & Scaling Dim \\ \hline
\( y^2 = x^3 + t^3 \) & \( 2 \) & \( y^2 = x^3 + t^3 \) & \( 2 \) \\ \hline
\( y^2 = x^3 + t^4 \) & \( 3 \) & \( y^2 = x^3 + t^2 \) & \( 3/2 \) \\ \hline
\( y^2 = x^3 + xt^3 \) & \( 4 \) & \( y^2 = x^3 + xt \) & \( 4/3 \) \\ \hline
\( y^2 = x^3 + t^5 \) & \( 6 \) & \( y^2 = x^3 + t \) & \( 6/5 \) \\ \hline
\end{tabular}
\label{tab:P2SCFTs}
\end{table}

\newpage
\section{Seiberg-Witten Geometry for Rank Two Theories}
\subsection{Liu's Algorithm}
\label{sec:liu-algorithm}

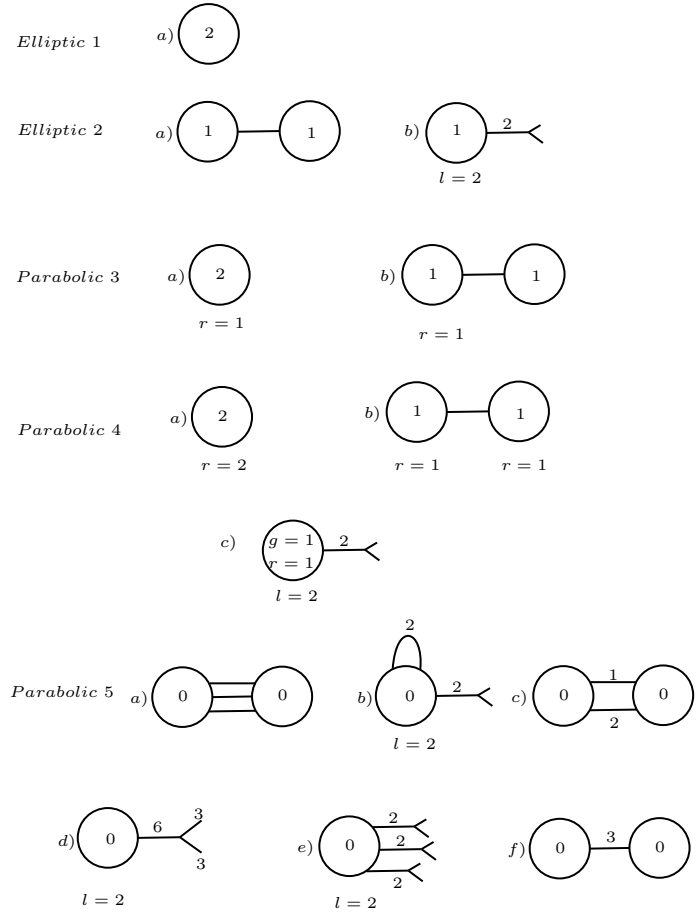
\begin{figure}[H]
 \begin{center}

\tikzset{every picture/.style={line width=0.75pt}} 

\begin{tikzpicture}[x=0.45pt,y=0.45pt,yscale=-1,xscale=1]

\draw   (176,72) .. controls (176,58.19) and (187.19,47) .. (201,47) .. controls (214.81,47) and (226,58.19) .. (226,72) .. controls (226,85.81) and (214.81,97) .. (201,97) .. controls (187.19,97) and (176,85.81) .. (176,72) -- cycle ;
\draw   (175,154) .. controls (175,140.19) and (186.19,129) .. (200,129) .. controls (213.81,129) and (225,140.19) .. (225,154) .. controls (225,167.81) and (213.81,179) .. (200,179) .. controls (186.19,179) and (175,167.81) .. (175,154) -- cycle ;
\draw    (225,154) -- (260,153.44) ;
\draw   (260,153.44) .. controls (260,139.63) and (271.19,128.44) .. (285,128.44) .. controls (298.81,128.44) and (310,139.63) .. (310,153.44) .. controls (310,167.24) and (298.81,178.44) .. (285,178.44) .. controls (271.19,178.44) and (260,167.24) .. (260,153.44) -- cycle ;
\draw   (382,155) .. controls (382,141.19) and (393.19,130) .. (407,130) .. controls (420.81,130) and (432,141.19) .. (432,155) .. controls (432,168.81) and (420.81,180) .. (407,180) .. controls (393.19,180) and (382,168.81) .. (382,155) -- cycle ;
\draw    (432,155) -- (467,154.44) ;
\draw    (467,154.44) -- (477,148.44) ;
\draw    (467,154.44) -- (479,163.44) ;
\draw   (362,274) .. controls (362,260.19) and (373.19,249) .. (387,249) .. controls (400.81,249) and (412,260.19) .. (412,274) .. controls (412,287.81) and (400.81,299) .. (387,299) .. controls (373.19,299) and (362,287.81) .. (362,274) -- cycle ;
\draw    (412,274) -- (447,273.44) ;
\draw   (447,273.44) .. controls (447,259.63) and (458.19,248.44) .. (472,248.44) .. controls (485.81,248.44) and (497,259.63) .. (497,273.44) .. controls (497,287.24) and (485.81,298.44) .. (472,298.44) .. controls (458.19,298.44) and (447,287.24) .. (447,273.44) -- cycle ;
\draw   (185,274) .. controls (185,260.19) and (196.19,249) .. (210,249) .. controls (223.81,249) and (235,260.19) .. (235,274) .. controls (235,287.81) and (223.81,299) .. (210,299) .. controls (196.19,299) and (185,287.81) .. (185,274) -- cycle ;
\draw   (187,393) .. controls (187,379.19) and (198.19,368) .. (212,368) .. controls (225.81,368) and (237,379.19) .. (237,393) .. controls (237,406.81) and (225.81,418) .. (212,418) .. controls (198.19,418) and (187,406.81) .. (187,393) -- cycle ;
\draw   (349,389) .. controls (349,375.19) and (360.19,364) .. (374,364) .. controls (387.81,364) and (399,375.19) .. (399,389) .. controls (399,402.81) and (387.81,414) .. (374,414) .. controls (360.19,414) and (349,402.81) .. (349,389) -- cycle ;
\draw    (399,389) -- (434,388.44) ;
\draw   (434,388.44) .. controls (434,374.63) and (445.19,363.44) .. (459,363.44) .. controls (472.81,363.44) and (484,374.63) .. (484,388.44) .. controls (484,402.24) and (472.81,413.44) .. (459,413.44) .. controls (445.19,413.44) and (434,402.24) .. (434,388.44) -- cycle ;
\draw   (246,505) .. controls (246,491.19) and (257.19,480) .. (271,480) .. controls (284.81,480) and (296,491.19) .. (296,505) .. controls (296,518.81) and (284.81,530) .. (271,530) .. controls (257.19,530) and (246,518.81) .. (246,505) -- cycle ;
\draw    (296,505) -- (331,504.44) ;
\draw    (331,504.44) -- (341,498.44) ;
\draw    (331,504.44) -- (343,513.44) ;
\draw   (154,628) .. controls (154,614.19) and (165.19,603) .. (179,603) .. controls (192.81,603) and (204,614.19) .. (204,628) .. controls (204,641.81) and (192.81,653) .. (179,653) .. controls (165.19,653) and (154,641.81) .. (154,628) -- cycle ;
\draw    (202,616.44) -- (240,616.44) ;
\draw   (237,627.44) .. controls (237,613.63) and (248.19,602.44) .. (262,602.44) .. controls (275.81,602.44) and (287,613.63) .. (287,627.44) .. controls (287,641.24) and (275.81,652.44) .. (262,652.44) .. controls (248.19,652.44) and (237,641.24) .. (237,627.44) -- cycle ;
\draw    (204,628) -- (237,627.44) ;
\draw    (201,640) -- (240,639.44) ;
\draw   (340,627) .. controls (340,613.19) and (351.19,602) .. (365,602) .. controls (378.81,602) and (390,613.19) .. (390,627) .. controls (390,640.81) and (378.81,652) .. (365,652) .. controls (351.19,652) and (340,640.81) .. (340,627) -- cycle ;
\draw    (390,627) -- (425,626.44) ;
\draw    (425,626.44) -- (435,620.44) ;
\draw    (425,626.44) -- (437,635.44) ;
\draw   (471,627) .. controls (471,613.19) and (482.19,602) .. (496,602) .. controls (509.81,602) and (521,613.19) .. (521,627) .. controls (521,640.81) and (509.81,652) .. (496,652) .. controls (482.19,652) and (471,640.81) .. (471,627) -- cycle ;
\draw    (519,615.44) -- (557,615.44) ;
\draw   (554,626.44) .. controls (554,612.63) and (565.19,601.44) .. (579,601.44) .. controls (592.81,601.44) and (604,612.63) .. (604,626.44) .. controls (604,640.24) and (592.81,651.44) .. (579,651.44) .. controls (565.19,651.44) and (554,640.24) .. (554,626.44) -- cycle ;
\draw    (518,639) -- (557,638.44) ;
\draw   (293,753) .. controls (293,739.19) and (304.19,728) .. (318,728) .. controls (331.81,728) and (343,739.19) .. (343,753) .. controls (343,766.81) and (331.81,778) .. (318,778) .. controls (304.19,778) and (293,766.81) .. (293,753) -- cycle ;
\draw   (468,755) .. controls (468,741.19) and (479.19,730) .. (493,730) .. controls (506.81,730) and (518,741.19) .. (518,755) .. controls (518,768.81) and (506.81,780) .. (493,780) .. controls (479.19,780) and (468,768.81) .. (468,755) -- cycle ;
\draw   (551,754.44) .. controls (551,740.63) and (562.19,729.44) .. (576,729.44) .. controls (589.81,729.44) and (601,740.63) .. (601,754.44) .. controls (601,768.24) and (589.81,779.44) .. (576,779.44) .. controls (562.19,779.44) and (551,768.24) .. (551,754.44) -- cycle ;
\draw    (518,755) -- (551,754.44) ;
\draw   (92,746) .. controls (92,732.19) and (103.19,721) .. (117,721) .. controls (130.81,721) and (142,732.19) .. (142,746) .. controls (142,759.81) and (130.81,771) .. (117,771) .. controls (103.19,771) and (92,759.81) .. (92,746) -- cycle ;
\draw    (142,746) -- (177,745.44) ;
\draw    (177,745.44) -- (195,731.44) ;
\draw    (177,745.44) -- (195,758.44) ;
\draw    (337,737) -- (372,736.44) ;
\draw    (372,736.44) -- (382,730.44) ;
\draw    (372,736.44) -- (384,745.44) ;
\draw    (343,756) -- (378,755.44) ;
\draw    (378,755.44) -- (388,749.44) ;
\draw    (378,755.44) -- (390,764.44) ;
\draw    (332,774) -- (367,773.44) ;
\draw    (367,773.44) -- (377,767.44) ;
\draw    (367,773.44) -- (379,782.44) ;
\draw    (354,604.44) .. controls (357,563.44) and (380,571.44) .. (377,604.44) ;

\draw (38,71.4) node [anchor=north west][inner sep=0.75pt]    [font=\tiny] {$Elliptic\ 1$};
\draw (195,64.4) node [anchor=north west][inner sep=0.75pt]    [font=\tiny] {$2$};
\draw (155,64.4) node [anchor=north west][inner sep=0.75pt]    [font=\tiny] {$a)$};

\draw (39,145.4) node [anchor=north west][inner sep=0.75pt]   [font=\tiny]  {$Elliptic\ 2$};
\draw (194,146.4) node [anchor=north west][inner sep=0.75pt]   [font=\tiny]  {$1$};
\draw (154,146.4) node [anchor=north west][inner sep=0.75pt]   [font=\tiny]  {$a)$};

\draw (280,148.4) node [anchor=north west][inner sep=0.75pt]   [font=\tiny]  {$1$};
\draw (401,145.4) node [anchor=north west][inner sep=0.75pt]   [font=\tiny]  {$1$};
\draw (361,145.4) node [anchor=north west][inner sep=0.75pt]   [font=\tiny]  {$b)$};

\draw (390,185.4) node [anchor=north west][inner sep=0.75pt]   [font=\tiny]  {$l=2$};
\draw (443,140.4) node [anchor=north west][inner sep=0.75pt]  [font=\tiny]  {$2$};
\draw (38,268.4) node [anchor=north west][inner sep=0.75pt]    [font=\tiny] {$Parabolic\ 3$};
\draw (39,396.4) node [anchor=north west][inner sep=0.75pt]   [font=\tiny]  {$Parabolic\ 4$};
\draw (33,616.4) node [anchor=north west][inner sep=0.75pt]   [font=\tiny]  {$Parabolic\ 5$};
\draw (381,266.4) node [anchor=north west][inner sep=0.75pt]  [font=\tiny]   {$1$};
\draw (341,266.4) node [anchor=north west][inner sep=0.75pt]  [font=\tiny]   {$b)$};

\draw (467,268.4) node [anchor=north west][inner sep=0.75pt]   [font=\tiny]  {$1$};
\draw (373,316.4) node [anchor=north west][inner sep=0.75pt]   [font=\tiny]  {$r=1$};
\draw (190,307.4) node [anchor=north west][inner sep=0.75pt]    [font=\tiny] {$r=1$};
\draw (204,266.4) node [anchor=north west][inner sep=0.75pt]   [font=\tiny]  {$2$};
\draw (164,266.4) node [anchor=north west][inner sep=0.75pt]   [font=\tiny]  {$a)$};

\draw (192,426.4) node [anchor=north west][inner sep=0.75pt]  [font=\tiny]   {$r=2$};
\draw (206,385.4) node [anchor=north west][inner sep=0.75pt]   [font=\tiny]  {$2$};
\draw (166,385.4) node [anchor=north west][inner sep=0.75pt]   [font=\tiny]  {$a)$};

\draw (368,381.4) node [anchor=north west][inner sep=0.75pt]   [font=\tiny]  {$1$};

\draw (328,381.4) node [anchor=north west][inner sep=0.75pt]   [font=\tiny]  {$b)$};

\draw (454,383.4) node [anchor=north west][inner sep=0.75pt]   [font=\tiny]  {$1$};
\draw (353,426.4) node [anchor=north west][inner sep=0.75pt]   [font=\tiny]  {$r=1$};
\draw (442,426.4) node [anchor=north west][inner sep=0.75pt]   [font=\tiny]  {$r=1$};
\draw (248,489.4) node [anchor=north west][inner sep=0.75pt]  [font=\tiny]   {$g=1$};
\draw (208,489.4) node [anchor=north west][inner sep=0.75pt]  [font=\tiny]   {$c)$};

\draw (254,535.4) node [anchor=north west][inner sep=0.75pt]   [font=\tiny]  {$l=2$};
\draw (307,490.4) node [anchor=north west][inner sep=0.75pt]  [font=\tiny]   {$2$};
\draw (248,508.4) node [anchor=north west][inner sep=0.75pt]  [font=\tiny]  {$r=1$};
\draw (173,620.4) node [anchor=north west][inner sep=0.75pt]   [font=\tiny]  {$0$};
\draw (133,620.4) node [anchor=north west][inner sep=0.75pt]   [font=\tiny]  {$a)$};

\draw (256,619.4) node [anchor=north west][inner sep=0.75pt]   [font=\tiny]  {$0$};
\draw (362,620.4) node [anchor=north west][inner sep=0.75pt]    [font=\tiny] {$0$};
\draw (322,620.4) node [anchor=north west][inner sep=0.75pt]    [font=\tiny] {$b)$};

\draw (352,659.4) node [anchor=north west][inner sep=0.75pt]  [font=\tiny]  {$l=2$};
\draw (401,612.4) node [anchor=north west][inner sep=0.75pt]  [font=\tiny]   {$2$};
\draw (490,619.4) node [anchor=north west][inner sep=0.75pt]   [font=\tiny]  {$0$};
\draw (450,619.4) node [anchor=north west][inner sep=0.75pt]   [font=\tiny]  {$c)$};

\draw (573,618.4) node [anchor=north west][inner sep=0.75pt]   [font=\tiny]  {$0$};
\draw (312,745.4) node [anchor=north west][inner sep=0.75pt]  [font=\tiny]   {$0$};
\draw (272,745.4) node [anchor=north west][inner sep=0.75pt]  [font=\tiny]   {$e)$};

\draw (487,747.4) node [anchor=north west][inner sep=0.75pt]  [font=\tiny]   {$0$};
\draw (447,747.4) node [anchor=north west][inner sep=0.75pt]  [font=\tiny]   {$f)$};

\draw (570,746.4) node [anchor=north west][inner sep=0.75pt]   [font=\tiny]  {$0$};
\draw (113,740.4) node [anchor=north west][inner sep=0.75pt]  [font=\tiny]   {$0$};
\draw (73,740.4) node [anchor=north west][inner sep=0.75pt]  [font=\tiny]   {$d)$};

\draw (93,790.4) node [anchor=north west][inner sep=0.75pt]  [font=\tiny]  {$l=2$};
\draw (152,730.4) node [anchor=north west][inner sep=0.75pt] [font=\tiny]  {$6$};
\draw (532,642.4) node [anchor=north west][inner sep=0.75pt]  [font=\tiny]   {$2$};
\draw (531,602.4) node [anchor=north west][inner sep=0.75pt] [font=\tiny]   {$1$};
\draw (529,738.4) node [anchor=north west][inner sep=0.75pt] [font=\tiny]   {$3$};
\draw (188,761.4) node [anchor=north west][inner sep=0.75pt]  [font=\tiny]  {$3$};
\draw (186,719.4) node [anchor=north west][inner sep=0.75pt]  [font=\tiny]  {$3$};
\draw (303,793.4) node [anchor=north west][inner sep=0.75pt]  [font=\tiny]  {$l=2$};
\draw (348,722.4) node [anchor=north west][inner sep=0.75pt]  [font=\tiny]   {$2$};
\draw (354,741.4) node [anchor=north west][inner sep=0.75pt] [font=\tiny]  {$2$};
\draw (351.5,777.12) node [anchor=north west][inner sep=0.75pt]  [font=\tiny]  {$2$};
\draw (362,560.4) node [anchor=north west][inner sep=0.75pt]  [font=\tiny]   {$2$};

\end{tikzpicture}
 
 \end{center}
 \caption{All possible weighted graphs for genus two Riemann surface. The number in the vertex represents the genus of the component, and the number $r$ denotes the number of internal cut. The general rule for finding the low energy theory is following: each vertex represents a matter system, which each edge would give a gauge group which is coupled with the matter system of its connecting node.}
  \label{weightedgenustwo}
 \end{figure}

In this section, we analyze the Seiberg-Witten geometry of rank two theories. Unlike rank one theories, where singular fibers are classified by affine ADE Dynkin diagrams \cite{barth2003compact}, rank two theories exhibit a vastly richer classification. The foundational work by Ueno and Namikawa \cite{namikawa1973complete} uses period integrals, while Matsumoto--Montesinos (MM) provide a more physically intuitive framework via mapping class group conjugacy classes \cite{matsumoto2011pseudo}. The MM formalism is particularly valuable for extracting low-energy physics: their combinatorial decomposition of Riemann surfaces yields gauge theory descriptions analogous to class $S$ constructions \cite{Gaiotto:2009we}. The weighted graphs encoding rank two singular fibers are depicted in Figure \ref{weightedgenustwo}.

\begin{figure}[htbp]
\centering
\includegraphics[totalheight=5cm]{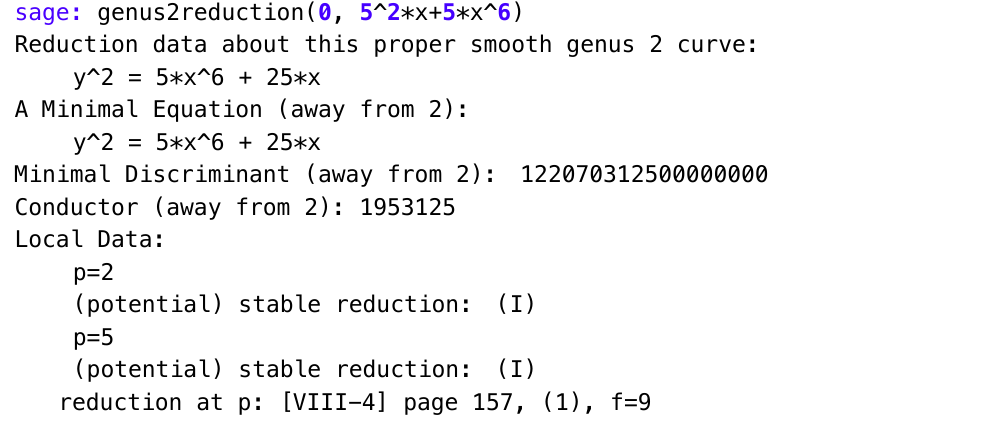}
\caption{Determining singular fibers for genus two families using Sage. The singular fiber is $VIII-4$ by looking at the result for characteristic $p=5$.}
\label{fig:genus2}
\end{figure}

MM's framework allows constructing smooth models of singular fibers via dual graphs, analogous to affine ADE diagrams in rank one. These dual graphs further encode 3d quiver gauge theories that serve as 3d mirrors of the infrared (IR) physics associated with local singularities \cite{Intriligator:1996ex}. 

However, directly determining the singular fiber type from a local equation with a degenerating central fiber remains challenging. While canonical resolutions \cite{horikawa1976deformations} or Tate-like algorithms exist for rank two cases, Liu's algorithm for genus two case \cite{liu1993courbes,liu1994conducteur} is  most useful for us. A practical alternative is provided by the \texttt{genus2reduction} function in Sage \cite{sagemath}, which simplifies computations using finite fields to reduce complexity.
See figure. \ref{fig:genus2}.

\paragraph{Example:} Consider the genus two family
\[
y^2 = t^2 x + t x^6.
\]
To analyze the singular fiber at $t=0$, we reduce the equation over $\mathbb{F}_5$:
\[
y^2 = \mathbf{5}^2 x + \mathbf{5} x^6 \quad (\text{in characteristic } p=5).
\]
Applying \texttt{genus2reduction} with $Q(x) = \mathbf{5}x^6$ and $P(x) = \mathbf{5}^2 x$, we find the singular fiber type at $t=0$ is $[VIII-4]$ (see \cite[p.~157]{namikawa1973complete}).

The singular fibers of rank two theories exhibit the following structural properties:
\begin{enumerate}
\item Each fiber is labeled using notation analogous to rank one theory \cite{namikawa1973complete}, and these labels are commonly used to denote singular fibers.
\item Two topological invariants $d_x$ and $\delta_x$ characterize each fiber \cite{kenji1988discriminants}. The quantity $d_x$ is computed from holomorphic data, while $\delta_x$ is determined via dual graphs. In contrast to rank one theories, $d_x \neq \delta_x$ in general. A local equation is called \emph{minimal} if the discriminant's order of vanishing satisfies $\mathrm{ord}_{t=0}(\Delta(t)) = d_x$.
\item Globally, the sum of all topological invariants satisfies $\sum d_x = 20$, a constraint arising from the geometry of the total space.
\end{enumerate}

\vspace{-0.3cm}

\subsection{4d theory}

\subsubsection{SCFT: non-degenerate family}

The Seiberg-Witten (SW) geometry for 4d $\mathcal{N}=2$ superconformal field theories (SCFTs) is constructed by finding a one-parameter family of curves of the form 
\[
y^2 = f(x, t),
\]
where $\mathbb{C}^*$ symmetry constraints dictate the structure of allowable families. This requirement ensures that the curve admits a $\mathbb{C}^*$ action. Results for specific families are summarized in Table~\ref{4da}, see the derivation in \cite{Xie:2023zxn}, where 
we derive the results by imposing: a): positive scaling dimension of $t$; b) isolated singularity at $x=t=0$ for $f$. Below are key observations:

\begin{itemize}
\item \textbf{Genus Increase via Substitutions:} 
Replacing $t$ with $P = vx + t$ (where $v$ and $t$ are Coulomb branch operators) can elevate the curve’s genus. For instance, substituting $P$ into $y^2 = t^2x + tx^6$ yields a genus-3 curve, corresponding to a rank-3 SCFT with scaling dimensions $\{10, 8, 6\}$. The associated 3d mirror theory arises from the singular fiber’s dual graph \cite{Xie:2023lko}. This theory can be engineered using a 6d $(2,0)$ $A_9$ theory on a sphere with punctures $[5^2]$, $[3^2, 2^2]$, and $[2^5]$ \footnote{The 6d $(2,0)$ $A_9$ theory is compactified on a sphere with the specified punctures.}. This means that the corresponding one-parameter family can not be used to engineer a rank two theory (it can be used to engineer a rank three model).

\item \textbf{Many interpretations:}
A single one-parameter family can describe both rank-2 and rank-3 theories. The dual graph of the singular fiber encodes the 3d mirror for rank-3 theories. To obtain a rank-2 theory, flavor symmetry must be frozen by truncating the dual graph’s tail. For example:
  \begin{itemize}
  \item \textit{Rank-2 Theory:} The family \( y^2 = t^2x + x^5t \) gives:
    \[
    y^2 = x^5(v_2x + t) + (v_2x + t)^2x,
    \]
    with scaling dimensions $\{8, 6\}$ and flavor symmetry $SO(20)$.
  \item \textit{Rank-3 Theory:} The same family can be extended to:
    \[
    y^2 = x^5(m^2x^3 + v_1x^2 + v_2x + t) + (m^2x^3 + v_1x^2 + v_2x + t)^2x,
    \]
    with scaling dimensions $\{8, 6, 4\}$. The flavor symmetry $SO(20) \times SU(2)$ arises from $m^2$’s coefficient in $x^3$, and the theory can be engineered using an $A_7$ $(2,0)$ theory with punctures $[4^2]$, $[3^2,1^2]$, and $[1^8]$.
  \end{itemize}

\item \textbf{Non-Genus $2$ Families:}
For non-genus-$2$ families (e.g., rank-2 theories with $\{4,3\}$ scaling dimensions, see table. \ref{4da}), additional deformations beyond $t$ are required to determine the singular fiber type at $\infty$. Consider $y^2 = x^4t + t^2$ (genus one for generic $t$). Introducing a mass deformation by substituting $t \to x^2 - t$ yields:
\[
y^2 = x^4(x^2 - t) + (x^2 - t)^2,
\]
which becomes genus two. Near $t = \infty$, the curve simplifies to:
\[
y^2 = t(x^2 - t) + x^2(x^2 - t)^2,
\]
and applying Liu’s algorithm identifies the fiber as $III-II^*_0$.
\end{itemize}

\begin{table}[H]
\caption{One parameter family for the SW geometry of 4d SCFT.}
\begin{center}
 \resizebox{5in}{!}{\begin{tabular}{|l|c| c | l|c| c|} \hline
 $0$& Physical Data & type  & $\infty$ & Physical Data & type  \\ \hline
 $y^2=x^5+t$ &$(\frac{10}{7}, \frac{8}{7}),~~No-flavor$ & $VIII1$ & $y^2=t^2x+tx^6$ & rank three $(10,8,6)$ & $VIII4$ \\ \hline
 $y^2=x^5+t^2$ &$(\frac{5}{2}, \frac{3}{2}),~~SU(5)$& $IX1$&$y^2=t^2x+x^6$ & $(5,3),~~~SO(14)\times U(1)$ & $IX4$\\ \hline 
 $y^2=x^6+t$ &$(\frac{6}{4}, \frac{5}{4}),~~U(1)$ & V& $y^2=t^2+tx^6$& rank three $(6,5,4)$ & $V^*$\\ \hline 
 $y^2=x^6+t^2$ & $(3,2),~~U(6)$&$III$ & $y^2=x^6+t^2$  &$(3,2),~~U(6)$ & $III$\\ \hline 
 $y^2= x^5+xt^2$ & $(4,2),~~SO(12)$&$VI$ & $y^2=x^5+xt^2$ & $(4,2),~~SO(12)$ & $VI$\\ \hline
 $y^2=x^5+xt$ & $(\frac{8}{5},\frac{6}{5}),~~SU(2)$&$VII$& $y^2=t^2x+x^5 t$ & $(8,6),~~SO(20)$ &$VII^*$ \\ \hline
 $y^2=x^6+xt$ & $(\frac{5}{3},\frac{4}{3}),~SU(2)\times U(1)$&$IX2$& $y^2=t^2+tx^5$ & $(5,4),~~SU(10)$ & $IX3$\\ \hline
 $y^2=x^4t+t^2$ &$(4,3),~~SU(8)\times SU(2)$ & $II_{1-4}+2I_1$& $y^2=t(x^2-t)+(x^2-t)^2x^2$ & ~&$III-II^*_0$ \\ \hline
 $y^2= x^4t+xt^2$ & $(6,4),~~SO(16)\times SU(2)$ & $\tilde{III}_2+2I_1$&  $y^2=t(x^2-t)+(x^2-t)^2x$ & ~&$II-II^*_0$\\ \hline
$y^2=tf_5(x)$& $(2,2),~~SU(2)^5$ & $I_{0-0-0}^*$&$y^2=tf_5(x)$& $(2,2)$ & $I_{0-0-0}^*$ \\ \hline
$y^2=x^5+t^3$& $(10,4),~~~E_8$ &$VIII2$& $y^2=xt^4+tx^6$& $N/A$ & $VIII3$ \\ \hline
\end{tabular}}
\end{center}
\label{4da}
\end{table}%

Once the one-parameter family is determined, one can compute the full Seiberg-Witten (SW) geometry, Coulomb branch spectrum, and flavor symmetry. Let us provide an example for completeness.

\textbf{Example}: Consider the one-parameter family \( y^2 = t^2 x + x^6 \). The physical information is derived as follows:

\begin{enumerate}
\item To determine the scaling dimensions \( [y] \), \( [t] \), and \( [x] \), we use two conditions:
    \begin{enumerate}
    \item The homogeneous condition on the one-parameter family yields:
        \[
        2[y] = 2[t] + [x],
        \]
        \[
        2[y] = 6[x].
        \]
    \item The SW differential \( \lambda_{\text{SW}} \) has scaling dimension \( 1 \), and its derivative obeys \( \partial_t \lambda_{\text{SW}} = \frac{dx}{y} \). This gives:
        \[
        1 - [t] = [x] - [y].
        \]
    \end{enumerate}
    Solving these equations yields the scaling dimensions: \( [t] = 5 \), \( [x] = 2 \), and \( [y] = 6 \). The Coulomb branch spectrum is encoded in the polynomial \( P(x) = ux + t \). Using the scaling dimensions of \( t \) and \( x \), we find \( [u] = [t] - [x] = 3 \).

\item The flavor symmetry is determined from the versal deformation of the singularity \( t^2x + x^6 \), which corresponds to a \( D_7 \) singularity. The flavor symmetry group is therefore \( SO(14) \). An additional \( U(1) \) symmetry arises from deformations in the polynomial \( P(x) = mx^2 + ux + t \), leading to the total flavor symmetry \( SO(14) \times U(1) \).

\item The full SW geometry is constructed by first deforming the one-parameter family with mass terms and substituting \( t \) with polynomials encoding Coulomb branch operators. The deformed family is:
\[
y^2 = x^6 + xt^2 + T_{12} + T_{10}x + T_8x^2 + T_6x^3 + T_4x^4 + T_2x^5 + T_7 t.
\]
The full geometry is obtained by substituting \( t \) with \( mx^2 + ux + t \), where \( m \) and \( u \) are parameters associated with mass and Coulomb branch operators, respectively.
\end{enumerate}

\subsubsection{SCFT:  degenerate family}

Another intriguing scenario arises when a one-parameter family degenerates into multiple points for generic \( t \), making it challenging to determine key properties such as the Coulomb branch scaling dimensions. This occurs for the case \( y^2 = t^3 \) \footnote{Other cases would be considered in next part.} so that there would only ADE singularity at total space. To analyze such theories, we introduce one extra deformation which seems to be the mass deformations. For instance, consider the family:
\[
y^2 = t^3 + m^6x^5,
\]
which degenerates when \( m = 0 \). By enforcing homogeneity conditions on the one-parameter family (assuming \( [m] = 1 \)), we derive three equations governing scaling dimensions:
\[
2[y] = 3[t], \quad 3[t] = 5[x] + 6.
\]
Additionally, the Seiberg-Witten differential is modified to:
\[
\lambda_{\text{SW}} = m R \frac{dx}{y},
\]
where \( R \) shares the same scaling as \( t \). This yields an extra constraint:
\[
1 + [t] + [x] - [y] = 1.
\]
Solving these equations yields the scaling dimensions:
\[
[y] = 18, \quad [x] = 6, \quad [t] = 12.
\]
This method applies to all rank-2 \( E_n \) theories. Specific examples are listed in Table~\ref{4db}. Note that since the branch curve $B$ is a cubic function in $t$,  its divisor class is chosen to be $6F+4S$.

\begin{table}[H]
\caption{SW geometry for the SCFT with degenerating one parameter family. To find the singular fiber at $t=\infty$, we need to first replace $t$ by the polynomial $x^2-t$ to get a genus two gamily and then change the coordinate near $t=\infty$.}
\begin{center}
 \resizebox{5in}{!}{\begin{tabular}{|l|c|c|l|c|c|} \hline
 $0$&Scaling   & Singularities& $\infty$ & type  \\ \hline
 $y^2=t^3+m^6x^5$ &$(12, 6)$ & $VIII2+I_1$& $y^2=t*(x^2-t)^3+t^4x$ & $2II-0$~\\ \hline
  $y^2=t^3+m^4tx^3$ &$(8, 4)$ & $II^*-III-\alpha+I_1$ & $y^2=t*(x^2-t)^3+t^3x(x^2-t)$ & $2III-0$ \\ \hline
  $y^2=t^3+m^6x^4$ &$(6, 3)$& $III^*-II_1+2I_1$& $y^2=t*(x^2-t)^3+t^4x^2$ & $2IV-0$ \\ \hline
  $y^2=t^3+m^6x^3$ &$(4, 2)$&$II_{1-1}+2I_1$ & $y^2=t*(x^2-t)^3+t^4x^{3}$ & $2I_0^*-0$\\ \hline
 $y^2=t^3+m^6x^2$ &$(3, \frac{3}{2})$& &$y^2=t*(x^2-t)^3+t^4x^{4}$ & $2IV^*-0$ \\ \hline
  $y^2=t^3+m^4tx$ &$(\frac{8}{3}, \frac{4}{3})$& &$y^2=t*(x^2-t)^3+t^3x(x^2-t)$ & $2III^*-0$ \\ \hline
  $y^2=t^3+m^6x$ &$(\frac{12}{5}, \frac{6}{5})$& &$y^2=t*(x^2-t)^3+t^4x^5$ & $2II^*-0$ \\ \hline
\end{tabular}}
\end{center}
\label{4db}
\end{table}%

\subsubsection{Asymptotically free theory}

The possible weighted graphs for describing non-abelian gauge theories were given in figure. \ref{weightedgenustwo}, see table \ref{weighted_af} for the data describing 4d asymptotical free theories. 
These correspond to the singular fibers at infinity. The task is to find a one-parameter family of curves such that the type of singular fibers at infinity matches the desired type. The difference with the superconformal field theory (SCFT) is the introduction of an extra scale due to the dynamically generated scale $\Lambda$. In the limit where $\Lambda$ vanishes, the curve must become homogeneous.

If the ultraviolet (UV) theory is an asymptotically free gauge theory, it contains an additional parameter—the dynamically generated scale $\Lambda$. By assigning a proper scaling dimension (commonly $[\Lambda] = 1$), one can determine the full Seiberg-Witten (SW) geometry. When $\Lambda = 0$, the curve becomes homogeneous. The following possibilities arise:
\[
y^2 = t^2, \quad y^2 = x t^2, \quad y^2 = f_n(x) t \quad (n \leq 6).
\]
where $f_n(x)$ is a degree $n$ polynomial in $x$.
Introducing the dynamically generated scale $\Lambda$, the curve takes the following forms:

(a) $y^2 = t^2 + \Lambda^n x^{6-n}$, \\

(b) $y^2 = x t^2 + \Lambda^n x^{5-n}$ and $y^2 = x t^2 + t^3 + \Lambda^{2-n} t x^n$, \\

(c) $y^2 = f_n(x) t + \Lambda^2 f_m(x)$.

The full SW curve can be constructed following the general approach outlined here. Results are summarized in table \ref{4dc}.

\textbf{Detailed analysis of linear $t$ dependence}: Consider the one-parameter geometry $y^2 = h_0(x) + h_1(x)t$. The SW differential is given by
\[
\lambda_{sw} = \frac{y \, dx}{h_1(x)} = \frac{y^2}{h_1(x)} \cdot \frac{dx}{y} = \left(t + \frac{h_0}{h_1}\right) \frac{dx}{y},
\]
which may have poles at the roots of $h_1(x)$. The number of poles corresponds to additional flavor symmetries, as these points must be excluded from the SW curve.
The pole structure on $\lambda_{sw}$ may link the class $S$ description. For example consider the cure $y^2=x^5+tx(x+1)$, so according to above formula, $\lambda_{sw}$ has 
a first order pole at $x=0, -1$, and may have an higher order pole at $x=\infty$ (which can be analyzed by changing the coordinate to $x=\infty$. The first order pole 
might be interpreted as the regular singularities in the class $S$ theory.  Indeed, the theory $H_1-SU(2)-2$ is described by a sphere with one irregular singularity (order $5/2$) and two regular singularities.

\begin{table}[H]
\caption{Singular fiber (at $\infty$) data for 4d asymptotical free theory.}
\begin{center}
 \resizebox{5in}{!}{\begin{tabular}{|c|c|c|c|} \hline
Type & Data & Name  &Gauge theory \\ \hline
Parabolic 3(a) & $f=(\frac{1}{2})+\frac{1}{4}+\frac{1}{4},~(\frac{1}{2})+\frac{3}{4}+\frac{3}{4}$ & $III-II_n^*,~III^*-II^*_n(178)$ & $H_2-SU(2)-(2-n)$ \\ \hline
& $f=(\frac{1}{3})+\frac{1}{2}+\frac{1}{6},~(\frac{2}{3})+\frac{1}{2}+\frac{5}{6}$ & $ II-II^*_n, II^*-II_n^* (176)$ &$H_1-SU(2)-(2-n)$  \\ \hline
& $f=(1)+\frac{1}{2}+\frac{1}{2}+\frac{1}{2}+\frac{1}{2}$ & $ I_{n-0-0}^* (171)$ & $(2-n)-su(2)-su(2)-0$ \\ \hline
Parabolic 4(a) & $f=(1)+(1)+\frac{1}{2}+\frac{1}{2}+\frac{1}{2}+\frac{1}{2}$ & $ I_{n-p-0}^* (180)$ & $(2-n)-su(2)-su(2)-(2-p)$\\ \hline
 & $f=(1)+\frac{3}{4}+\frac{1}{4}$ & $ \tilde{III}_n(182)$ & $sp(4)-(6-n)F$ \\ \hline
 Parabolic 4(c) & $f=(1)+\frac{1}{2}+\bf{\frac{1}{2}}~$ & $ 2I_n^*-0(181)$ &$V-sp(4)-nF$ \\ \hline
Parabolic 5(f) & $f_1=f_2=\frac{1}{3}+\frac{2}{3}+(1)~$ & $ III_n$ & $su(3)-(6-n)F$ \\ \hline
\end{tabular}}
\end{center}
\label{weighted_af}
\end{table}%

\begin{table}[H]
\caption{SW geometry for 4d Asymototical free theories. For the last entry, the branch locus $B$ of the double covering is in the divisor class $6F+4S$.}
\begin{center}
 \resizebox{5in}{!}{\begin{tabular}{|l|c|l|  l | c|} \hline
$\infty$ &  $type$ & $0$ &  Singularities & theory\\ \hline
$x(x^2+t)^2+t^2x^{n+1}$  & $\tilde{III}_{n} (182),~d_x=\delta_x=10+n$ & $x(x^2+t)^2+\Lambda^nx^{5-n}$ & $SO(12-2n)$ &$sp(4)-(6-n)F$ \\ \hline
$(x^3+ t)^2+t^2 x^{n}$ & $III_{n},~d_x=\delta_x=10+n$ & $(x^3+t)^2+\Lambda^nx^{6-n}$ & $U(6-n)$ & $su(3)-(6-n)F$  \\ \hline
$t^2+txf_2^{'}(x)(x-1)^2$ &  $I^*_{1-0-0},~d_x=\delta_x=11$ &$txf_2(x)(x-1)^2+\Lambda^2x^6$ & $IX2+3 I_1$ &  $1-su(2)-su(2)-2$\\ \hline
$t^2+tx(x+1)^2(x-1)^2$ &  $I^*_{1-1-0},~d_x=\delta_x=12$ &$tx(x+1)^2(x-1)^2+\Lambda^2x^6$ & $IX2+2I_1$ &  $1-su(2)-su(2)-1$\\ \hline
$xt^2+tf_3^{'}(x)x^2$ &  $I^*_{2-0-0},~d_x=\delta_x=12$ &$txf_3(x)+\Lambda^2x^5$ & $VII+3 I_1$  & $0-su(2)-su(2)-2$\\ \hline
$xt^2+tx^2(x+1)^2(x-1)$ &  $I^*_{1-2-0},~d_x=\delta_x=13$ &$tx(x+1)^2(x-1)+\Lambda^2x^5$ & $VII+2I_1$ & $0-su(2)-su(2)-1$ \\ \hline
$tx(x-1)[t+x(x-1)f_2(x)]$ &  $I^*_{2-2-0}~d_x=\delta_x=14$ &$tf_2(x)(x-1)^2+\Lambda^2x^4(x-1)$ & $VIII_1+2I_1$ & $0-su(2)-su(2)-0$ \\ \hline
$xt^2+tx^4(x+1)$ &  $\makecell {II^*-II^*_0(176), \\ d_x=\delta_x=14}$& $ x^5+tx(x+1)$ &   &$H_1-SU(2)-2$  \\ \hline
$xt^2+tx^4(x-1)^2$ & $\makecell {II^*-II^*_1(176), \\ d_x=\delta_x=15}$& $ x^5+t(x-1)^2$ &  &$H_1-SU(2)-1$  \\ \hline
$x(x-1)t^2+tx^4(x-1)^2$ &  $\makecell {II^*-II^*_2(176), \\ d_x=\delta_x=16}$& $ x^4(x-1)+t(x-1)^2$ &. & $H_1-SU(2)$  \\ \hline
$t^2+tx^4(x+1)$ &  $\makecell{III^*-II^*_0(178), \\ d_x=\delta_x=13}$& $ x^6+tx(x+1)$ &  & $H_2-SU(2)-2$  \\ \hline
$t^2+tx^4(x+1)^2$ &   $\makecell{III^*-II^*_1(178), \\ d_x=\delta_x=14}$& $ x^6+t(x-1)^2$ & & $H_2-SU(2)-1$  \\ \hline
$(x+1)t^2+tx^4(x+1)^2$ &   $\makecell{III^*-II^*_2(178), \\ d_x=\delta_x=15}$& $ x^5(x-1)+t(x-1)^2$ & & $H_2-SU(2)$  \\ \hline
$t(x^2+t)^3+t^2(x^2+t)^2x+x^{4-n}(x^2+t)t^3$ &   $\makecell{2I_{4-2n}^*-0(181), \\ d_x=\delta_x=15-2n}$& $ t^3+t^2x+tx^n,~~n=0,1$ & & $V-sp(4)-nF$  \\ \hline
\end{tabular}}
\end{center}
\label{4dc}
\end{table}%

\newpage
\subsection{5d  theory}
The possible singular fibers at $\infty$ for the 5D theory are listed in Table~\ref{5dsingular}. 
The corresponding one-parameter families are presented in Table~\ref{5da} and \ref{5db}. 
The search strategy relies on the following key conditions:
\begin{enumerate}
    \item The singular fiber at $\infty$ must coincide with one of the cases listed in Table~\ref{5dsingular}.
    \item The maximal power of $t$ is constrained to three, with bulk singularities restricted to ADE type.
\end{enumerate}

For theories with small flavor symmetry, constructing a simple family with a minimal number of bulk singularities becomes progressively more challenging. 
In such cases, we content ourselves with providing a basic family where the fiber at $\infty$ satisfies the desired properties.

\textit{Example}: To illustrate the main idea, consider the SW geometry defined by $f(x,t) = x^5 t + t^2 x (x + 1)$. 
Performing a coordinate change at $(x, t) = (\infty, \infty)$, we obtain the local equation 
\[
    y^2 = x t + x^4 (x + 1),
\]
where $x$ and $t$ are local coordinates near $\infty$. 
Using Sage, we verify that the singularity is of type $\text{IV-II}(0)$\footnote{The dual graph of this singularity can also be determined via the canonical resolution method.}, 
whose dual graph consists of a chain of rational curves forming a loop, see the diagram in page 175 of \cite{namikawa1973complete}.

The physical implications are extracted as follows:
\begin{itemize}
    \item[(a)] The discriminant is $\Delta = (256 - 27 t) t^{13}$, indicating two singularities on the finite $t$-plane.
    \item[(b)] The singularity type at $(x, t) = (0, 0)$ is $D_{10}$, implying that the flavor symmetry of this theory is $\mathfrak{so}(20)$.
\end{itemize}

\begin{table}[htp]
\caption{Singular fiber type at $\infty$ for a 5d theory. We use $\tilde{II}$ for the singular fiber $II$ in page 182 to distinguish it from that in page 183.}
\begin{center}
 \resizebox{4in}{!}{\begin{tabular}{|c|c|c|} \hline
Type & Data & Name \\ \hline
Parabolic 3(a) & $(\frac{1}{3})+(\frac{1}{3})+\frac{1}{3},~~(\frac{2}{3})+(\frac{2}{3})+\frac{2}{3} $ & $ (IV-II_n,~~IV^*-II_n) (175)$ \\ \hline
~& $(\frac{1}{4})+(\frac{1}{4})+\frac{2}{4},~~(\frac{3}{4})+(\frac{3}{4})+\frac{2}{4} $ & $ (III-II_n,~~III^*-II_n) (176)$ \\ \hline
~&$(\frac{1}{2})+(\frac{1}{2})+\frac{1}{2}+\frac{1}{2} $ & $II_{n-0} (171)$ \\ \hline
Parabolic 3(b) & $f_1=id,~~f_2=rank~one~T$ & $ I_n-T-m$ \\ \hline
Parabolic 4(a) & $f=(\frac{1}{2})+(\frac{1}{2})+\tilde{(1)}+\frac{1}{2}$ & $ \tilde{II}_{n-p} (182)$ \\ \hline
Parabolic 4(b) & $f_1=id,~~f_2=(1)+\frac{1}{2}+\frac{1}{2}$ & $ I_n-I_p^*-m (180)$ \\ \hline
Parabolic 4(c) & $f=id$ & $ 2I_n-m (181)$ \\ \hline
Parabolic 5(b) & $f=id~$ & $ II^*_{n-p}(184)$ \\ \hline
Parabolic 5(c) & $f_1=f_2=\frac{1}{2}+\frac{1}{2}+\bf{1}$ & $ II_{n-p} (183)$ \\ \hline
\end{tabular}}
\end{center}
\label{5dsingular}
\end{table}%

\begin{table}[H]
\caption{SW geometry for some 5d $\mathcal{N}=1$ theories}
\begin{center}
 \resizebox{5in}{!}{
\begin{tabular}{| l | l | l | l | l |} \hline
$\infty$ &  $Type$ & $0$ &  Flavor & Singularities\\ \hline
$xt+x^4(x+1)$ & $IV-II(0) (175),d_x=4$& $x^5 t+t^2x(x+1)$ & $SO(20)$ & $VII^*+I_1$  \\ \hline
$xt+x^4(x+1)^2$ & $IV-II(1)(175),~ d_x=5$& $x^5 t+t^2(x+1)^2$ & $SU(10)$ & $IX3+I_1$ \\ \hline
$x(x+1)t+x^4(x+1)^2$ & $IV-II(2)(175),~d_x=6$& $x^4(x+1) t+t^2(x+1)^2$ & $SU(8)\times SU(2)$ & $II^*(0)-III^*+I_1$  \\ \hline
$t(x^2+t)+(x^2+t)^2(x+1)$ & $II_{(1-0)}$ $(183),~d_x=\delta_x=6$& $x^{4} t+t^2x(x+1)$ & $SO(16)\times SU(2)$ & $\tilde{III}_2+2 I_1$  \textcolor{red}{T} \\ \hline
$t^2 +(x^2+t)^2(x+1)$& $II_{(1-0)}$ $(183),~d_x=\delta_x=6~$& $x^{6}+t^2x(x+1)$ & $SO(14)\times U(1)$ & $IX4+2I_1$  \\ \hline
~\\ \hline
$xt^2 +(x^2+t)^2(x+1)$& $II_{(1-1)}$ $(183),~d_x=\delta_x=7~$& $x^{5}+t^2x(x+1)$ & $SO(12)\times SU(2)$ & $IV-III^*-(-1)+2I_1$  \\ \hline
$tx(x^2+t)+(x^2+t)^2(x+1)$ & $II_{(1-2)}$ $(183),~d_x=\delta_x=8$& $x^{3}(x^2+t)+(x^2+t)^2x(x+1)$ & $SO(12)\times U(1)$ & $VI+2I_1$ \textcolor{red}{T} \\ \hline

$t^2 +x^4(x+1)^2$& $II_{(3-0)}$ $(183),~d_x=\delta_x=8~$& $x^{6}+t^2(x+1)^2$ & $SU(6)\times SU(2) \times SU(2)$ & $III+2I_1$  \\ \hline


$(x+1)t^2 +x^4(x+1)^2$& $II_{(5-0)}$ $(183),~d_x=\delta_x=10~$& $x^{5}(x+1)+t^2(x+1)^2$ & $SU(5)\times SU(2)$ & $IX1+2I_1$  \\ \hline

$tx^2(x^2+t)+(x^2+t)^2(1+2x)$ & $II_{(1-4)}$ $(183),~d_x=\delta_x=10$& $x^{2} (x^2+t)+(x^2+t)^2x(x+2)$ & $SU(5)\times SU(2)$  &$II_{2-1}+ 2I_1$\\ \hline
$tx^2(x^2+t)+(x^2+t)^2(x+a)^2$ & $II_{(2-4)}$ $(183),~d_x=\delta_x=11$& $x^{2} t+t^2(x+\sqrt{-1})^2$ & $SU(4)\times SU(2)$ & $II-II^*_0+ 2I_1$  \\ \hline
~\\ \hline
$x t^4+t(x^2+t)^3+(x^2+t)^2t^2 $& $2I_1-0$ $(181),~d_x=6,\delta_x=5$& $ x^5+(x^2+t)^3+(x^2+t)^2x^2$ & $E_8\times SU(2)$ & $VIII2+2I_1$  \\ \hline
$x t^4+t x^6+t^2 x^4$& $IV^*-II(0)$ $(178),~d_x=7,\delta_x=6$& $ x^5+t^3+t^2x^2$ & $E_8$ &$VIII2+I_1$  \\ \hline
$t(x^2+t)^3+(x^2+t)^2t^2+ t^3(x^2+t)x$& $2I_2-0$ $(181),~d_x=7,\delta_x=6$& $ (x^2+t)^3+(x^2+t)^2x^2+(x^2+t)x^3$ & $E_7\times SU(2)$ & $II^*-III-\alpha+2I_1$  \\ \hline
$t(x^2+t)^3+(x^2+t)^2t^2 +x^3t^4$& $2I_3-0$ $(181),~d_x=8,\delta_x=7$& $ (x^2+t)^3+(x^2+t)^2x^2+x^3$ & $E_6\times SU(2)$ & $II^*-III-\alpha+2I_1$  \\ \hline

\end{tabular}}
\end{center}
\label{5da}
\end{table}%

\begin{table}[H]
\caption{More SW families for 5d $\mathcal{N}=1$ theories. Here $G_l$ is the  exceptional flavor symmetry determined by the singularity $t^3+x^{6-l}$ or $t^3+tx^{4-l}$.}
\begin{center}
 \resizebox{5in}{!}{\begin{tabular}{| l | l | l | l | l |} \hline
$\infty$ &  $Type$ & $0$ &  Flavor\\ \hline


$tx^l(x^2+t)+(x^2+t)^2(x+1)$ & $II_{(1-2l)}$ $(183),~d_x=\delta_x=6+2l$& $x^{4-l} t+t^2x(x+1)$ & $SO(16-4l)\times SU(2)$  \\ \hline
$tx^l(x^2+t)+(x^2+t)^2(x+1)^2$ & $II_{(2-2l)}$ $(183),~d_x=\delta_x=7+2l$& $x^{4-l} t+t^2(x+1)^2$ & $SU(8-2l)\times SU(2)$  \\ \hline

$x^{l}t^2 +(x^2+t)^2(x+1)$& $II_{(1-l)}$ $(183),~d_x=\delta_x=6+l~$& $x^{6-l}+t^2x(x+1)$ & $SO(14-2l)\times SU(2)$  \\ \hline
$t^2 x^l+(x^2+t)^2(x+1)^2$& $II_{(3-l)}$ $(183),~d_x=\delta_x=8+l~$& $x^{6-l}+t^2(x+1)^2$ & $SU(6-l)\times SU(2)\times SU(2)$  \\ \hline
$t^2 x^l(x+1)+(x^2+t)^2(x+1)^2$& $II_{(5-l)}$ $(183),~d_x=\delta_x=10+l~$& $x^{5-l}(x+1)+t^2(x+1)^2$ & $SU(5-l)\times SU(2)$  \\ \hline
$t(x^2+t)^3+(x^2+t)^2t^2+t^4x^{l}$& $2I_l-0$ $(181),~d_x=5+l,\delta_x=4+l~$& $t^3+t^2x^2+x^{6-l}$ & $G_l\times SU(2)$  \\ \hline
$t(x^2+t)^3+(x^2+t)^2t^2+t^3(x^2+t)x^{4-l}$& $2I_{2l}-0$ $(181),~d_x=5+2l,\delta_x=4+2l~$& $t^3+t^2x^2+tx^{4-l}$ & $G_l\times SU(2)$  \\ \hline




\end{tabular}}
\end{center}
\label{5db}
\end{table}%

\begin{table}[htp]
\caption{The SW geometry of some known 5d  $\mathcal{N}=1$ theories constructed from 5d gauge theories.}
\begin{center}
 \resizebox{4in}{!}{\begin{tabular}{|c|c|c|}\hline
IR theory & Flavor & SW geometry $f(x,t)$ \\  \hline
$SU(3)$ with $n_f=9, |k|=\frac{1}{2}$ & $SO(20)$ &  $x^5 t+t^2x(x+1)$ \\ \hline
$SU(3)$ with $n_f=8, |k|=0$ & $SU(10)$ &$x^5 t+t^2(x+1)^2$ \\ \hline
$SU(3)$ with $n_f=8, |k|=1$ & $SO(16)\times SU(2)$ & $x^{4} t+t^2x(x+1)$  \\ \hline
$SU(3)$ with $n_f=7, |k|=\frac{1}{2}$ & $SU(8)\times SU(2)$ & $x^4(x+1) t+t^2(x+1)^2$ \\ \hline
$SU(3)$ with $n_f=7, |k|=\frac{3}{2}$ & $SO(14)\times U(1)$ & $x^{6}+t^2x(x+1)$ \\ \hline
\end{tabular}}
\end{center}
\label{5dgauge}
\end{table}%

Here are some remarks about our results:

\begin{enumerate}
    \item Comparison with existing theories: Our results should be compared with those studied in \cite{Yonekura:2015ksa}. It was argued in \cite{Yonekura:2015ksa} that there should exist 5D SCFTs whose infrared (IR) theory is described by an $SU(3)$ gauge theory coupled with $n_f \geq 6$ hypermultiplets. Indeed, the bound for fundamental hypermultiplets in the $SU(3)$ gauge theory is given by $n_f + 2|k| < 10$, where $k$ denotes the Chern-Simons level of the IR theory. The possible flavor symmetries were also discussed in \cite{Yonekura:2015ksa}, and we have reproduced these results in Table~\ref{5dgauge}. From our Table~\ref{5da}, one can verify that our families of theories recover these predictions, providing strong evidence for the existence of such 5D theories. Let's emphasize that the low energy theory in our case is actually 4d IR free gauge theory, which can not detect the 5d CS levels from the IR theory. However, it is possible to tune the parameters so that the IR theory at one large singularity is $SU(3)$ gauge theory  coupled with $n_f \leq 9$ flavors.

    \item 5D theories whose IR limit is not IR free in 4d sense: How can we identify 5D theories associated with gauge theories that are not IR-free in the 4D sense? These cannot be detected by examining only the IR singularity and the associated IR theory, since the IR gauge theory must be either IR-free or conformal (in 4d sense). Instead, we must consider combinations of multiple singularities: specifically, we analyze a large singularity together with additional $I_1$ singularities and compare the resulting geometry with the Seiberg-Witten (SW) geometry of the corresponding 4D gauge theory. 

    The relationship between singularities is as follows: for 5D theories, there are two additional singularities—one arising from the flavor symmetry associated with the 5D gauge coupling, and another from the Kaluza-Klein (KK) charge of the 5D theory. However, the presence of enhanced flavor symmetry in the 5D theory can complicate these computations.

    \textit{Example}: Consider the theory described by the geometry $f(x,t) = x^4 + t^2(x+1)^2$, whose bulk singularity is of type $II_{1-0} + 4I_1$. For the 4D gauge theory $SU(3)-4F$, the SW geometry is $f(x,t) = (x^3 + t)^2 + x^4$ with bulk singularity $II_{1-0} + 2I_1$. This suggests that the corresponding 5D theory has an IR limit described by the $SU(3)-4F$ gauge theory. More generally, this implies a relationship between 5D theories defined by curves of the form $f(x,t) = x^l + t^2(x+1)^2$ and the IR (5D) gauge theories $SU(3)-lF$.

    \item Non-uniqueness of the fiber at $\infty$: In general, the fiber at $\infty$ does not uniquely determine the theory, as the global geometry may differ even when the fiber at $\infty$ is identical. We have already encountered such examples in rank-one theories, and there are also numerous rank-two cases where this phenomenon occurs. We hope to investigate them further elsewhere.
\end{enumerate}

\newpage
\subsection{6d  theory}
The dual graph for the fiber at $\infty$ must consist of a chain of rational curves with exactly two loops. The possible configurations are listed in Table~\ref{6dsingular}. Some particularly interesting one-parameter families are presented in Table~\ref{6da}; remarkably, there exists a surprisingly large number of possible configurations in this case.

One immediate correspondence can be made with the UV limit of IR-free gauge theories as studied in \cite{Seiberg:1996qx}. For rank-two theories, we must consider rank-one gauge theories coupled with tensor multiplets to ensure gauge anomaly cancellation. 

A concrete example arises in 6D theories: the IR description involves an $SU(2)$ gauge theory coupled with $n_f=10$ hypermultiplets, where the presence of a tensor multiplet cancels the anomaly. The corresponding 6d theory has flavor symmetry $SO(20)$ and whose SW geometry can be described by $f(x,t)=x^5 t+t^2(x-1)^2(x+1)x$.

It would be interesting to study other SW geometries listed in table. \ref{6da} and \ref{6db}.

\begin{table}[htp]
\caption{Singular fiber configuration at $\infty$ for 6d $KK$ theories.}
\begin{center}
\begin{tabular}{|c|c|c|} \hline
Type & Data & Name \\ \hline
Parabolic 4(a) & $f=id$ & $ I_{n-p-0} (179)$ \\ \hline
Parabolic 4(b) & $f_1=f_2=id$ & $ I_n-I_p-m (179)$ \\ \hline
Parabolic 5(a) & $f=id~$ & $ I_{n-p-q}(182)$ \\ \hline
\end{tabular}
\end{center}
\label{6dsingular}
\end{table}%

\begin{table}[H]
\caption{SW geometry for 6d $(1,0)$ theories}
\begin{center}
 \resizebox{5in}{!}{\begin{tabular}{|c|c|c|c|c|} \hline
$\infty$ &  $type$ & $0$ &  flavor & Singularities\\ \hline
$xt+x^2(x-1)^2(x+1)$ & $I_{2-1-0}$ $(182),~d_x=\delta_x=3$& $x^5 t+t^2(x-1)^2(x+1)x$ & $SO(20)$ & $VII^*+2 I_1$  \\ \hline
$xt+x^2(x-1)^2(x+1)^2$ & $I_{1-1-2}$ $(182),~d_x=\delta_x=4$& $x^5 t+t^2(x-1)^2(x+1)^2$ & $SU(10)$ &$IX3+2I_1$  \\ \hline
$x(x-1)t+x^2(x-1)^2(x+1)$ & $I_{2-2-0}$ $(182),~d_x=\delta_x=4$& $x^4(x-1) t+t^2(x-1)^2(x+1)x$ & $SO(16)\times SU(2)$ & $II^*-II^*_0+2I_1$  \\ \hline
$x(x-1)t+x^2(x-1)^2(x+1)^2$ & $I_{1-2-2}$ $(182),~d_x=\delta_x=5$& $x^4(x-1) t+t^2(x-1)^2(x+1)^2$ & $SU(8)\times SU(2)\times SU(2)$ &$III^*-II_0^*+2I_1$  \\ \hline
$x(x-1)(x+1)t+x^2(x-1)^2(x+1)^2$ & $I_{2-2-2}$ $(182),~d_x=\delta_x=6$& $x^3(x-1)(x+1) t+t^2(x-1)^2(x+1)^2$ & $SU(6)\times SU(2)^3$ & $I_0^*-IV^*-\alpha+2I_1$  \\ \hline
~
\\ \hline
$xt^2+x^2(x-1)^2(x+1)$ & $I_{2-4-0}$ $(182),~d_x=\delta_x=6$& $x^5 +t^2(x+1)(x-1)^2x$ & $~$ & $VI+4I_1$ \\ \hline
$t^2+x^2(x-1)^2(x+1)$ & $I_{2-2-0}$ $(182),~d_x=\delta_x=4$& $x^6+t^2(x-1)^2x(x+1)$ & &$IX4+4I_1$  \\ \hline
$t^2+x^2(x-1)^2(x+1)^2$ & $I_{2-2-2}$ $(182),~d_x=\delta_x=6$& $x^6+t^2(x-1)^2(x+1)^2$ & & $V+4I_1$  \\ \hline
$xt^2+x^2(x-1)^2(x+1)^2$ & $I_{2-4-2}$ $(182),~d_x=\delta_x=8$& $x^5+t^2(x-1)^2(x+1)^2$ & & $IX1+4I_1$  \\ \hline
\end{tabular}}
\end{center}
\label{6da}
\end{table}%

\begin{table}[H]
\caption{More SW geometry for 6d $(1,0)$ theories}
\begin{center}
 \resizebox{5in}{!}{\begin{tabular}{|c|c|c|c|} \hline
$\infty$ &  $type$ & $0$ &  flavor\\ \hline
$x^{l}t^2 +(x-t)^2(x-1)^2(x+1)$& $I_{(2-2+l-0)}$ $(183),~d_x=\delta_x=4+l~$& $x^{6-l}+t^2(x-1)^2(x+1)x$ & $SU(6-l)$  \\ \hline
$x^{l}t^2 +(x-t)^2(x-1)^2(x+1)^2$& $I_{(2-2-2+l)}$ $(183),~d_x=\delta_x=6+l~$& $x^{6-l}+t^2(x-1)^2(x+1)^2$ & $SU(6-l)$  \\ \hline
$x^{l}t(x-t) +(x-t)^2(x-1)^2(x+1)$& $I_{(1-2l-0)}$ $(183),~d_x=\delta_x=2l+1~$& $tx^{6-l}+t^2(x-1)^2(x+1)x$ & $SO(2l)$  \\ \hline
$x^{l}t(x-t) +(x-t)^2(x-1)^2(x+1)^2$& $I_{(1-1-2l)}$ $(183),~d_x=\delta_x=2l+2~$& $(x-t) x^{6-l}+(x-t)^2(x-1)^2(x+1)^2$ & $SU(6-l)$  \\ \hline
\end{tabular}}
\end{center}
\label{6db}
\end{table}%

\section{Conclusion}
In this work, we present a systematic construction of complete Seiberg-Witten (SW) geometries for rank-two theories across four, five, and six dimensions. The SW geometries are determined by the following form
\[
y^2 = f(x,t),
\]
where the central challenge reduces to determining the appropriate functional form of $f(x,t)$. 
The construction is constrained by two principal considerations:
\begin{itemize}
    \item The structure of singular fibers at $t=\infty$, which can be determined either through Liu's algorithm or via canonical resolution.
    \item In the 4D context, the $\mathbb{C}^*$ action plays a fundamental role in classifying possible SW geometries.
    \item For 5D and 6D theories, we employ an exploratory approach where systematic pattern recognition emerges from carefully constrained parameter spaces. 
\end{itemize}
The crucial connection between our singular models and previous topological approaches is established through the canonical resolution procedure, which we have thoroughly investigated in this work.

The particularly elegant form of the full Seiberg-Witten geometry presented in this paper, combined with the method developed in \cite{Xie:2022aad,Xie:2023wqx}, provides the complete Coulomb branch solution for these theories at every vacuum. This achievement represents the ultimate goal in solving an $\mathcal{N}=2$ theory \cite{Seiberg:1994aj, Seiberg:1994rs}, which is in general a very difficult problem. We will use the SW geometry to further explore the low energy physics of the vacua.

Our results establish an alternative systematic framework for constructing 5D and 6D quantum field theories. While existing approaches rely either on conjectural ultraviolet (UV) limits of infrared (IR)-free gauge theories or on string-theoretic constructions, our geometric method significantly expands the landscape of accessible superconformal field theories (SCFTs) in higher dimensions.

The methodology developed here naturally extends to higher-rank theories whose SW geometries are described by hyperelliptic curves; we will present these results in forthcoming work. More generally, our approach applies to SW geometries represented by arbitrary algebraic curves, and we anticipate reporting further progress in this direction in subsequent publications.

All models considered in this paper feature flavor symmetries of ADE type. To incorporate non-simply laced flavor groups, a quotient construction of our current framework is required. We will provide a comprehensive treatment of this generalization in the next paper of this series.

~

\appendix
\newpage

\section{Algebraic surface}
In this section, we collect some useful facts about the algebraic surfaces.

\subsection{Numerical Invariants}
The following invariants play a crucial role in characterizing a smooth compact algebraic surface $S$:

\begin{itemize}
    \item \textbf{Hodge numbers}: Let $\mathcal{O}_S = \Omega_S^0$ denote the structure sheaf, and $\Omega_S^p$ the sheaf of $(p,0)$-forms. The Hodge numbers are defined via sheaf cohomology groups:
    \[
    h^{p,q}(S) = \dim H^q(S, \Omega_S^p)
    \]
    These satisfy the symmetry relations:
    \[
    h^{p,q} = h^{q,p} \quad \text{and} \quad h^{p,q} = h^{2-p,2-q}.
    \]
    For a connected compact surface, the non-trivial Hodge numbers are $h^{0,1}$, $h^{0,2}$, and $h^{1,1}$, since $h^{0,0} = h^{2,2} = 1$ and $h^{p,q} = 0$ for $p > 2$ or $q > 2$. Two important special cases are:
    \begin{align*}
        q(S) &= h^{0,1}(S) \quad \text{(irregularity)} \\
        p_g(S) &= h^{0,2}(S) \quad \text{(geometric genus)}
    \end{align*}

    \item \textbf{Chern numbers}: Using the Chern classes $c_1$ and $c_2$, we define:
    \[
    c_1^2 = (K^2) \quad \text{(where $K$ is the canonical divisor)}
    \]
    and $c_2$ (the topological Euler characteristic).

    \item \textbf{Holomorphic Euler characteristic}:
    \[
    \chi(\mathcal{O}_S) = 1 - q + p_g
    \]
    This gives the Euler characteristic of the structure sheaf $\mathcal{O}_S$, defined using sheaf cohomology groups.
\end{itemize}

These invariants are related by several fundamental formulas. First, Noether's theorem states:
\[
\boxed{c_1^2 + c_2 = 12\chi(\mathcal{O}_S) = 12(1 - q + p_g)}
\]
Moreover, the topological Euler characteristic can be expressed in terms of Hodge numbers:
\[
\boxed{c_2 = 2 - 2q + 2p_g + h^{1,1}}
\]
Consequently, a compact algebraic surface is completely characterized by three independent numerical invariants $(q, p_g, h^{1,1})$.

The pair $(\chi, c_1^2)$ typically satisfies non-trivial inequalities. A notable example is the Miyaoka-Yau inequality:
\[
c_1^2 \leq 3c_2 \quad \text{(equivalent to $c_1^2 \leq 9\chi$ when combined with Noether's formula)}
\]

\textbf{Hirzebruch surface}: $F_n$ is the total space of a $P^1$ bundle over $P^1$.  $F^0$ is just the product of $P^1\times P^1$. There are four coordinate patches for $F_n$.

\subsection{Blow-up of Algebraic Surfaces}
We collect fundamental formulas for the blow-up $\sigma \colon Y_1 \to Y$ of an algebraic surface $Y$ at a point $p$. 

\textbf{Local Construction}:

Choose local coordinates $(x,y)$ centered at $p$ (so $x(p) = y(p) = 0$). The blow-up $Y_1$ is realized as the subvariety of $Y \times \mathbb{P}^1$ defined by the equation
\begin{equation*}
x\zeta_1 - y\zeta_0 = 0,
\end{equation*}
where $[\zeta_0:\zeta_1]$ are homogeneous coordinates on $\mathbb{P}^1$. The projection $\sigma \colon Y_1 \to Y$ has the following properties:
\begin{itemize}
    \item Over $p$, the fiber $\sigma^{-1}(p) \cong \mathbb{P}^1$ (the \emph{exceptional divisor} $E$)
    \item Over $q \neq p$, $\sigma^{-1}(q)$ consists of a single point
\end{itemize}

\textbf{Coordinate Charts}:

The surface $Y_1$ is covered locally by two affine charts:
\begin{align*}
U_0 &= \{(x, u_0) \mid u_0 = \zeta_1/\zeta_0\} \\
U_1 &= \{(y, u_1) \mid u_1 = \zeta_0/\zeta_1\}
\end{align*}
with transition relations:
\begin{equation*}
y = x u_0 \quad \text{and} \quad u_1 = u_0^{-1} \quad \text{on} \quad U_0 \cap U_1
\end{equation*}

\textbf{Strict and Total Transforms}:

For a curve $C \subset Y$ defined locally by $f(x,y) = 0$, its \emph{total transform} $\sigma^*(C)$ decomposes as:
\begin{itemize}
    \item In chart $U_0$: $f(x, x u_0) = 0$
    \item In chart $U_1$: $f(y u_1, y) = 0$
\end{itemize}
The \emph{strict transform} $\sigma'(C)$ is the proper transform of $C$ in $Y_1$, while the \emph{total transform} includes any exceptional components.

\textbf{Divisor Relations}:

Let $K_Y$ and $K_{Y_1}$ denote canonical divisors, and $E$ the exceptional divisor. Then:
\begin{align*}
\operatorname{Pic}(Y_1) &\cong \sigma^* \operatorname{Pic}(Y) \oplus \mathbb{Z}E \\
K_{Y_1} &= \sigma^* K_Y + E
\end{align*}

For a curve $C \subset Y$ with multiplicity $r$ at $p$:
\begin{equation*}
\sigma^*(C) = \sigma'(C) + rE
\end{equation*}

\textbf{Intersection Theory}:

The intersection numbers satisfy:
\begin{align*}
\sigma'(C) \cdot E &= r \\
\sigma'(C_1) \cdot \sigma'(C_2) &= C_1 \cdot C_2 - r_1 r_2 \\
\sigma'(C) \cdot K_{Y_1} &= C \cdot K_Y + r
\end{align*}
where $r_i$ denotes the multiplicity of $C_i$ at $p$. The self-intersection number of $E$ is given as $E^2=-1$.

\textbf{Example (Resolution of Singularities via Blow-up)}:  
Consider the plane curve $C \subset Y$ defined by $f(x,y) = x^p + y^q = 0$ with $\gcd(p,q) = 1$ and $p > q$. This curve has a singularity at the origin of multiplicity $q$. 

\vspace{0.2cm}

\noindent \textit{Blow-up analysis}:
\begin{itemize}
    \item In the coordinate chart $U_0 = \{(x,u_0)\}$ where $y = xu_0$, the total transform becomes:
    \begin{equation*}
        f(x,xu_0) = x^p + (xu_0)^q = x^q(x^{p-q} + u_0^q) = 0
    \end{equation*}
    This factorization reveals:
    \begin{itemize}
        \item The exceptional divisor $E$ (with multiplicity $q$) corresponding to $x^q=0$
        \item The strict transform $C'$ given by $x^{p-q} + u_0^q = 0$
    \end{itemize}

    \item In the chart $U_1 = \{(y,u_1)\}$ where $x = yu_1$, we have:
    \begin{equation*}
        f(yu_1,y) = (yu_1)^p + y^q = y^q(y^{p-q}u_1^p + 1) = 0
    \end{equation*}
    Here $y^q=0$ again represents $E$ with multiplicity $q$, while $y^{p-q}u_1^p + 1 = 0$ is non-singular in this chart.
\end{itemize}

\noindent \textit{Resolution process}:
\begin{itemize}
    \item The strict transform $C'$ has a new singularity at $(x,u_0) = (0,0)$ in $U_0$ defined by $x^{p-q} + u_0^q = 0$, which is simpler than the original singularity.
    \item Iterating this blow-up procedure eventually resolves all singularities, yielding a smooth proper transform of the original curve.
\end{itemize}

\subsection{Double Coverings of Algebraic Surfaces}
\label{double}

Let $\pi \colon X \to Y$ be a double covering of algebraic surfaces with branch locus $B \subset Y$. There exists a line bundle $L$ on $Y$ satisfying $L^{\otimes 2} \cong \mathcal{O}_Y(B)$. The covering induces pullback maps on divisors and Picard groups:
\[
\pi^* \colon \operatorname{Div}(Y) \to \operatorname{Div}(X) \quad \text{and} \quad \pi^* \colon \operatorname{Pic}(Y) \to \operatorname{Pic}(X).
\]
For any divisor $D$ on $Y$, we have the following fundamental relations:
\begin{enumerate}
    \item $\pi_*\pi^*D \cong \mathcal{O}_Y(D) \oplus \mathcal{O}_Y(D \otimes L^{-1})$
    \item $h^i(X, \pi^*D) = h^i(Y, D) + h^i(Y, D \otimes L^{-1})$ (Leray spectral sequence)
    \item $K_X \cong \pi^*(K_Y \otimes L)$ (canonical bundle formula)
    \item $(\pi^*D \cdot \pi^*D') = 2(D \cdot D')$ (intersection pairing)
\end{enumerate}

\textbf{Resolution of Singularities}:

When $B$ has only simple singularities, we obtain a smooth surface $\bar{X}$ via canonical resolution. The invariants relate as:
\begin{align*}
\chi(\mathcal{O}_{\bar{X}}) &= 2\chi(\mathcal{O}_Y) + \tfrac{1}{2}L \cdot K_Y + \tfrac{1}{2}L^2 \\
p_g(\bar{X}) &= p_g(Y) + h^0(Y, K_Y \otimes L) \\
c_1^2(\bar{X}) &= 2c_1^2(Y) + 4L \cdot K_Y + 2L^2 \\
c_2(\bar{X}) &= 2c_2(Y) + 2L \cdot K_Y + 4L^2
\end{align*}

\textbf{Examples}:

\textbf{Example 1 (Rational Double Cover)}: 
Let $Y = \mathbb{P}^1 \times \mathbb{P}^1 = F_0$ with basis $S,F$ of $\operatorname{Pic}(Y)$ satisfying $S^2 = F^2 = 0$, $S \cdot F = 1$. Then:
\begin{itemize}
    \item $K_Y = -2S - 2F$, $c_1^2(Y) = 8$, $\chi(\mathcal{O}_Y) = 1$, $c_2(Y) = 4$
    \item Branch locus $B \sim 6F + 2S \Rightarrow L \sim 3F + S$
    \item Computations yield:
    \[
    \chi(\mathcal{O}_{\bar{X}}) = 1,\; p_g(\bar{X}) = 0,\; c_1^2(\bar{X}) = -4,\; c_2(\bar{X}) = 16
    \]
\end{itemize}

\textbf{Example 2 (Generalized Branch Locus)}: 
For $B \sim (2g+2)F + 2S$ (hence $L \sim (g+1)F + S$):
\[
\chi(\mathcal{O}_{\bar{X}}) = 1,\; p_g(\bar{X}) = 0,\; c_1^2(\bar{X}) = -4g+4,\; c_2(\bar{X}) = 4g+8
\]

\textbf{Example 3 (Non-simple singularities)}: 
For $B \sim 6F + 2aS$ and $L \sim 3F + aS$, with singularities of multiplicity $d_i = 2m_i$ or $2m_i+1$, the topological data for resolved surface reads \cite{barth2003compact}:
\begin{align*}
\chi(\mathcal{O}_{\bar{X}}) &= 2a - 1 - \tfrac{1}{2}\sum m_i(m_i-1) \\
c_1^2(\bar{X}) &= -8 + 4a - 2\sum (m_i-1)^2
\end{align*}
Rational surface cases might occur ($\chi(\mathcal{O}_{\bar{X}})=1$ )when:
\begin{itemize}
    \item[(a)] $a=1$ with $B$ having only simple singularities ($m_i=0$).
    \item[(b)] $a=2$ requiring two blow-ups for points with $m_i=2$ (multiplicities = 5 or 6)
\end{itemize}

\subsection{Canonical Resolution and Normal Model for Singular Fibers}
\label{canonical}

Let $\pi: S \to Y$ be a double covering of an algebraic surface with branch curve $B$, and let $f: S \to \mathbb{P}^1$ be a fibration induced by a projection to a base curve. We describe the process of resolving singularities via blow-ups to obtain a smooth model $S_n$ 
of $S$ fibered over $P^1$ so that the singular fiber is normal-crossing.

\textbf{Singular Fibers and Bad Points}:

Let $t$ be a local coordinate on $\mathbb{P}^1$ and $x$ a coordinate on $Y$. The double covering is locally given by:
\[ y^2 = f(x,t) \]
The fiber at $t=0$ is singular if and only if $f(x,0)$ has multiple roots. The corresponding points $(x_i,0)$ are singular points of the branch curve $B$.

A point $P \in B$ is called \emph{bad} if either:
\begin{enumerate}
    \item It is a singular point of $B$, or
    \item For the strict transform $\widetilde{B}$ and reduced total transform $\Gamma_0^{\text{red}}$ of $\Gamma_0 = \{t=0\}$, the intersection number satisfies:
    \[ I_P(\sigma_r \circ \cdots \circ \sigma_1(\Gamma_0)^{\text{red}}, \widetilde{B}_r) \geq 2 \]
    where $\widetilde{B}_r$ is the strict transform of:
    \[ \widetilde{B}_0 = \begin{cases}
        B & \text{if } \Gamma_0 \not\subset B \\
        B - \Gamma_0 & \text{if } \Gamma_0 \subset B
    \end{cases} \]
\end{enumerate}

\textbf{Resolution Process}:

For a bad point $P$ of multiplicity $m$:
\begin{enumerate}
    \item Blow up $P$ to get $\sigma: Y_1 \to Y$ with exceptional curve $E$ ($E^2 = -1$)
    \item Define the new branch locus:
    \[ B_1 = \sigma^*(B) - 2\left\lfloor \frac{m}{2} \right\rfloor E \]
    where $\sigma^*(B)$ is the total transform of $B$
    \item The new double cover $S_1 \to Y_1$ has branch locus $B_1$
\end{enumerate}

After iterating this process until no bad points remain, we obtain a smooth model of the singular fiber.

\textbf{Normal crossing model from canonical resolution}:

For $\pi: X \to Y$ branched along smooth reduced $B$:
\begin{enumerate}
    \item If $C \subset B$ is irreducible, then:
    \[ \pi^*(C) = 2D \quad \text{with} \quad D^2 = \frac{C^2}{2} \]
    
    \item For a smooth rational curve $E \not\subset B$, let $\Gamma = B|_E$:
    \begin{enumerate}
        \item If $\Gamma$ is even ($\Gamma = 2Q$, including $\Gamma = 0$):
        \[ \pi^{-1}(E) = D_1 + D_2 \quad \text{with} \quad D_i^2 = E^2 - \deg Q \]
        and $D_1 \cdot D_2 = \deg Q$.
        
        \item If $\Gamma$ is not even (e.g., $\Gamma$ is a single point):
        \[ \pi^{-1}(E) = D \quad \text{irreducible with} \quad D^2 = 2E^2 \]
    \end{enumerate}
\end{enumerate}


\begin{figure}
\begin{center}

\tikzset{every picture/.style={line width=0.75pt}} 

\begin{tikzpicture}[x=0.45pt,y=0.45pt,yscale=-1,xscale=1]

\draw   (234,127) .. controls (234,113.19) and (245.19,102) .. (259,102) .. controls (272.81,102) and (284,113.19) .. (284,127) .. controls (284,140.81) and (272.81,152) .. (259,152) .. controls (245.19,152) and (234,140.81) .. (234,127) -- cycle ;
\draw   (285,127) .. controls (285,113.19) and (296.19,102) .. (310,102) .. controls (323.81,102) and (335,113.19) .. (335,127) .. controls (335,140.81) and (323.81,152) .. (310,152) .. controls (296.19,152) and (285,140.81) .. (285,127) -- cycle ;
\draw    (381,120) -- (478,119.72) ;
\draw [shift={(480,119.72)}, rotate = 179.84] [color={rgb, 255:red, 0; green, 0; blue, 0 }  ][line width=0.75]    (10.93,-3.29) .. controls (6.95,-1.4) and (3.31,-0.3) .. (0,0) .. controls (3.31,0.3) and (6.95,1.4) .. (10.93,3.29)   ;
\draw   (526,119) .. controls (526,105.19) and (537.19,94) .. (551,94) .. controls (564.81,94) and (576,105.19) .. (576,119) .. controls (576,132.81) and (564.81,144) .. (551,144) .. controls (537.19,144) and (526,132.81) .. (526,119) -- cycle ;
\draw   (628,121) .. controls (628,107.19) and (639.19,96) .. (653,96) .. controls (666.81,96) and (678,107.19) .. (678,121) .. controls (678,134.81) and (666.81,146) .. (653,146) .. controls (639.19,146) and (628,134.81) .. (628,121) -- cycle ;
\draw   (577,121) .. controls (577,107.19) and (588.19,96) .. (602,96) .. controls (615.81,96) and (627,107.19) .. (627,121) .. controls (627,134.81) and (615.81,146) .. (602,146) .. controls (588.19,146) and (577,134.81) .. (577,121) -- cycle ;
\draw  [color={rgb, 255:red, 0; green, 0; blue, 0 }  ,draw opacity=1 ][fill={rgb, 255:red, 0; green, 0; blue, 0 }  ,fill opacity=1 ] (282,124.5) .. controls (282,123.12) and (283.12,122) .. (284.5,122) .. controls (285.88,122) and (287,123.12) .. (287,124.5) .. controls (287,125.88) and (285.88,127) .. (284.5,127) .. controls (283.12,127) and (282,125.88) .. (282,124.5) -- cycle ;
\draw  [color={rgb, 255:red, 0; green, 0; blue, 0 }  ,draw opacity=1 ][fill={rgb, 255:red, 0; green, 0; blue, 0 }  ,fill opacity=1 ] (625,119.5) .. controls (625,118.12) and (626.12,117) .. (627.5,117) .. controls (628.88,117) and (630,118.12) .. (630,119.5) .. controls (630,120.88) and (628.88,122) .. (627.5,122) .. controls (626.12,122) and (625,120.88) .. (625,119.5) -- cycle ;

\draw (250,74.4) node [anchor=north west][inner sep=0.75pt]    {$\Gamma $};
\draw (308,73.4) node [anchor=north west][inner sep=0.75pt]    {$B$};
\draw (542,66.4) node [anchor=north west][inner sep=0.75pt]    {$\Gamma $};
\draw (653,67.4) node [anchor=north west][inner sep=0.75pt]    {$B$};
\draw (599,69.4) node [anchor=north west][inner sep=0.75pt]    {$E$};

\end{tikzpicture}

\end{center}
\caption{Blow-up a point at the intersection of $B$ and $\Gamma$. After the blow-up, one get an exceptional divisor and two proper transformed curve.}
\label{resolve}
\end{figure}
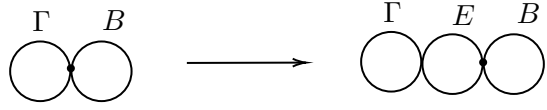

\begin{figure}[H]
\begin{center}

\tikzset{every picture/.style={line width=0.75pt}} 

\begin{tikzpicture}[x=0.40pt,y=0.40pt,yscale=-1,xscale=1]

\draw  [fill={rgb, 255:red, 208; green, 2; blue, 27 }  ,fill opacity=1 ] (285,103) .. controls (285,89.19) and (296.19,78) .. (310,78) .. controls (323.81,78) and (335,89.19) .. (335,103) .. controls (335,116.81) and (323.81,128) .. (310,128) .. controls (296.19,128) and (285,116.81) .. (285,103) -- cycle ;
\draw  [color={rgb, 255:red, 208; green, 2; blue, 27 }  ,draw opacity=1 ][fill={rgb, 255:red, 208; green, 2; blue, 27 }  ,fill opacity=1 ] (387,105) .. controls (387,91.19) and (398.19,80) .. (412,80) .. controls (425.81,80) and (437,91.19) .. (437,105) .. controls (437,118.81) and (425.81,130) .. (412,130) .. controls (398.19,130) and (387,118.81) .. (387,105) -- cycle ;
\draw   (336,105) .. controls (336,91.19) and (347.19,80) .. (361,80) .. controls (374.81,80) and (386,91.19) .. (386,105) .. controls (386,118.81) and (374.81,130) .. (361,130) .. controls (347.19,130) and (336,118.81) .. (336,105) -- cycle ;
\draw  [fill={rgb, 255:red, 255; green, 255; blue, 255 }  ,fill opacity=1 ] (438,106) .. controls (438,92.19) and (449.19,81) .. (463,81) .. controls (476.81,81) and (488,92.19) .. (488,106) .. controls (488,119.81) and (476.81,131) .. (463,131) .. controls (449.19,131) and (438,119.81) .. (438,106) -- cycle ;
\draw  [fill={rgb, 255:red, 208; green, 2; blue, 27 }  ,fill opacity=1 ] (489,107) .. controls (489,93.19) and (500.19,82) .. (514,82) .. controls (527.81,82) and (539,93.19) .. (539,107) .. controls (539,120.81) and (527.81,132) .. (514,132) .. controls (500.19,132) and (489,120.81) .. (489,107) -- cycle ;
\draw   (593,110) .. controls (593,96.19) and (604.19,85) .. (618,85) .. controls (631.81,85) and (643,96.19) .. (643,110) .. controls (643,123.81) and (631.81,135) .. (618,135) .. controls (604.19,135) and (593,123.81) .. (593,110) -- cycle ;
\draw   (231,261) .. controls (231,247.19) and (242.19,236) .. (256,236) .. controls (269.81,236) and (281,247.19) .. (281,261) .. controls (281,274.81) and (269.81,286) .. (256,286) .. controls (242.19,286) and (231,274.81) .. (231,261) -- cycle ;
\draw   (641,262) .. controls (641,248.19) and (652.19,237) .. (666,237) .. controls (679.81,237) and (691,248.19) .. (691,262) .. controls (691,275.81) and (679.81,287) .. (666,287) .. controls (652.19,287) and (641,275.81) .. (641,262) -- cycle ;
\draw   (308,258) .. controls (308,244.19) and (319.19,233) .. (333,233) .. controls (346.81,233) and (358,244.19) .. (358,258) .. controls (358,271.81) and (346.81,283) .. (333,283) .. controls (319.19,283) and (308,271.81) .. (308,258) -- cycle ;
\draw   (550,259) .. controls (550,245.19) and (561.19,234) .. (575,234) .. controls (588.81,234) and (600,245.19) .. (600,259) .. controls (600,272.81) and (588.81,284) .. (575,284) .. controls (561.19,284) and (550,272.81) .. (550,259) -- cycle ;
\draw   (389,259) .. controls (389,245.19) and (400.19,234) .. (414,234) .. controls (427.81,234) and (439,245.19) .. (439,259) .. controls (439,272.81) and (427.81,284) .. (414,284) .. controls (400.19,284) and (389,272.81) .. (389,259) -- cycle ;
\draw   (467,259) .. controls (467,245.19) and (478.19,234) .. (492,234) .. controls (505.81,234) and (517,245.19) .. (517,259) .. controls (517,272.81) and (505.81,284) .. (492,284) .. controls (478.19,284) and (467,272.81) .. (467,259) -- cycle ;
\draw   (721,223) .. controls (721,209.19) and (732.19,198) .. (746,198) .. controls (759.81,198) and (771,209.19) .. (771,223) .. controls (771,236.81) and (759.81,248) .. (746,248) .. controls (732.19,248) and (721,236.81) .. (721,223) -- cycle ;
\draw    (282,260) -- (309,259.72) ;
\draw    (686,276) -- (727,297.72) ;
\draw    (359,260) -- (388,259.72) ;
\draw    (602,260) -- (641,260.72) ;
\draw    (439,259) -- (467,258.72) ;
\draw    (517,261) -- (549,260.72) ;
\draw  [fill={rgb, 255:red, 255; green, 255; blue, 255 }  ,fill opacity=1 ] (541,109) .. controls (541,95.19) and (552.19,84) .. (566,84) .. controls (579.81,84) and (591,95.19) .. (591,109) .. controls (591,122.81) and (579.81,134) .. (566,134) .. controls (552.19,134) and (541,122.81) .. (541,109) -- cycle ;
\draw   (543,58) .. controls (543,44.19) and (554.19,33) .. (568,33) .. controls (581.81,33) and (593,44.19) .. (593,58) .. controls (593,71.81) and (581.81,83) .. (568,83) .. controls (554.19,83) and (543,71.81) .. (543,58) -- cycle ;
\draw   (725,306) .. controls (725,292.19) and (736.19,281) .. (750,281) .. controls (763.81,281) and (775,292.19) .. (775,306) .. controls (775,319.81) and (763.81,331) .. (750,331) .. controls (736.19,331) and (725,319.81) .. (725,306) -- cycle ;
\draw    (685,245) -- (723,231.72) ;

\draw (607,98.4) node [anchor=north west][inner sep=0.75pt]    {$\tilde{\Gamma }$};
\draw (354,95.4) node [anchor=north west][inner sep=0.75pt]   [font=\tiny] {$E_{2}$};
\draw (301,94.4) node [anchor=north west][inner sep=0.75pt]    [font=\tiny] {$E_{1}$};
\draw (402,97.4) node [anchor=north west][inner sep=0.75pt]    [font=\tiny] {$E_{3}$};
\draw (562,45.4) node [anchor=north west][inner sep=0.75pt]    [font=\tiny] {$\tilde{B}$};
\draw (250,248.4) node [anchor=north west][inner sep=0.75pt]     [font=\tiny]{$2$};
\draw (489,251.4) node [anchor=north west][inner sep=0.75pt]   [font=\tiny]  {$4$};
\draw (328,251.4) node [anchor=north west][inner sep=0.75pt]    [font=\tiny] {$2$};
\draw (408,249.4) node [anchor=north west][inner sep=0.75pt]    [font=\tiny] {$6$};
\draw (566,251.4) node [anchor=north west][inner sep=0.75pt]   [font=\tiny]  {$10$};
\draw (738,212.4) node [anchor=north west][inner sep=0.75pt]    [font=\tiny] {$1$};
\draw (504,100.4) node [anchor=north west][inner sep=0.75pt]   [font=\tiny]  {$E_{5}$};
\draw (453,97.4) node [anchor=north west][inner sep=0.75pt]    [font=\tiny] {$E_{4}$};
\draw (555,99.4) node [anchor=north west][inner sep=0.75pt]    [font=\tiny] {$E_{6}$};
\draw (656,251.4) node [anchor=north west][inner sep=0.75pt]   [font=\tiny]  {$6$};
\draw (746,296.4) node [anchor=north west][inner sep=0.75pt]   [font=\tiny]  {$1$};

\end{tikzpicture}

\end{center}
\caption{The smooth model for the singular fiber defined by $x^6+t=0$.}
\label{resolve1}
\end{figure}
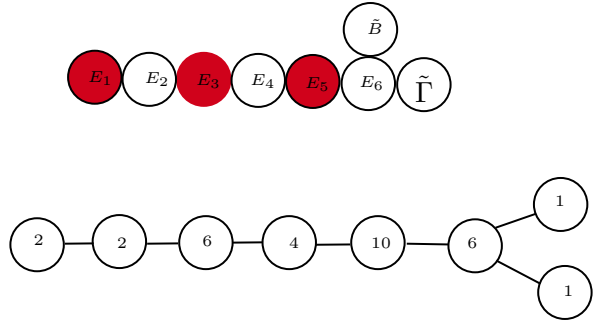

\textbf{Example 1 (Resolution of $y^2=x^6 + t$)}: 
Consider the local branch curve $B$ defined by $x^6 + t = 0$. After canonical resolution (see figure. \ref{resolve1}), we obtain the following data:

\begin{align*}
\Gamma &= \widetilde{\Gamma} + E_1 + 2E_2 + 3E_3 + 4E_4 + 5E_5 + 6E_6 \\
B &= \widetilde{B} + E_1 + E_3 + E_5
\end{align*}

with intersection properties:
\begin{itemize}
    \item Self-intersections: $E_i^2 = -2$ for $i=1,\dots,5$, $E_6^2 = -1$, $\widetilde{\Gamma}^2 = -6$
    \item Preimages of exceptional divisors:
    \begin{itemize}
        \item $\pi^{-1}(E_{2i+1}) = 2D_{2i+1}$ for $i=0,1,2$ (ramified cover)
        \item $\pi^{-1}(E_{2i}) = D_{2i}$ for $i=1,2,3$ (unramified cover)
    \end{itemize}
    \item Self-intersections of preimages:
    \begin{itemize}
        \item $D_{2i}^2 = -4$ (even indices)
        \item $D_{2i+1}^2 = -1$ (odd indices)
    \end{itemize}
    \item $\pi^{-1}(\widetilde{\Gamma}) = D_7 + D_7'$ (splitting into two components)
\end{itemize}

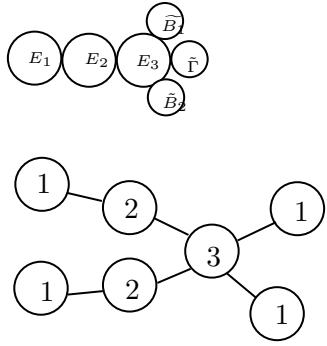
\begin{figure}[H]
\begin{center}

\tikzset{every picture/.style={line width=0.75pt}} 

\begin{tikzpicture}[x=0.40pt,y=0.40pt,yscale=-1,xscale=1]

\draw  [fill={rgb, 255:red, 255; green, 255; blue, 255 }  ,fill opacity=1 ] (208,110) .. controls (208,96.19) and (219.19,85) .. (233,85) .. controls (246.81,85) and (258,96.19) .. (258,110) .. controls (258,123.81) and (246.81,135) .. (233,135) .. controls (219.19,135) and (208,123.81) .. (208,110) -- cycle ;
\draw  [color={rgb, 255:red, 0; green, 0; blue, 0 }  ,draw opacity=1 ][fill={rgb, 255:red, 255; green, 255; blue, 255 }  ,fill opacity=1 ] (310,112) .. controls (310,98.19) and (321.19,87) .. (335,87) .. controls (348.81,87) and (360,98.19) .. (360,112) .. controls (360,125.81) and (348.81,137) .. (335,137) .. controls (321.19,137) and (310,125.81) .. (310,112) -- cycle ;
\draw   (259,112) .. controls (259,98.19) and (270.19,87) .. (284,87) .. controls (297.81,87) and (309,98.19) .. (309,112) .. controls (309,125.81) and (297.81,137) .. (284,137) .. controls (270.19,137) and (259,125.81) .. (259,112) -- cycle ;
\draw  [fill={rgb, 255:red, 255; green, 255; blue, 255 }  ,fill opacity=1 ] (338,75) .. controls (338,65.61) and (345.61,58) .. (355,58) .. controls (364.39,58) and (372,65.61) .. (372,75) .. controls (372,84.39) and (364.39,92) .. (355,92) .. controls (345.61,92) and (338,84.39) .. (338,75) -- cycle ;
\draw   (215,229) .. controls (215,215.19) and (226.19,204) .. (240,204) .. controls (253.81,204) and (265,215.19) .. (265,229) .. controls (265,242.81) and (253.81,254) .. (240,254) .. controls (226.19,254) and (215,242.81) .. (215,229) -- cycle ;
\draw   (214,327) .. controls (214,313.19) and (225.19,302) .. (239,302) .. controls (252.81,302) and (264,313.19) .. (264,327) .. controls (264,340.81) and (252.81,352) .. (239,352) .. controls (225.19,352) and (214,340.81) .. (214,327) -- cycle ;
\draw   (297,251) .. controls (297,237.19) and (308.19,226) .. (322,226) .. controls (335.81,226) and (347,237.19) .. (347,251) .. controls (347,264.81) and (335.81,276) .. (322,276) .. controls (308.19,276) and (297,264.81) .. (297,251) -- cycle ;
\draw   (374,292) .. controls (374,278.19) and (385.19,267) .. (399,267) .. controls (412.81,267) and (424,278.19) .. (424,292) .. controls (424,305.81) and (412.81,317) .. (399,317) .. controls (385.19,317) and (374,305.81) .. (374,292) -- cycle ;
\draw    (413,313) -- (440,335.72) ;
\draw    (423,282) -- (459,264.72) ;
\draw    (263,237) -- (296,244.72) ;
\draw  [fill={rgb, 255:red, 255; green, 255; blue, 255 }  ,fill opacity=1 ] (339,147) .. controls (339,137.61) and (346.61,130) .. (356,130) .. controls (365.39,130) and (373,137.61) .. (373,147) .. controls (373,156.39) and (365.39,164) .. (356,164) .. controls (346.61,164) and (339,156.39) .. (339,147) -- cycle ;
\draw  [fill={rgb, 255:red, 255; green, 255; blue, 255 }  ,fill opacity=1 ] (361,111) .. controls (361,101.61) and (368.61,94) .. (378,94) .. controls (387.39,94) and (395,101.61) .. (395,111) .. controls (395,120.39) and (387.39,128) .. (378,128) .. controls (368.61,128) and (361,120.39) .. (361,111) -- cycle ;
\draw   (454,252) .. controls (454,238.19) and (465.19,227) .. (479,227) .. controls (492.81,227) and (504,238.19) .. (504,252) .. controls (504,265.81) and (492.81,277) .. (479,277) .. controls (465.19,277) and (454,265.81) .. (454,252) -- cycle ;
\draw   (435,351) .. controls (435,337.19) and (446.19,326) .. (460,326) .. controls (473.81,326) and (485,337.19) .. (485,351) .. controls (485,364.81) and (473.81,376) .. (460,376) .. controls (446.19,376) and (435,364.81) .. (435,351) -- cycle ;
\draw   (297,324) .. controls (297,310.19) and (308.19,299) .. (322,299) .. controls (335.81,299) and (347,310.19) .. (347,324) .. controls (347,337.81) and (335.81,349) .. (322,349) .. controls (308.19,349) and (297,337.81) .. (297,324) -- cycle ;
\draw    (264,333) -- (297,329.72) ;
\draw    (348,322) -- (379,308.72) ;
\draw    (345,261) -- (377,276.72) ;

\draw (373,106.4) node [anchor=north west][inner sep=0.75pt]  [font=\tiny]  {$\tilde{\Gamma }$};
\draw (277,102.4) node [anchor=north west][inner sep=0.75pt]    [font=\tiny] {$E_{2}$};
\draw (224,101.4) node [anchor=north west][inner sep=0.75pt]    [font=\tiny] {$E_{1}$};
\draw (325,104.4) node [anchor=north west][inner sep=0.75pt]    [font=\tiny] {$E_{3}$};
\draw (349,66.4) node [anchor=north west][inner sep=0.75pt]  [font=\tiny]  {$\widetilde{B_{1}}$};
\draw (314,240.4) node [anchor=north west][inner sep=0.75pt]    {$2$};
\draw (232,216.4) node [anchor=north west][inner sep=0.75pt]    {$1$};
\draw (391,285.4) node [anchor=north west][inner sep=0.75pt]    {$3$};
\draw (350,140.4) node [anchor=north west][inner sep=0.75pt]  [font=\tiny]  {$\tilde{B}_{2}$};
\draw (315,316.4) node [anchor=north west][inner sep=0.75pt]    {$2$};
\draw (235,318.4) node [anchor=north west][inner sep=0.75pt]    {$1$};
\draw (473,243.4) node [anchor=north west][inner sep=0.75pt]    {$1$};
\draw (455,340.4) node [anchor=north west][inner sep=0.75pt]    {$1$};

\end{tikzpicture}

\end{center}
\caption{The smooth model for the singular fiber defined by $x^6+t^2$.}
\label{resolve2}
\end{figure}

\textbf{Example 2 (Resolution of $y^2 = x^6 + t^2$)}: 
Consider the double covering defined by $y^2 = x^6 + t^2$. The resolution process yields the following divisor structure (see figure. \ref{resolve2}):

\begin{align*}
\Gamma &= \widetilde{\Gamma} + E_1 + 2E_2 + 3E_3 \\
B &= \widetilde{B}_1 + \widetilde{B}_2
\end{align*}

where $B = B_1 + B_2$ is the branch locus. The self-intersection numbers are:
\begin{itemize}
    \item $E_1^2 = E_2^2 = -2$
    \item $E_3^2 = -1$
    \item $\widetilde{\Gamma}^2 = -3$
\end{itemize}

\vspace{0.2cm}

\noindent The preimages under the double cover $\pi$ behave as follows:
\begin{itemize}
    \item For $E_1$, $E_2$, and $\widetilde{\Gamma}$ (which don't intersect $B$):
    \begin{itemize}
        \item Each lifts to two disjoint divisors in the singular fiber
        \item All have self-intersection $-2$
    \end{itemize}
    
    \item For $E_3$ (which intersects $B$):
    \begin{itemize}
        \item $\pi^*(E_3)$ is an irreducible curve
        \item Self-intersection number $-2$
    \end{itemize}
\end{itemize}

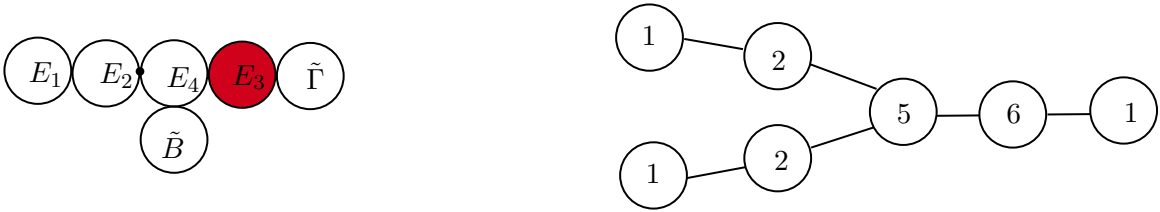
\begin{figure}[H]
\begin{center}

\tikzset{every picture/.style={line width=0.75pt}} 

\begin{tikzpicture}[x=0.50pt,y=0.50pt,yscale=-1,xscale=1]

\draw   (117,107) .. controls (117,93.19) and (128.19,82) .. (142,82) .. controls (155.81,82) and (167,93.19) .. (167,107) .. controls (167,120.81) and (155.81,132) .. (142,132) .. controls (128.19,132) and (117,120.81) .. (117,107) -- cycle ;
\draw   (219,109) .. controls (219,95.19) and (230.19,84) .. (244,84) .. controls (257.81,84) and (269,95.19) .. (269,109) .. controls (269,122.81) and (257.81,134) .. (244,134) .. controls (230.19,134) and (219,122.81) .. (219,109) -- cycle ;
\draw   (168,109) .. controls (168,95.19) and (179.19,84) .. (193,84) .. controls (206.81,84) and (218,95.19) .. (218,109) .. controls (218,122.81) and (206.81,134) .. (193,134) .. controls (179.19,134) and (168,122.81) .. (168,109) -- cycle ;
\draw  [color={rgb, 255:red, 0; green, 0; blue, 0 }  ,draw opacity=1 ][fill={rgb, 255:red, 0; green, 0; blue, 0 }  ,fill opacity=1 ] (216,107.5) .. controls (216,106.12) and (217.12,105) .. (218.5,105) .. controls (219.88,105) and (221,106.12) .. (221,107.5) .. controls (221,108.88) and (219.88,110) .. (218.5,110) .. controls (217.12,110) and (216,108.88) .. (216,107.5) -- cycle ;
\draw  [fill={rgb, 255:red, 208; green, 2; blue, 27 }  ,fill opacity=1 ] (270,110) .. controls (270,96.19) and (281.19,85) .. (295,85) .. controls (308.81,85) and (320,96.19) .. (320,110) .. controls (320,123.81) and (308.81,135) .. (295,135) .. controls (281.19,135) and (270,123.81) .. (270,110) -- cycle ;
\draw   (321,111) .. controls (321,97.19) and (332.19,86) .. (346,86) .. controls (359.81,86) and (371,97.19) .. (371,111) .. controls (371,124.81) and (359.81,136) .. (346,136) .. controls (332.19,136) and (321,124.81) .. (321,111) -- cycle ;
\draw   (219,159) .. controls (219,145.19) and (230.19,134) .. (244,134) .. controls (257.81,134) and (269,145.19) .. (269,159) .. controls (269,172.81) and (257.81,184) .. (244,184) .. controls (230.19,184) and (219,172.81) .. (219,159) -- cycle ;
\draw   (575,82) .. controls (575,68.19) and (586.19,57) .. (600,57) .. controls (613.81,57) and (625,68.19) .. (625,82) .. controls (625,95.81) and (613.81,107) .. (600,107) .. controls (586.19,107) and (575,95.81) .. (575,82) -- cycle ;
\draw   (578,186) .. controls (578,172.19) and (589.19,161) .. (603,161) .. controls (616.81,161) and (628,172.19) .. (628,186) .. controls (628,199.81) and (616.81,211) .. (603,211) .. controls (589.19,211) and (578,199.81) .. (578,186) -- cycle ;
\draw   (671,96) .. controls (671,82.19) and (682.19,71) .. (696,71) .. controls (709.81,71) and (721,82.19) .. (721,96) .. controls (721,109.81) and (709.81,121) .. (696,121) .. controls (682.19,121) and (671,109.81) .. (671,96) -- cycle ;
\draw   (671,173) .. controls (671,159.19) and (682.19,148) .. (696,148) .. controls (709.81,148) and (721,159.19) .. (721,173) .. controls (721,186.81) and (709.81,198) .. (696,198) .. controls (682.19,198) and (671,186.81) .. (671,173) -- cycle ;
\draw   (765,138) .. controls (765,124.19) and (776.19,113) .. (790,113) .. controls (803.81,113) and (815,124.19) .. (815,138) .. controls (815,151.81) and (803.81,163) .. (790,163) .. controls (776.19,163) and (765,151.81) .. (765,138) -- cycle ;
\draw   (847,140) .. controls (847,126.19) and (858.19,115) .. (872,115) .. controls (885.81,115) and (897,126.19) .. (897,140) .. controls (897,153.81) and (885.81,165) .. (872,165) .. controls (858.19,165) and (847,153.81) .. (847,140) -- cycle ;
\draw   (930,137) .. controls (930,123.19) and (941.19,112) .. (955,112) .. controls (968.81,112) and (980,123.19) .. (980,137) .. controls (980,150.81) and (968.81,162) .. (955,162) .. controls (941.19,162) and (930,150.81) .. (930,137) -- cycle ;
\draw    (628,188) -- (672,179.72) ;
\draw    (626,83) -- (670,90.72) ;
\draw    (721,165) -- (768,149.72) ;
\draw    (720,102) -- (770,120.72) ;
\draw    (815,141) -- (847,140.72) ;
\draw    (898,140) -- (930,139.72) ;

\draw (341,98.4) node [anchor=north west][inner sep=0.75pt]    {$\tilde{\Gamma }$};
\draw (186,99.4) node [anchor=north west][inner sep=0.75pt]    {$E_{2}$};
\draw (133,98.4) node [anchor=north west][inner sep=0.75pt]    {$E_{1}$};
\draw (236,102.4) node [anchor=north west][inner sep=0.75pt]    {$E_{4}$};
\draw (285,100.4) node [anchor=north west][inner sep=0.75pt]    {$E_{3}$};
\draw (232,150.4) node [anchor=north west][inner sep=0.75pt]    {$\tilde{B}$};
\draw (592,71.4) node [anchor=north west][inner sep=0.75pt]    {$1$};
\draw (595,175.4) node [anchor=north west][inner sep=0.75pt]    {$1$};
\draw (691,165.4) node [anchor=north west][inner sep=0.75pt]    {$2$};
\draw (689,90.4) node [anchor=north west][inner sep=0.75pt]    {$2$};
\draw (783,130.4) node [anchor=north west][inner sep=0.75pt]    {$5$};
\draw (865,130.4) node [anchor=north west][inner sep=0.75pt]    {$6$};
\draw (953,129.4) node [anchor=north west][inner sep=0.75pt]    {$1$};

\end{tikzpicture}

\end{center}
\caption{The smooth model for the singular fiber defined by $x^5+t^2=0$.}
\label{resolve3}
\end{figure}

\textbf{Example 3 (Resolution of $y^2=x^5 + t^2$ )}: 
Consider the double cover with branch locus $B$ defined by $x^5 + t^2 = 0$. After performing four blow-ups (see figure. \ref{resolve3}), we obtain the following resolution data:

\begin{align*}
\Gamma &= \widetilde{\Gamma} + E_1 + 2E_2 + 3E_3 + 5E_4 \\
B &= \widetilde{B} + E_3
\end{align*}

with intersection properties:
\begin{itemize}
    \item Self-intersections:
    \begin{itemize}
        \item $E_1^2 = -2$, $E_2^2 = -3$, $E_3^2 = -2$, $E_4^2 = -1$
    \end{itemize}
    
    \item Preimages of exceptional divisors:
    \begin{itemize}
        \item $\pi^{-1}(E_1) = D_1 + D_1'$ (two disjoint copies, $E_1$ doesn't meet $B$)
        \item $\pi^{-1}(E_2) = D_2 + D_2'$ (two disjoint copies, $E_2$ doesn't meet $B$)
        \item $\pi^{-1}(E_3) = 2D_3$ (ramified cover, $D_3^2 = -1$)
        \item $\pi^{-1}(E_4) = D_4$ (irreducible, $D_4^2 = -2$)
    \end{itemize}
    
    \item Proper transforms:
    \begin{itemize}
        \item $\pi^{-1}(\widetilde{\Gamma}) = D_5$ with $D_5^2 = -6$
    \end{itemize}
\end{itemize}

\begin{figure}[H]
\begin{center}

\tikzset{every picture/.style={line width=0.75pt}} 

\begin{tikzpicture}[x=0.45pt,y=0.45pt,yscale=-1,xscale=1]

\draw  [fill={rgb, 255:red, 208; green, 2; blue, 27 }  ,fill opacity=1 ] (305,123) .. controls (305,109.19) and (316.19,98) .. (330,98) .. controls (343.81,98) and (355,109.19) .. (355,123) .. controls (355,136.81) and (343.81,148) .. (330,148) .. controls (316.19,148) and (305,136.81) .. (305,123) -- cycle ;
\draw  [color={rgb, 255:red, 208; green, 2; blue, 27 }  ,draw opacity=1 ][fill={rgb, 255:red, 208; green, 2; blue, 27 }  ,fill opacity=1 ] (407,125) .. controls (407,111.19) and (418.19,100) .. (432,100) .. controls (445.81,100) and (457,111.19) .. (457,125) .. controls (457,138.81) and (445.81,150) .. (432,150) .. controls (418.19,150) and (407,138.81) .. (407,125) -- cycle ;
\draw   (356,125) .. controls (356,111.19) and (367.19,100) .. (381,100) .. controls (394.81,100) and (406,111.19) .. (406,125) .. controls (406,138.81) and (394.81,150) .. (381,150) .. controls (367.19,150) and (356,138.81) .. (356,125) -- cycle ;
\draw  [fill={rgb, 255:red, 255; green, 255; blue, 255 }  ,fill opacity=1 ] (458,126) .. controls (458,112.19) and (469.19,101) .. (483,101) .. controls (496.81,101) and (508,112.19) .. (508,126) .. controls (508,139.81) and (496.81,151) .. (483,151) .. controls (469.19,151) and (458,139.81) .. (458,126) -- cycle ;
\draw  [fill={rgb, 255:red, 208; green, 2; blue, 27 }  ,fill opacity=1 ] (509,127) .. controls (509,113.19) and (520.19,102) .. (534,102) .. controls (547.81,102) and (559,113.19) .. (559,127) .. controls (559,140.81) and (547.81,152) .. (534,152) .. controls (520.19,152) and (509,140.81) .. (509,127) -- cycle ;
\draw   (560,133.22) .. controls (560,121.9) and (569.18,112.72) .. (580.5,112.72) .. controls (591.82,112.72) and (601,121.9) .. (601,133.22) .. controls (601,144.54) and (591.82,153.72) .. (580.5,153.72) .. controls (569.18,153.72) and (560,144.54) .. (560,133.22) -- cycle ;
\draw   (251,281) .. controls (251,267.19) and (262.19,256) .. (276,256) .. controls (289.81,256) and (301,267.19) .. (301,281) .. controls (301,294.81) and (289.81,306) .. (276,306) .. controls (262.19,306) and (251,294.81) .. (251,281) -- cycle ;
\draw   (661,282) .. controls (661,268.19) and (672.19,257) .. (686,257) .. controls (699.81,257) and (711,268.19) .. (711,282) .. controls (711,295.81) and (699.81,307) .. (686,307) .. controls (672.19,307) and (661,295.81) .. (661,282) -- cycle ;
\draw   (328,278) .. controls (328,264.19) and (339.19,253) .. (353,253) .. controls (366.81,253) and (378,264.19) .. (378,278) .. controls (378,291.81) and (366.81,303) .. (353,303) .. controls (339.19,303) and (328,291.81) .. (328,278) -- cycle ;
\draw   (570,279) .. controls (570,265.19) and (581.19,254) .. (595,254) .. controls (608.81,254) and (620,265.19) .. (620,279) .. controls (620,292.81) and (608.81,304) .. (595,304) .. controls (581.19,304) and (570,292.81) .. (570,279) -- cycle ;
\draw   (409,279) .. controls (409,265.19) and (420.19,254) .. (434,254) .. controls (447.81,254) and (459,265.19) .. (459,279) .. controls (459,292.81) and (447.81,304) .. (434,304) .. controls (420.19,304) and (409,292.81) .. (409,279) -- cycle ;
\draw   (487,279) .. controls (487,265.19) and (498.19,254) .. (512,254) .. controls (525.81,254) and (537,265.19) .. (537,279) .. controls (537,292.81) and (525.81,304) .. (512,304) .. controls (498.19,304) and (487,292.81) .. (487,279) -- cycle ;
\draw    (302,280) -- (329,279.72) ;
\draw    (379,280) -- (408,279.72) ;
\draw    (622,280) -- (661,280.72) ;
\draw    (459,279) -- (487,278.72) ;
\draw    (537,281) -- (569,280.72) ;
\draw   (554,49) .. controls (554,37.95) and (562.95,29) .. (574,29) .. controls (585.05,29) and (594,37.95) .. (594,49) .. controls (594,60.05) and (585.05,69) .. (574,69) .. controls (562.95,69) and (554,60.05) .. (554,49) -- cycle ;
\draw   (534,86.22) .. controls (534,74.9) and (543.18,65.72) .. (554.5,65.72) .. controls (565.82,65.72) and (575,74.9) .. (575,86.22) .. controls (575,97.54) and (565.82,106.72) .. (554.5,106.72) .. controls (543.18,106.72) and (534,97.54) .. (534,86.22) -- cycle ;
\draw   (595,197) .. controls (595,183.19) and (606.19,172) .. (620,172) .. controls (633.81,172) and (645,183.19) .. (645,197) .. controls (645,210.81) and (633.81,222) .. (620,222) .. controls (606.19,222) and (595,210.81) .. (595,197) -- cycle ;
\draw    (603,256) -- (613,220.72) ;

\draw (575,122.4) node [anchor=north west][inner sep=0.75pt]     [font=\tiny] {$\tilde{\Gamma }$};
\draw (374,115.4) node [anchor=north west][inner sep=0.75pt]  [font=\tiny]  {$E_{2}$};
\draw (321,114.4) node [anchor=north west][inner sep=0.75pt]   [font=\tiny]   {$E_{1}$};
\draw (422,117.4) node [anchor=north west][inner sep=0.75pt]    [font=\tiny]  {$E_{3}$};
\draw (568,35.4) node [anchor=north west][inner sep=0.75pt]    [font=\tiny]  {$\tilde{B}$};
\draw (270,268.4) node [anchor=north west][inner sep=0.75pt]   [font=\tiny]   {$2$};
\draw (509,271.4) node [anchor=north west][inner sep=0.75pt]   [font=\tiny]   {$4$};
\draw (348,271.4) node [anchor=north west][inner sep=0.75pt]   [font=\tiny]   {$2$};
\draw (428,269.4) node [anchor=north west][inner sep=0.75pt]   [font=\tiny]   {$6$};
\draw (586,271.4) node [anchor=north west][inner sep=0.75pt]   [font=\tiny]   {$10$};
\draw (524,120.4) node [anchor=north west][inner sep=0.75pt]   [font=\tiny]   {$E_{5}$};
\draw (473,117.4) node [anchor=north west][inner sep=0.75pt]    [font=\tiny]  {$E_{4}$};
\draw (680,271.4) node [anchor=north west][inner sep=0.75pt]    [font=\tiny]  {$1$};
\draw (546,77.4) node [anchor=north west][inner sep=0.75pt]    [font=\tiny]  {$E_{6}$};
\draw (613,189.4) node [anchor=north west][inner sep=0.75pt]    [font=\tiny]  {$5$};

\end{tikzpicture}

\end{center}
\caption{The smooth model for the singular fiber defined by $x^5+t=0$.}
\label{resolve4}
\end{figure}
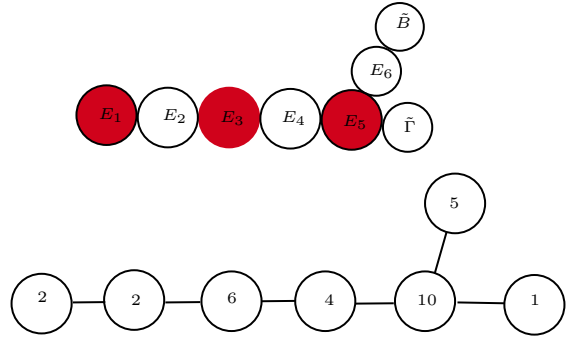

\textbf{Example 4 (Resolution of $x^5 + t$ Branch Locus)}: 
Consider the double cover with branch locus $B$ defined by $x^5 + t = 0$. After performing a sequence of blow-ups at the bad points (see figure. \ref{resolve4}), we obtain the following resolution data:

\begin{align*}
\Gamma &= \widetilde{\Gamma} + E_1 + 2E_2 + 3E_3 + 4E_4 + 5E_5 + 5E_6 \\
B &= \widetilde{B} + E_1 + E_3 + E_5
\end{align*}

where:
\begin{itemize}
    \item $\widetilde{\Gamma}$ is the proper transform of the original fiber
    \item $E_i$ are the exceptional divisors from the blow-up sequence
    \item The coefficients in $\Gamma$ represent the multiplicities
    \item $E_1$, $E_3$, and $E_5$ appear in the branch locus $B$
\end{itemize}
The dual graph for the singular fiber is shown in figure. \ref{resolve4}.

\begin{figure}[H]
\begin{center}

\tikzset{every picture/.style={line width=0.75pt}} 

\begin{tikzpicture}[x=0.45pt,y=0.45pt,yscale=-1,xscale=1]

\draw  [fill={rgb, 255:red, 208; green, 2; blue, 27 }  ,fill opacity=1 ] (305,123) .. controls (305,109.19) and (316.19,98) .. (330,98) .. controls (343.81,98) and (355,109.19) .. (355,123) .. controls (355,136.81) and (343.81,148) .. (330,148) .. controls (316.19,148) and (305,136.81) .. (305,123) -- cycle ;
\draw  [color={rgb, 255:red, 208; green, 2; blue, 27 }  ,draw opacity=1 ][fill={rgb, 255:red, 208; green, 2; blue, 27 }  ,fill opacity=1 ] (407,125) .. controls (407,111.19) and (418.19,100) .. (432,100) .. controls (445.81,100) and (457,111.19) .. (457,125) .. controls (457,138.81) and (445.81,150) .. (432,150) .. controls (418.19,150) and (407,138.81) .. (407,125) -- cycle ;
\draw   (356,125) .. controls (356,111.19) and (367.19,100) .. (381,100) .. controls (394.81,100) and (406,111.19) .. (406,125) .. controls (406,138.81) and (394.81,150) .. (381,150) .. controls (367.19,150) and (356,138.81) .. (356,125) -- cycle ;
\draw  [fill={rgb, 255:red, 255; green, 255; blue, 255 }  ,fill opacity=1 ] (458,126) .. controls (458,112.19) and (469.19,101) .. (483,101) .. controls (496.81,101) and (508,112.19) .. (508,126) .. controls (508,139.81) and (496.81,151) .. (483,151) .. controls (469.19,151) and (458,139.81) .. (458,126) -- cycle ;
\draw  [fill={rgb, 255:red, 208; green, 2; blue, 27 }  ,fill opacity=1 ] (508,126) .. controls (508,112.19) and (519.19,101) .. (533,101) .. controls (546.81,101) and (558,112.19) .. (558,126) .. controls (558,139.81) and (546.81,151) .. (533,151) .. controls (519.19,151) and (508,139.81) .. (508,126) -- cycle ;
\draw   (510,173.22) .. controls (510,161.9) and (519.18,152.72) .. (530.5,152.72) .. controls (541.82,152.72) and (551,161.9) .. (551,173.22) .. controls (551,184.54) and (541.82,193.72) .. (530.5,193.72) .. controls (519.18,193.72) and (510,184.54) .. (510,173.22) -- cycle ;
\draw   (193,337) .. controls (193,323.19) and (204.19,312) .. (218,312) .. controls (231.81,312) and (243,323.19) .. (243,337) .. controls (243,350.81) and (231.81,362) .. (218,362) .. controls (204.19,362) and (193,350.81) .. (193,337) -- cycle ;
\draw   (593,336) .. controls (593,324.95) and (601.95,316) .. (613,316) .. controls (624.05,316) and (633,324.95) .. (633,336) .. controls (633,347.05) and (624.05,356) .. (613,356) .. controls (601.95,356) and (593,347.05) .. (593,336) -- cycle ;
\draw   (270,334) .. controls (270,320.19) and (281.19,309) .. (295,309) .. controls (308.81,309) and (320,320.19) .. (320,334) .. controls (320,347.81) and (308.81,359) .. (295,359) .. controls (281.19,359) and (270,347.81) .. (270,334) -- cycle ;
\draw   (512,335) .. controls (512,321.19) and (523.19,310) .. (537,310) .. controls (550.81,310) and (562,321.19) .. (562,335) .. controls (562,348.81) and (550.81,360) .. (537,360) .. controls (523.19,360) and (512,348.81) .. (512,335) -- cycle ;
\draw   (351,335) .. controls (351,321.19) and (362.19,310) .. (376,310) .. controls (389.81,310) and (401,321.19) .. (401,335) .. controls (401,348.81) and (389.81,360) .. (376,360) .. controls (362.19,360) and (351,348.81) .. (351,335) -- cycle ;
\draw   (429,335) .. controls (429,321.19) and (440.19,310) .. (454,310) .. controls (467.81,310) and (479,321.19) .. (479,335) .. controls (479,348.81) and (467.81,360) .. (454,360) .. controls (440.19,360) and (429,348.81) .. (429,335) -- cycle ;
\draw    (244,336) -- (271,335.72) ;
\draw    (321,336) -- (350,335.72) ;
\draw    (563,337) -- (592,336.72) ;
\draw    (401,335) -- (429,334.72) ;
\draw    (479,337) -- (511,336.72) ;
\draw   (506,213) .. controls (506,201.95) and (514.95,193) .. (526,193) .. controls (537.05,193) and (546,201.95) .. (546,213) .. controls (546,224.05) and (537.05,233) .. (526,233) .. controls (514.95,233) and (506,224.05) .. (506,213) -- cycle ;
\draw   (558,127.72) .. controls (558,115.57) and (567.85,105.72) .. (580,105.72) .. controls (592.15,105.72) and (602,115.57) .. (602,127.72) .. controls (602,139.87) and (592.15,149.72) .. (580,149.72) .. controls (567.85,149.72) and (558,139.87) .. (558,127.72) -- cycle ;
\draw   (663,336) .. controls (663,325.51) and (671.51,317) .. (682,317) .. controls (692.49,317) and (701,325.51) .. (701,336) .. controls (701,346.49) and (692.49,355) .. (682,355) .. controls (671.51,355) and (663,346.49) .. (663,336) -- cycle ;
\draw    (537,391.72) -- (537,360.72) ;
\draw   (649,132.22) .. controls (649,120.9) and (658.18,111.72) .. (669.5,111.72) .. controls (680.82,111.72) and (690,120.9) .. (690,132.22) .. controls (690,143.54) and (680.82,152.72) .. (669.5,152.72) .. controls (658.18,152.72) and (649,143.54) .. (649,132.22) -- cycle ;
\draw  [fill={rgb, 255:red, 208; green, 2; blue, 27 }  ,fill opacity=1 ] (600,130) .. controls (600,116.19) and (611.19,105) .. (625,105) .. controls (638.81,105) and (650,116.19) .. (650,130) .. controls (650,143.81) and (638.81,155) .. (625,155) .. controls (611.19,155) and (600,143.81) .. (600,130) -- cycle ;
\draw   (725,335.5) .. controls (725,325.28) and (733.28,317) .. (743.5,317) .. controls (753.72,317) and (762,325.28) .. (762,335.5) .. controls (762,345.72) and (753.72,354) .. (743.5,354) .. controls (733.28,354) and (725,345.72) .. (725,335.5) -- cycle ;
\draw    (634,336) -- (663,335.72) ;
\draw    (701,337) -- (726,336.72) ;
\draw   (516,411) .. controls (516,399.95) and (524.95,391) .. (536,391) .. controls (547.05,391) and (556,399.95) .. (556,411) .. controls (556,422.05) and (547.05,431) .. (536,431) .. controls (524.95,431) and (516,422.05) .. (516,411) -- cycle ;

\draw (664,118.4) node [anchor=north west][inner sep=0.75pt]   [font=\tiny]   {$\tilde{\Gamma }$};
\draw (374,115.4) node [anchor=north west][inner sep=0.75pt]    [font=\tiny]  {$E_{5}$};
\draw (321,114.4) node [anchor=north west][inner sep=0.75pt]   [font=\tiny]   {$E_{1}$};
\draw (422,117.4) node [anchor=north west][inner sep=0.75pt]   [font=\tiny]   {$E_{3}$};
\draw (521,202.4) node [anchor=north west][inner sep=0.75pt]   [font=\tiny]   {$\tilde{B}$};
\draw (212,324.4) node [anchor=north west][inner sep=0.75pt]   [font=\tiny]   {$2$};
\draw (451,327.4) node [anchor=north west][inner sep=0.75pt]   [font=\tiny]   {$8$};
\draw (290,327.4) node [anchor=north west][inner sep=0.75pt]    [font=\tiny]  {$4$};
\draw (370,325.4) node [anchor=north west][inner sep=0.75pt]    [font=\tiny]  {$6$};
\draw (528,327.4) node [anchor=north west][inner sep=0.75pt]   [font=\tiny]   {$10$};
\draw (522,116.4) node [anchor=north west][inner sep=0.75pt]   [font=\tiny]   {$E_{4}$};
\draw (472,116.4) node [anchor=north west][inner sep=0.75pt]    [font=\tiny]  {$E_{6}$};
\draw (736,330.4) node [anchor=north west][inner sep=0.75pt]   [font=\tiny]   {$1$};
\draw (570,117.4) node [anchor=north west][inner sep=0.75pt]    [font=\tiny]  {$E_{8}$};
\draw (672,330.4) node [anchor=north west][inner sep=0.75pt]    [font=\tiny]  {$4$};
\draw (520,165.4) node [anchor=north west][inner sep=0.75pt]   [font=\tiny]   {$E_{7}$};
\draw (617,119.4) node [anchor=north west][inner sep=0.75pt]   [font=\tiny]   {$E_{2}$};
\draw (607,329.4) node [anchor=north west][inner sep=0.75pt]   [font=\tiny]   {$7$};
\draw (532,401.4) node [anchor=north west][inner sep=0.75pt]    [font=\tiny]  {$5$};

\end{tikzpicture}

\end{center}
\caption{The smooth model for the singular fiber defined by $x^5+t^3=0$.}
\label{resolve5}
\end{figure}
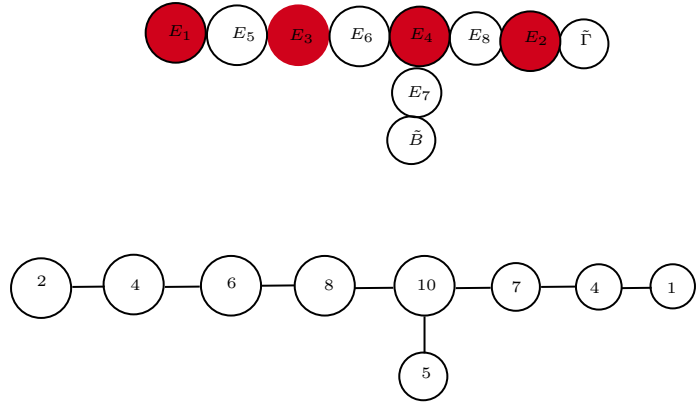

\textbf{Example 5 (Resolution of $x^5 + t^3$ Branch Locus)}: 
Consider the double cover with singular branch locus $B$ defined by $x^5 + t^3 = 0$. The minimal resolution process yields the following divisor configuration (see figure. \ref{resolve5}):

\begin{align*}
\Gamma &= \widetilde{\Gamma} + E_1 + 2E_2 + 3E_3 + 5E_4 + 4E_5 + 8E_6 + 5E_7 + 7E_8 \\
B &= \widetilde{B} + E_1 + E_2 + E_3 + E_4
\end{align*}

where:
\begin{itemize}
    \item $\widetilde{\Gamma}$ is the proper transform of the initial fiber
    \item $E_1,\ldots,E_8$ are exceptional divisors from successive blow-ups
    \item Multiplicities in $\Gamma$ reflect the weighted blow-up sequence
    \item $E_1$ through $E_4$ appear in the transformed branch locus
\end{itemize}
The dual graph for the singular fiber is shown in figure. \ref{resolve5}.

\section{Igusa invariants}

The singular fiber type of a genus one fibration is determined by its discriminant and $j$-invariant. For genus two fibrations, the set of \emph{Igusa invariants} plays an analogous role in classifying singular fibers.
A genus two curve is given by the equation:
\[
y^2 = \sum_{i=0}^6 a_i x^{6-i} = a_0x^6 + a_1x^5 + a_2x^4 + a_3x^3 + a_4x^2 + a_5x + a_6
\]
The Igusa invariants for this curve are defined as follows:
\begin{align*}
j_2 &= \frac{1}{4}(-120a_0a_6 + 20a_1a_5 - 8a_2a_4 + 3a_3^2) \\
j_4 &= \frac{1}{128}\bigl(2640a_0^2a_6^2 - 880a_0a_1a_5a_6 + 1312a_0a_2a_4a_6 \\
    &\quad - 400a_0a_2a_5^2 - 672a_0a_3^2a_6 + 240a_0a_3a_4a_5 - 64a_0a_4^3 \\
    &\quad - 400a_1^2a_4a_6 + 240a_1^2a_5^2 + 240a_1a_2a_3a_6 - 112a_1a_2a_4a_5 \\
    &\quad - 8a_1a_3^2a_5 + 16a_1a_3a_4^2 - 64a_2^3a_6 + 16a_2^2a_3a_5 \\
    &\quad + 16a_2^2a_4^2 - 16a_2a_3^2a_4 + 3a_3^4\bigr) \\
I_4 &= j_2^2 - 24j_4 \\
j_6 &= -\frac{1}{1024}\Bigl(
    4000a_0^2a_3a_5^3 - 1600a_0^2a_4^2a_5^2 - 1600a_0a_1a_2a_5^3 \\
    &\quad - 640a_0a_1a_3a_4a_5^2 + 384a_0a_1a_4^3a_5 + 640a_0a_2^2a_4a_5^2 \\
    &\quad - 80a_0a_2a_3^2a_5^2 - 192a_0a_2a_3a_4^2a_5 + 48a_0a_3^3a_4a_5 \\
    &\quad + 4000a_1^3a_3a_6^2 - 1600a_1^3a_4a_5a_6 + 320a_1^3a_5^3 \\
    &\quad - 1600a_1^2a_2^2a_6^2 - 640a_1^2a_2a_3a_5a_6 + 640a_1^2a_2a_4^2a_6 \\
    &\quad - 64a_1^2a_2a_4a_5^2 - 80a_1^2a_3^2a_4a_6 - 176a_1^2a_3^2a_5^2 \\
    &\quad + 224a_1^2a_3a_4^2a_5 - 64a_1^2a_4^4 + 384a_1a_2^3a_5a_6 \\
    &\quad - 192a_1a_2^2a_3a_4a_6 + 224a_1a_2^2a_3a_5^2 - 128a_1a_2^2a_4^2a_5 \\
    &\quad + 48a_1a_2a_3^3a_6 - 112a_1a_2a_3^2a_4a_5 + 64a_1a_2a_3a_4^3 \\
    &\quad + 28a_1a_3^4a_5 - 16a_1a_3^3a_4^2 - 64a_2^4a_5^2 \\
    &\quad + 64a_2^3a_3a_4a_5 - 16a_2^2a_3^3a_5 - 16a_2^2a_3^2a_4^2 \\
    &\quad + 8a_2a_3^4a_4 - a_3^6
\Bigr) \\
&\quad + 5a_0^3a_6^3 - \frac{159}{64}a_0^2a_6^2a_3^2 + \frac{39}{128}a_0a_3^4a_6 \\
&\quad - 4a_0a_6(a_0a_4^3 + a_2^3a_6) - a_0a_2a_4a_6(14a_0a_6 - a_2a_4) \\
&\quad - \frac{5}{16}a_0a_1a_5a_6(8a_0a_6 + 7a_1a_5) + \frac{25}{4}a_0a_6(a_0a_2a_5^2 + a_1^2a_4a_6) \\
&\quad - \frac{51}{8}a_0a_1a_2a_4a_5a_6 + \frac{165}{16}a_0a_3a_6(a_0a_4a_5 + a_1a_2a_6) \\
&\quad - \frac{49}{32}a_0a_2a_3^2a_4a_6 + \frac{41}{16}a_0a_3a_6(a_1a_4^2 + a_2^2a_5) \\
&\quad - \frac{277}{64}a_0a_1a_3^2a_5a_6 \\
j_8 &= \frac{1}{4}(j_2j_6 - j_4^2) \\
j_{10} &= \begin{cases}
          \ \Delta/2^{12} & \text{if } a_0 \neq 0 \\
          a_1^2\Delta/2^{12} & \text{otherwise}
         \end{cases} \\
I_{12} &= -8j_4^3 + 9j_2j_4j_6 - 27j_6^2 - j_2^2j_8
\end{align*}
The singular fiber type can be determined by using the pole and zero orders of the Igusa invariants \cite{liu1993courbes,liu1994conducteur}.

\section{Holomorphic Differentials for Hyperelliptic Curves}
\label{holodifferential}

Let $C$ be a genus two hyperelliptic curve defined by $y^2 = f(x)$, which represents a branched double covering $\pi: C \to \mathbb{P}^1$ of the Riemann sphere. Here $x$ and $y$ are meromorphic functions on $C$ whose poles and zeros are characterized as follows.

\textbf{Case 1: Six Distinct Branch Points}
Consider the curve $y^2 = \prod_{i=1}^6 (x-a_i)$ with $a_i \neq a_j$ for $i \neq j$. The covering has:
\begin{itemize}
    \item Six branch points $P_i = \pi^{-1}(a_i) \in C$ (zeros of $y$)
    \item Two points at infinity $Q_1, Q_2 = \pi^{-1}(\infty)$
    \item Two preimages of zero $M_1, M_2 = \pi^{-1}(0)$
\end{itemize}

The divisors are given by:
\begin{align*}
(y) &= \frac{P_1P_2\cdots P_6}{Q_1^3 Q_2^3}, \\
(x) &= \frac{M_1 M_2}{Q_1 Q_2}, \\
(dx) &= \frac{P_1 P_2\cdots P_6}{Q_1^2 Q_2^2}
\end{align*}

The holomorphic differentials are precisely:
\[
\left\{ \frac{dx}{y}, \frac{x\,dx}{y} \right\}
\]

\textbf{Case 2: Five Distinct Branch Points}
For the curve $y^2 = \prod_{i=1}^5 (x-a_i)$ with $a_i \neq a_j$, we have:
\begin{itemize}
    \item Five branch points $P_i = \pi^{-1}(a_i)$
    \item One point at infinity $Q = \pi^{-1}(\infty)$ (since $\deg f(x) = 5$)
    \item Two preimages of zero $M_1, M_2 = \pi^{-1}(0)$
\end{itemize}

The corresponding divisors are:
\begin{align*}
(y) &= \frac{P_1P_2\cdots P_5}{Q^5}, \\
(x) &= \frac{M_1 M_2}{Q^2}, \\
(dx) &= \frac{P_1P_2\cdots P_5}{Q^3}
\end{align*}

The holomorphic differentials are precisely:
\[
\left\{ \frac{dx}{y}, \frac{x\,dx}{y} \right\}
\]

One can follow the similar computation to conclude that the basis of holomorphic differential for a hyperellitpic 
curve of genus $g$ is 
The holomorphic differentials are precisely:
\[
\left\{ \frac{dx}{y}, \frac{x\,dx}{y}, \ldots,~ \frac{x^{g-1}\,dx}{y}\right\}
\]

\bibliographystyle{utphys}

\bibliography{ADhigher}

\end{document}